\newif\ifsubmode
\submodefalse

\newif\ifprintfig
\printfigtrue

\ifsubmode
  \documentstyle[12pt,aasms4,epsf]{article}
  \received{}
  \accepted{}
  \journalid{}{}
  \articleid{}{}
\else
  \documentstyle[11pt,aaspp4,epsf]{article}
  \slugcomment{{\it Astronomical Journal, submitted.}}
\fi

\lefthead{van der Marel et al.}
\righthead{Large Magellanic Cloud Structure}

\newcommand{\etal}{{et al.~}}

\newcommand{\lta}{\lesssim}
\newcommand{\gta}{\gtrsim}

\newcommand{\masyr}{\>{\rm mas}\,{\rm yr}^{-1}}
\newcommand{\kms}{\>{\rm km}\,{\rm s}^{-1}}
\newcommand{\Gyr}{\>{\rm Gyr}}

\newcommand{\kpc}{\>{\rm kpc}}

\newcommand{\Msun}{\>{\rm M_{\odot}}}
\newcommand{\Lsun}{\>{\rm L_{\odot}}}

\newcommand{\chour}{^{\rm h}\>}
\newcommand{\cmin}{^{\rm m}\>}

\begin{document}

\title{New Understanding of Large Magellanic Cloud Structure, 
Dynamics and Orbit from Carbon Star Kinematics}

\author{Roeland P.~van der Marel}
\affil{Space Telescope Science Institute, 3700 San Martin Drive, 
       Baltimore, MD 21218}

\author{David R.~Alves}
\affil{Columbia Astrophysics Laboratory, New York, NY 10027}

\author{Eduardo Hardy}
\affil{National Radio Astronomy Observatory, Casilla 36-D, Santiago, Chile}

\author{Nicholas B.~Suntzeff}
\affil{Cerro-Tololo Inter-American Observatory, Casilla 603, La Serena, Chile}



\ifsubmode\else
\clearpage\fi


\ifsubmode\else
\baselineskip=14pt
\fi


\begin{abstract}
We formulate a new, revised and coherent understanding of the
structure and dynamics of the Large Magellanic Cloud (LMC), and its
orbit around and interaction with the Milky Way. Much of our
understanding of these issues hinges on studies of the LMC
line-of-sight kinematics. The observed velocity field includes
contributions from the LMC rotation curve $V(R')$, the LMC transverse
velocity vector ${\vec v}_t$, and the rate of inclination change
$di/dt$. All previous studies have assumed $di/dt = 0$. We show that
this is incorrect, and that combined with uncertainties in ${\vec
v}_t$ this has led to incorrect estimates of many important structural
parameters of the LMC. We derive general expressions for the velocity
field which we fit to kinematical data for 1041 carbon stars. We
calculate ${\vec v}_t$ by compiling and improving LMC proper motion
measurements from the literature, and we show that for known ${\vec
v}_t$ all other model parameters are uniquely determined by the data.
The position angle of the line of nodes is $\Theta = 129.9^{\circ} \pm
6.0^{\circ}$, consistent with the value determined geometrically by
van der Marel \& Cioni (2001). The rate of inclination change is
$di/dt = -0.37 \pm 0.22 \masyr = -103 \pm 61$ degrees/Gyr. This is
similar in magnitude to predictions from $N$-body simulations by
Weinberg (2000), which predict LMC disk precession and nutation due to
Milky Way tidal torques. The LMC rotation curve $V(R')$ has amplitude
$49.8 \pm 15.9 \kms$. This is $40$\% lower than what has previously
(and incorrectly) been inferred from studies of HI, carbon stars, and
other tracers. The line-of-sight velocity dispersion has an average
value $\sigma = 20.2 \pm 0.5 \kms$, with little variation as function
of radius. The dynamical center of the carbon stars is consistent with
the center of the bar and the center of the outer isophotes, but it is
offset by $1.2^{\circ} \pm 0.6^{\circ}$ from the kinematical center of
the HI. The enclosed mass inside the last data point is $M_{\rm LMC}
(8.9 \kpc) = (8.7 \pm 4.3) \times 10^9 \Msun$, more than half of which
is due to a dark halo. The LMC has a considerable vertical thickness;
its $V/\sigma = 2.9 \pm 0.9$ is less than the value for the Milky
Way's {\it thick} disk ($V/\sigma \approx 3.9$). Simple arguments for
models stratified on spheroids indicate that the (out-of-plane) axial
ratio could be $\sim 0.3$ or larger. Isothermal disk models for the
observed velocity dispersion profile confirm the finding of Alves \&
Nelson (2000) that the scale height must increase with radius. A
substantial thickness for the LMC disk is consistent with the
simulations of Weinberg, which predict LMC disk thickening due to
Milky Way tidal forces. These affect LMC structure even inside the LMC
tidal radius, which we calculate to be $r_t = 15.0 \pm 4.5
\kpc$ (i.e., $17.1^{\circ} \pm 5.1^{\circ}$). The new insights into
LMC structure need not significantly alter existing predictions for
the LMC self-lensing optical depth, which to lowest order depends only
on $\sigma$. The compiled proper motion data imply an LMC transverse
velocity $v_t = 406 \kms$ in the direction of position angle
$78.7^{\circ}$ (with errors of $\sim 40 \kms$ in each
coordinate). This can be combined with the observed systemic velocity,
$v_{\rm sys} = 262.2 \pm 3.4 \kms$, to calculate the LMC velocity in
the Galactocentric rest frame. This yields $v_{\rm LMC} = 293 \pm 39
\kms$, with radial and tangential components $v_{\rm LMC,rad} = 84 \pm
7\kms$ and $v_{\rm LMC,tan} = 281 \pm 41 \kms$, respectively. This is
consistent with the range of velocities that has been predicted by
models for the Magellanic Stream. The implied orbit of the LMC has an
apocenter to pericenter distance ratio $\sim 2.5:1$, a perigalactic
distance $\sim 45 \kpc$, and a present orbital period around the Milky
Way $\sim 1.5 \Gyr$. The constraint that the LMC is bound to the Milky
Way provides a robust limit on the minimum mass and extent of the
Milky Way dark halo: $M_{\rm MW} \geq 4.3 \times 10^{11} \Msun$ and
$r_h \geq 39 \kpc$ (68.3\% confidence). Finally, we present
predictions for the LMC proper motion velocity field, and we discuss
how measurements of this may lead to kinematical distance estimates of
the LMC.
\end{abstract}



\clearpage


\section{Introduction}
\label{s:intro}

The Large and Small Magellanic Clouds (LMC and SMC) are close
companion galaxies of our Milky Way. Their interaction with the Milky
Way has produced the Magellanic Stream, which spans more than
$100^{\circ}$ across the sky (e.g., Westerlund 1997; Wakker 2002). It
consists of gas that trails the Magellanic Clouds as they orbit the
Milky Way. A less prominent leading gas component was recently
discovered as well (Lu \etal 1998; Putman 1998). Models of the
Magellanic Stream hold the promise of providing important constraints
on the mass and gravitational potential of the Milky Way dark halo
(e.g., Lin, Jones \& Klemola 1995). However, this requires an accurate
understanding of the three-dimensional velocity of the Magellanic
Clouds. While the systemic line-of-sight velocities of both the LMC
and SMC have been accurately determined using Doppler shifts, a
determination of their transverse velocities in the plane of the sky
has proven to be much more difficult. We restrict our attention in the
present paper to the LMC, which has been studied more and is
understood better than the SMC. Several proper motion studies of the
LMC have been published (Jones, Klemola \& Lin 1994; Kroupa, R\"oser
\& Bastian 1994; Kroupa \& Bastian 1997; Drake \etal 2002; Pedreros, 
Anguita \& Maza 2002), but the uncertainties in the measurements are
considerable. Accurate knowledge of the LMC transverse velocity has
therefore remained elusive.

The uncertainty in the transverse velocity of the LMC has not only
restricted models of the Magellanic Stream, but also the
interpretation of the observed LMC line-of-sight velocity field. As
one moves away from the LMC center, the transverse velocity component
of the LMC center of mass (CM) ceases to be perpendicular to the line
of sight. This causes a spurious solid-body rotation component in the
observed line-of-sight velocity field (Feast, Thackeray \& Wesselink
1961). Initially this effect was used to obtain a rough estimate of
the transverse velocity of the LMC under the assumption (appropriate
for a circular disk) that its kinematic line of nodes (defined as the
line of maximum velocity gradient) should be equal to the photometric
major axis (Feitzinger, Schmidt-Kaler \& Isserstedt 1977;
Meatheringham \etal 1988; Gould 2000). However, this assumption is now
known to be invalid. Robust geometric methods show that the line of
nodes of the LMC (defined as the intersection of the galaxy-plane and
the sky-plane) does not coincide with the photometric major axis (van
der Marel \& Cioni 2001, hereafter Paper~I), and hence, that the LMC
is intrinsically elongated (van der Marel 2001, hereafter
Paper~II). The most recent analyses of the LMC velocity field have
corrected for the spurious solid-body rotation component using the
available LMC proper motion measurements (e.g., Kim \etal 1998; Alves
\& Nelson 2000). However, the corrections involved are large (because
the spurious component is comparable to the intrinsic LMC rotation
component) and highly uncertain (because the LMC proper motion is not
known accurately). In addition, all previous studies have assumed that
the LMC viewing angles have no explicit time dependence, contrary to
the results of $N$-body simulations which predict precession and
nutation of the LMC disk due to Milky Way tidal torques (Weinberg
2000). All this probably explains why analyses of the LMC velocity
field have yielded somewhat puzzling results. For example, Alves \&
Nelson (2000) obtained from a study of carbon-star velocities that the
position angle of the kinematic line of nodes rises from $\Theta_{\rm
kin} = 143^{\circ} \pm 7^{\circ}$ at a mean distance of $4.5^{\circ}$
from the galaxy center to $\Theta_{\rm max} = 183^{\circ} \pm
8^{\circ}$ at $9.2^{\circ}$. The results obtained from HI kinematics
by Kim \etal (1998) were qualitatively not very different. A kinematic
twist of $40^{\circ} \pm 11^{\circ}$ is quite unexpected, given that,
over the same range of distances, the LMC has no twist in the stellar
number density contours (to an accuracy of $\sim 3^{\circ}$;
Paper~II). More importantly, the inferred kinematic line of nodes
differs considerably from the true line of nodes, for which a position
angle $\Theta = 122.5 \pm 8.3^{\circ}$ was determined geometrically
from DENIS and 2MASS near-IR stellar catalogs (Paper~I). While the LMC
is not intrinsically circular, the implied difference between $\Theta$
and $\Theta_{\rm kin}$ is too large to plausibly attribute to
non-circular orbits (Paper~II).

In the present paper we address these issues, and their consequences,
by presenting the most detailed and sophisticated analysis to date of
the LMC line-of-sight velocity field. In Section~\ref{s:vmath} we
derive general expressions for the velocity fields of a rotating disk
galaxy that has a non-negligible angular extent on the sky, both for
the line-of-sight velocity component and for the velocity components
in the plane of the sky. In Section~\ref{s:vfieldconcept} we present a
conceptual discussion of the line-of-sight velocity field. We address
which quantities are uniquely constrained by the observations, and
which ones are not. In Section~\ref{s:data} we fit the general
equations for the line-of-sight velocity field to the velocities
available for 1041 carbon stars, and we discuss the results. In
Section~\ref{s:litreview} we review what is known about the proper
motion and distance of the LMC from other sources. In
Section~\ref{s:distdidt} we combine this knowledge with the
information obtained from the analysis of the line-of-sight velocity
field to determine $di/dt$, the rate at which the LMC inclination
changes with time. In Section~\ref{s:kindisk} we discuss the
kinematical properties of the LMC disk as implied by our analysis. We
address the position of the dynamical center, the disk rotation curve
and the velocity dispersion profile, the position angle of the line of
nodes, and the influence of non-circular orbits on the results of the
analysis. In Section~\ref{s:structure} we discuss the implications of
the results for our understanding of the structure of the LMC,
including its mass, tidal radius, scale height, dark halo and
self-lensing optical depth. In Section~\ref{s:disc} we discuss the
implications of the results for the transverse velocity of the LMC,
the three-dimensional velocity of the LMC in the Galactocentric rest
frame, the orbit of the LMC around the Milky Way, the mass and extent
of the Milky Way, and the origin of the Magellanic Stream. In
Section~\ref{s:propfield} we present predictions for the proper motion
velocity field of the LMC, which may be observationally accessible
with future astrometric missions. In Section~\ref{s:distprospects} we
discuss a new method for kinematical determination of the LMC
distance, and its prospects for yielding results that are competitive
with other methods. Conclusions are presented in
Section~\ref{s:conc}. Appendix~\ref{s:app} discusses an improved
analysis of the published proper measurements of Jones
\etal (1994) and Pedreros \etal (2002), which were obtained for fields
that are offset from the LMC CM.

\section{General Expressions for the Velocity Field of a Rotating Disk}
\label{s:vmath}

Many previous studies have given equations for the description of the
velocity field of an external galaxy. However, it is customary to make
the assumption that the angular size subtended by the galaxy is
negligible. This is equivalent to assuming that the galaxy is at
infinite distance and that `the sky is flat' over the area of the
galaxy. However, the LMC subtends a very large angle on the sky (more
than $20^{\circ}$ end-to-end) and this makes the usual equations
inadequate for describing its kinematics. We therefore start by
deriving general expressions for the velocity field of a rotating disk
which are valid for a galaxy of arbitrary angular size. We address not
only the line-of-sight velocities, but also the velocity components
perpendicular to the line of sight. Readers interested mostly in the
application to the LMC may wish to skip directly to
Section~\ref{s:vfieldconcept}.
 
\subsection{Coordinate Systems and Velocity Components}
\label{ss:vmathobs}

We adopt the same coordinate systems as in Paper~I. The position of
any point in space is uniquely determined by its right ascension and
declination on the sky, $(\alpha,\delta)$, and its distance $D$. The
point ${\cal O}$ with coordinates $(\alpha_0,\delta_0,D_0)$ is the
origin of the analysis, and is chosen to be the galaxy center of mass
(CM). Angular coordinates $(\rho,\phi)$ are defined on the celestial
sphere, where $\rho$ is the angular distance between the points
$(\alpha,\delta)$ and $(\alpha_0,\delta_0)$, and $\phi$ is the
position angle of the point $(\alpha,\delta)$ with respect to
$(\alpha_0,\delta_0)$; see Figure~\ref{f:drawskyview}. In particular,
$\phi$ is the angle at $(\alpha_0,\delta_0)$ between the tangent to
the great circle on the celestial sphere through $(\alpha,\delta)$ and
$(\alpha_0,\delta_0)$, and the circle of constant declination
$\delta_0$. By convention, $\phi$ is measured counterclockwise
starting from the axis that runs in the direction of decreasing right
ascension at constant declination $\delta_0$. Equations~(1)--(3) of
Paper~I allow $(\rho,\phi)$ to be calculated for any
$(\alpha,\delta)$.

At any given position $(D,\rho,\phi)$, a velocity vector can be
decomposed into a sum of three orthogonal components:
\begin{equation}
\label{vonetwothreedef}
  v_1 \equiv {{dD}\over{dt}} , \qquad
  v_2 \equiv D {{d\rho}\over{dt}} , \qquad
  v_3 \equiv D \sin \rho {{d\phi}\over{dt}} .
\end{equation}
Here, $v_1$ is the line-of-sight velocity, and $v_2$ and $v_3$ are the
velocity components in the plane of the sky; see
Figure~\ref{f:drawsideview}. The goal of the present analysis is to
obtain expressions for $(v_1,v_2,v_3)$ as a function of $(\rho,\phi)$.

We introduce a Cartesian coordinate system $(x,y,z)$ that has its
origin at ${\cal O}$, with the $x$-axis anti-parallel to the RA axis,
the $y$-axis parallel to the declination axis, and the $z$-axis
towards the observer; see Figures~\ref{f:drawskyview}
and~\ref{f:drawsideview}. The transformation equations from
$(D,\rho,\phi)$ to $(x,y,z)$ are given by equation~(5) of Paper~I. The
inverse transformation equations are:
\begin{eqnarray}
\label{rhophiDdef}
   D    & = & [x^2 + y^2 + (D_0 - z)^2]^{1/2} , \nonumber \\
   \rho & = & \arctan [ (x^2 + y^2) / (D_0-z)^2 ]^{1/2} , \\
   \phi & = & \arctan(y/x) . \nonumber
\end{eqnarray}
Substitution of equation~(\ref{rhophiDdef}) in
equation~(\ref{vonetwothreedef}) yields after some manipulations
\begin{equation}
\label{vxyz}
  \left ( \begin{array}{c} v_1 \\ v_2 \\ v_3 
  \end{array} \right ) 
     =
  \left ( \begin{array}{ccc}
     \sin \rho \cos \phi & \sin \rho \sin \phi & -\cos \rho \\
     \cos \rho \cos \phi & \cos \rho \sin \phi & \sin \rho \\
     -\sin \phi & \cos \phi & 0 
  \end{array} \right )
     \>
  \left ( \begin{array}{c} v_x \\ v_y \\ v_z
  \end{array} \right ) ,
\end{equation}
where $(v_x,v_y,v_z)$ is the three-dimensional velocity in the
$(x,y,z)$ coordinate system. To describe the internal kinematics of
the galaxy it is useful to adopt a second Cartesian coordinate system
$(x',y',z')$ that is obtained from the system $(x,y,z)$ by
counterclockwise rotation around the $z$-axis by an angle $\theta$,
followed by a clockwise rotation around the new $x'$-axis by an angle
$i$; see Figure~\ref{f:drawviewang}. With this definition, the
$(x',y')$ plane is inclined with respect to the sky by the angle $i$
(with face-on viewing corresponding to $i=0$). The angle $\theta$ is
the position angle of the line of nodes (the intersection of the
$(x',y')$-plane and the $(x,y)$-plane of the sky), measured
counterclockwise from the $x$-axis. In practice, $i$ and $\theta$ will
be chosen such that the $(x',y')$ plane coincides with the equatorial
plane of the galaxy. The transformation equations from $(x,y,z)$ to
$(x',y',z')$ are given by equation~(6) of Paper~I. The inverse
transformations are
\begin{equation}
\label{xyzprimetransf}
  \left ( \begin{array}{c} x \\ y \\ z
  \end{array} \right )
     =
  \left ( \begin{array}{ccc}
     \cos \theta & - \sin \theta \cos i & - \sin \theta \sin i \\
     \sin \theta &   \cos \theta \cos i &   \cos \theta \sin i \\
     0           & - \sin i             &   \cos i \\  
  \end{array} \right )
     \>
  \left ( \begin{array}{c} x' \\ y' \\ z'
  \end{array} \right ) .
\end{equation}
Upon taking the time derivative on both sides this yields
transformation equations from $(v_x',v_y',v_z')$ to $(v_x,v_y,v_z)$.
These can be substituted in equation~(\ref{vxyz}), which after some
further manipulations yields
{\scriptsize
\begin{equation}
\label{vxyzprimeone}
  \left ( \begin{array}{c} v_1 \\ v_2 \\ v_3
  \end{array} \right )
     = 
  \left ( \begin{array}{ccc}
     \sin \rho \cos (\phi-\theta) & 
         [\sin \rho \cos i \sin (\phi-\theta) + \cos \rho \sin i ] &
         [\sin \rho \sin i \sin (\phi-\theta) - \cos \rho \cos i ] \\
     \cos \rho \cos (\phi-\theta) & 
         [\cos \rho \cos i \sin (\phi-\theta) - \sin \rho \sin i ] &
         [\cos \rho \sin i \sin (\phi-\theta) + \sin \rho \cos i ] \\
     -\sin (\phi-\theta) &
         \cos i \cos (\phi-\theta) &
         \sin i \cos (\phi-\theta) \\
  \end{array} \right )
     \>
  \left ( \begin{array}{c} v_x' \\ v_y' \\ v_z'
  \end{array} \right ) . 
\end{equation}
}

For the analysis of the line-of-sight velocity field one needs only
$v_1$, but to interpret proper motion measurements one needs to know
in addition how the velocities $v_2$ and $v_3$ relate to the local
directions of West and North. The proper motions in these directions
are defined as
\begin{equation}
\label{propdef}
   \mu_W = - \cos \delta ({{d\alpha}/{dt}}) , \qquad
   \mu_N = ({{d\delta}/{dt}}) .
\end{equation}
To obtain expressions for these proper motions we start with
equations~(1)--(3) of Paper~I, which relate $(\rho,\phi)$ to the right
ascension and declination $(\alpha,\delta)$ on the sky. Upon taking
the time derivative of these equations one obtains relations between
${{d\rho}/{dt}}$ and ${{d\phi}/{dt}}$ on the one hand, and
${{d\alpha}/{dt}}$ and ${{d\delta}/{dt}}$ on the other hand. These
can be solved to obtain:
\begin{equation}
\label{propmone}
  \left ( \begin{array}{c} \mu_W \\ \mu_N 
  \end{array} \right )
     =
  (1/D) \times 
  \left ( \begin{array}{cc}
     - \sin \Gamma & - \cos \Gamma \\
       \cos \Gamma & - \sin \Gamma \\
  \end{array} \right )
     \>
  \left ( \begin{array}{c} v_2 \\ v_3
  \end{array} \right ) .
\end{equation}
The angle $\Gamma$ determines the rotation angle of the $(v_2,v_3)$
frame on the sky\footnote{The angle $\Gamma$ is close to the position
angle on the sky, $\Phi \equiv \phi - 90^{\circ}$ (see
Section~\ref{ss:vmathposang} below), but is not identical to it. The
two are explicitly related to each other through the relation $\sin
\Gamma = (\cos \delta_0 / \cos \delta) \sin \Phi$.}; see 
Figure~\ref{f:drawskyview}. It is determined by
\begin{eqnarray}
\label{Gammadef}
   \cos \Gamma & = & [\sin \delta \cos \delta_0 \cos (\alpha-\alpha_0) -
                      \cos \delta \sin \delta_0] \> 
                      / \sin \rho , \nonumber \\    
   \sin \Gamma & = & [\cos \delta_0 \sin (\alpha-\alpha_0)] \> / \sin \rho .
\end{eqnarray}
In practice we restrict the discussion to a planar system with all
tracers in the $(x',y')$ plane. In this case the distance $D$ to a
tracer is related to the CM distance $D_0$ through equation~(8) of
Paper~I. Substitution of that equation yields
\begin{equation}
\label{propmtwo}
  \left ( \begin{array}{c} \mu_W \\ \mu_N 
  \end{array} \right )
     = { {[\cos i \cos \rho - \sin i \sin \rho \sin (\phi-\theta)]} 
         \over {D_0 \cos i} } \times
  \left ( \begin{array}{cc}
     - \sin \Gamma & - \cos \Gamma \\
       \cos \Gamma & - \sin \Gamma \\
  \end{array} \right )
     \>
  \left ( \begin{array}{c} v_2 \\ v_3
  \end{array} \right ) .
\end{equation}
In practice one needs to express this in observable units, which is
achieved by using the relation 
\begin{equation}
\label{pmtovel}
  v_i / D_0 = \left \lbrace (v_i / [\kms]) \> / \> 
                      (4.7403885 \> D_0 / \kpc) 
              \right \rbrace {\rm mas/yr} .
\end{equation}

\subsection{Velocity Field Model}
\label{ss:vmathmodel}

For a planar system, the velocity of a tracer can be written as a sum
of three components:
\begin{equation}
\label{vectot}
  \left ( \begin{array}{c} v_1 \\ v_2 \\ v_3
  \end{array} \right )
     =
  \left ( \begin{array}{c} v_1 \\ v_2 \\ v_3
  \end{array} \right )_{{\rm CM}}
     +
  \left ( \begin{array}{c} v_1 \\ v_2 \\ v_3
  \end{array} \right )_{{\rm pn}}
     +
  \left ( \begin{array}{c} v_1 \\ v_2 \\ v_3
  \end{array} \right )_{{\rm int}} .
\end{equation}
The first component, ${\vec v}_{{\rm CM}}$, is the velocity
contributed by the space motion of the CM. The second component,
${\vec v}_{{\rm pn}}$, is the velocity contributed by time-variations
in the viewing angles $i$ and $\theta$ of the disk plane. Such
variations arise if the orientation of the disk plane is not static in
an inertial frame. For the case of the LMC, this occurs naturally as a
result of the precession and nutation of the disk plane induced by
external tidal torques (Weinberg 2000). The third component, ${\vec
v}_{{\rm int}}$, is the velocity contributed by the internal motion of
the tracer in the disk plane.  We proceed by deriving expressions for
each of these contributions.

\subsubsection{Center-of-Mass Motion}
\label{sss:vmathmodelCM}

The velocity of the center of mass can be described by quantities
$v_{\rm sys}$, $v_t$ and $\theta_t$, such that $v_{\rm sys}$ is the
systemic velocity of the CM along the line of sight (positive when
receding), $v_t$ is the transverse velocity of the CM, and $\theta_t$
is the direction of the transverse motion on the sky; see
Figure~\ref{f:drawvel}. This implies
\begin{equation}
\label{vCMdef}
  \left ( \begin{array}{c} v_x \\ v_y \\ v_z
  \end{array} \right )_{{\rm CM}}
     =
  \left ( \begin{array}{c}
     v_t \cos \theta_t \\   
     v_t \sin \theta_t \\
     - v_{\rm sys} 
  \end{array} \right ) .
\end{equation}
From equations~(\ref{vxyz}) and~(\ref{vCMdef}) one obtains
for the observable velocity components
\begin{equation}
\label{vecCM}
  \left ( \begin{array}{c} v_1 \\ v_2 \\ v_3
  \end{array} \right )_{{\rm CM}}
     =
  \left ( \begin{array}{c} 
     v_t \sin \rho \cos (\phi -\theta_t) + v_{\rm sys} \cos \rho \\
     v_t \cos \rho \cos (\phi -\theta_t) - v_{\rm sys} \sin \rho \\
    -v_t \sin (\phi -\theta_t) 
  \end{array} \right ) .
\end{equation}

\subsubsection{Precession and Nutation}
\label{sss:vmathmodelpn}

To assess the effect of changes in $i$ and $\theta$, consider a point
that has fixed coordinates in the $(x',y',z')$ frame. By taking the
time derivative of equation~(\ref{xyzprimetransf}) one obtains:
\begin{eqnarray}
\label{dangles}
  \left ( \begin{array}{c} v_x \\ v_y \\ v_z
          \end{array} \right )
     = & (d\theta/dt) &
          \left ( \begin{array}{ccc}
            - \sin \theta & - \cos \theta \cos i & - \cos \theta \sin i \\
              \cos \theta & - \sin \theta \cos i & - \sin \theta \sin i \\
              0           & 0                    & - \sin \theta \\
                  \end{array} \right )     \>
          \left ( \begin{array}{c} x' \\ y' \\ z'
                  \end{array} \right ) + \nonumber \\
       & (di/dt) &
          \left ( \begin{array}{ccc}
              0 &   \sin \theta \sin i & - \sin \theta \cos i \\
              0 & - \cos \theta \sin i &   \cos \theta \cos i \\
              0 & - \cos i             &   0 \\
                  \end{array} \right )     \>
          \left ( \begin{array}{c} x' \\ y' \\ z'
                  \end{array} \right ) .
\end{eqnarray}
For a planar disk ($z'=0$), one has from equation~(2) of Paper~II:
\begin{equation}
\label{xpypzp}
  \left ( \begin{array}{c} x' \\ y' \\ z'
          \end{array} \right ) = 
  {{D_0 \sin \rho} \over 
   {\cos i \cos \rho - \sin i \sin \rho \sin (\phi-\theta)}}
  \left ( \begin{array}{c} 
             \cos i \cos (\phi -\theta) \\ 
             \sin (\phi -\theta) \\ 0
          \end{array} \right ) .
\end{equation} 
These equations can be substituted in equation~(\ref{vxyz}), which
yields after some manipulations
{\scriptsize
\begin{equation}
\label{vecpn}
  \left ( \begin{array}{c} v_1 \\ v_2 \\ v_3
  \end{array} \right )_{{\rm pn}}
     =
  {{D_0 \sin \rho} \over
   {\cos i \cos \rho - \sin i \sin \rho \sin (\phi-\theta)}}
  \left ( \begin{array}{c} 
          (di/dt) \sin (\phi-\theta) [\cos i \cos \rho -
                   \sin i \sin \rho \sin (\phi-\theta)] \\
          (di/dt) \sin (\phi-\theta) [-\cos i \sin \rho -
                   \sin i \cos \rho \sin (\phi-\theta)] \\
          (di/dt) \sin (\phi-\theta) [-\sin i \cos (\phi-\theta)] + 
          (d\theta/dt) [\cos i] 
  \end{array} \right ) .
\end{equation}
}
For future reference it is useful to note that the line-of-sight
component of this vector simplifies to
\begin{equation}
\label{vonepn}
  v_{1,{\rm pn}} = D_0 \> (di/dt) \> \sin \rho \> \sin (\phi-\theta) .
\end{equation}
Note that the quantity $d\theta/dt$ does not affect the line-of-sight
velocity; it does affect the predicted proper motions, as discussed in
Section~\ref{s:propfield} below.

\subsubsection{Internal Rotation}
\label{sss:vmathmodelint}

We consider the case in which the mean streaming (i.e., the rotation)
in the disk plane can be approximated as being circular. This is for
simplicity only; there is no reason why the streamlines couldn't be
non-circular, and why this couldn't potentially be
important. Section~\ref{ss:noncirc} discusses the extent to which this
may influence our results for the LMC. For circular motion one has
\begin{equation}
\label{vprimeone}
  \left ( \begin{array}{c} v_x' \\ v_y' \\ v_z' 
  \end{array} \right )
     =
  \left ( \begin{array}{c}
     s V(R') y' / R' \\ 
    -s V(R') x' / R' \\
     0
  \end{array} \right ) ,
\end{equation}   
where $R'$ is the polar radius in the $(x',y')$ plane, $V(R')$ is the
mean streaming velocity at radius $R'$, and $s = \pm 1$ is the `spin
sign' that determines in which of the two possible directions the disk
rotates. With help of equation~(\ref{xpypzp}) one obtains
\begin{equation}
\label{Racc}
  R' = { {D_0 \sin \rho \> [\cos^2 i \cos^2 (\phi-\theta) +
                                 \sin^2 (\phi-\theta)]^{1/2}} \over
         {\cos i \cos \rho -
                     \sin i \sin \rho \sin (\phi-\theta)} } 
\end{equation}
and
\begin{equation}
\label{vprime}
  \left ( \begin{array}{c} v_x' \\ v_y' \\ v_z'
  \end{array} \right )
     = { {s V(R')} \over {[\cos^2 i \cos^2 (\phi-\theta) +
                                 \sin^2 (\phi-\theta)]^{1/2}} }
  \times
  \left ( \begin{array}{c}
     \sin (\phi-\theta) \\
    -\cos i \cos (\phi-\theta) \\
     0
  \end{array} \right ) .
\end{equation}
Substitution in equation~(\ref{vxyzprimeone}) yields 
{\scriptsize
\begin{equation}
\label{vecint}
  \left ( \begin{array}{c} v_1 \\ v_2 \\ v_3
  \end{array} \right )_{{\rm int}}
     = { {s V(R')} \over {[\cos^2 i \cos^2 (\phi-\theta) +
                                 \sin^2 (\phi-\theta)]^{1/2}} }
  \times
  \left ( \begin{array}{c}
      - \sin i \cos (\phi-\theta)
          [\cos i \cos \rho -
           \sin i \sin \rho \sin (\phi-\theta)] \\
        \sin i \cos (\phi-\theta)
          [\cos i \sin \rho +
           \sin i \cos \rho \sin (\phi-\theta)] \\
      -   [\cos^2 i \cos^2 (\phi-\theta) +
           \sin^2 (\phi-\theta)]
  \end{array} \right ) .
\end{equation}
}

\subsection{Position Angles}
\label{ss:vmathposang}

In practice it is useful to employ positions angles $\Phi$, $\Theta$
and $\Theta_t$ that are measured from North over East in the usual
astronomical convention:
\begin{equation}
\label{posangfromnorth}
  \Phi     \equiv \phi     - 90^{\circ} , \qquad
  \Theta   \equiv \theta   - 90^{\circ} , \qquad
  \Theta_t \equiv \theta_t - 90^{\circ} .
\end{equation}
These are the angles that we use henceforth; see
Figures~\ref{f:drawskyview}, \ref{f:drawviewang}
and~\ref{f:drawvel}. The equations for the velocity field depend on
the position angles $\phi$, $\theta$ and $\theta_t$ only through the
differences $\phi - \theta$ and $\phi -
\theta_t$, for which
\begin{equation}
\label{newanglesdiffs}
  \phi - \theta   = \Phi - \Theta , \qquad
  \phi - \theta_t = \Phi - \Theta_t .
\end{equation}

\section{Information Content and Degeneracies of the Line-of-Sight Velocity 
Field}
\label{s:vfieldconcept}

The model described in Section~\ref{ss:vmathmodel} provides a closed
expression for the line-of-sight velocity field of a planar disk with
circular streamlines. Combination of equations~(\ref{vectot},
\ref{vecCM}, \ref{vonepn}, \ref{Racc}, \ref{vecint}, \ref{newanglesdiffs}) 
yields
{\small
\begin{equation}
\label{vfield}
   v_{\rm los}(\rho,\Phi) = v_{\rm sys} \cos \rho +
                 v_t \sin \rho \cos (\Phi -\Theta_t) +
                 D_0 (di/dt) \sin \rho \sin (\Phi-\Theta) -
                 \> s \, V(R') f \sin i \cos (\Phi-\Theta) ,
\end{equation}
}
with 
\begin{equation}
\label{Rfdef}
  R' = D_0 \sin \rho / f , \qquad 
  f        \equiv { {\cos i \cos \rho -
                     \sin i \sin \rho \sin (\Phi-\Theta)} \over 
                    { {[\cos^2 i \cos^2 (\Phi-\Theta) + 
                                 \sin^2 (\Phi-\Theta)]^{1/2}} } } .
\end{equation}
The quantities that feature in these equations have been defined
previously in Section~\ref{s:vmath}, but it is useful to recap them
briefly. The velocity $v_{\rm los} \equiv v_1$ is the component of the
velocity along the line of sight. The quantities $(\rho,\Phi)$
identify the position on the sky: $\rho$ is the angular distance from
the CM, and $\Phi$ is the position angle with respect to the CM
(measured from North over East). The quantities $(v_{\rm sys}, v_t,
\Theta_t)$ describe the velocity of the CM in an inertial frame in
which the sun is at rest: $v_{\rm sys}$ is the systemic velocity along
the line of sight, $v_t$ is the transverse velocity, and $\Theta_t$ is
the position angle of the transverse velocity on the sky. The angles
$(i,\Theta)$ describe the direction from which the plane of the galaxy
is viewed: $i$ is the inclination angle ($i=0$ for a face-on disk),
and $\Theta$ is the position angle of the line of nodes. The velocity
$V(R')$ is the rotation velocity at cylindrical radius $R'$ in the
disk plane. $D_0$ is the distance to the CM, and $f$ is a geometrical
factor. The quantity $s = \pm 1$ is the `spin sign' that determines in
which of the two possible directions the disk rotates. In the
following we will set $s=+1$, with the understanding that $V(R')$ can
be both positive and negative.

The first two terms on the right hand side of equation~(\ref{vfield})
describe the line-of-sight component of the CM velocity of the
LMC. The third term describes the line-of-sight component of the
velocities induced by the precession and nutation of the LMC disk
plane. The last term describes the line-of-sight component of the LMC
rotation (in the limit $\rho \rightarrow 0$, the last term reduces to
the rotation field model used by, e.g., Alves \& Nelson (2000) and
many other authors). The first term in equation~(\ref{vfield}) differs
from $v_{\rm sys}$ only by an amount of order $\rho^2$. For $\rho =
10^{\circ}$ one has for the LMC $v_{\rm sys} \cos \rho \approx v_{\rm
sys} - 4 \kms$. Hence, the large size of the LMC has only a minor
influence on the systemic velocity component of the observed velocity
field. The second term in equation~(\ref{vfield}) is proportional to
$\sin \rho$, which corresponds to a solid-body rotation component. At
$\rho = 10^{\circ}$, this component has an amplitude between 50--$100
\kms$, depending on the exact value of $v_t$. This exceeds the
projected amplitude of the intrinsic velocity field of the LMC (the
fourth term in eq.~[\ref{vfield}]), which is only $\sim 30$--$40
\kms$. The third term also corresponds to a solid-body rotation 
component. Its amplitude is comparable to that of the second term if
$di/dt \approx v_t / D_0$, which is in the range 1--2 mas/yr,
depending on the exact value of $v_t$. Expressed in more relevant
units, this corresponds to $\sim 275$--$550$ degrees/Gyr. This is
larger than what one would expect on the basis of $N$-body
simulations, but only by a factor of a few (Weinberg 2000). The third
term in equation~(\ref{vfield}) therefore should not be neglected,
although this has in fact been done by all previous authors.

To gain insight into the interplay between the different components
that contribute to the observed velocity field it is useful to
decompose the transverse velocity vector of the CM into a sum of two
orthogonal vectors, one along the position angle of the line of nodes
$\Theta$, and one along $\Theta+90^{\circ}$; see
Figure~\ref{f:drawvel}. These vectors have lengths
\begin{equation}
\label{vtsc}  
   v_{tc} = v_t \cos(\Theta_t - \Theta) , \qquad 
   v_{ts} = v_t \sin(\Theta_t - \Theta) , 
\end{equation}
respectively. The proper motion vector $(\mu_W,\mu_N)$ of the CM (see
eq.~[\ref{propdef}]) can be similarly decomposed into a sum of vectors
along $\Theta$ and $\Theta+90^{\circ}$, respectively. These vectors
have lengths
\begin{equation}
\label{mucsdef}
   \mu_c = - \mu_W \sin \Theta + \mu_N \cos \Theta , \qquad
   \mu_s = - \mu_W \cos \Theta - \mu_N \sin \Theta .
\end{equation}
The transverse velocity and proper motion components of the CM are
related through
\begin{equation}
\label{vtmurel}
   v_x    = D_0 \mu_W , \qquad
   v_y    = D_0 \mu_N , \qquad
   v_{tc} = D_0 \mu_c , \qquad
   v_{ts} = D_0 \mu_s ,
\end{equation}
where $v_x$ and $v_y$ are the components of the transverse velocity in
the $(x,y,z)$ coordinate system defined in
Section~\ref{ss:vmathobs}. In the following it is useful to define
\begin{equation} 
\label{wtsdef}
   w_{ts} = v_{ts} + D_0 (di/dt) = D_0 [\mu_s + (di/dt)] 
\end{equation}
and
\begin{equation}
   {\hat v}_{\rm los} \equiv v_{\rm los} - v_{\rm sys} \cos \rho .
\end{equation}
The equation~(\ref{vfield}) for the line-of-sight velocity field
then reduces to:
\begin{equation} 
\label{vlosorth}
   {\hat v}_{\rm los} = 
        [w_{ts} \sin \rho] \sin (\Phi -\Theta) + 
        [v_{tc} \sin \rho - f \> V(R') \sin i] \cos (\Phi-\Theta) .
\end{equation}
Along the line of nodes one has that $\sin (\Phi-\Theta) = 0$
and $\cos (\Phi-\Theta) = \pm 1$. This yields
\begin{equation}
\label{vlosalong}
   {\hat v}_{\rm los} ({\rm along}) = 
          \pm [v_{tc} \sin \rho - 
               V(D_0 \tan \rho) \sin i \cos \rho] .
\end{equation}
Perpendicular to the line of nodes one has that $\cos (\Phi-\Theta) =
0$ and $\sin (\Phi-\Theta) = \pm 1$, and therefore
\begin{equation}
\label{vlosperp}
   {\hat v}_{\rm los} ({\rm perpendicular}) = \pm w_{ts} \sin \rho .
\end{equation}
This implies that perpendicular to the line of nodes ${\hat v}_{\rm
los}$ is linearly proportional to $\sin \rho$. By contrast, along the
line of nodes this is true only if $V(R')$ is a linear function of
$R'$. This is not expected to be the case, because galaxies do not
generally have solid-body rotation curves; disk galaxies tend to have
flat rotation curves, at least outside the very center. This implies
that, at least in principle, both the position angle $\Theta$ of the
line of nodes and the quantity $w_{ts}$ are uniquely determined by the
observed velocity field: $\Theta$ is the angle along which the
observed ${\hat v}_{\rm los}$ are best fit by a linear proportionality
with $\sin \rho$, and $w_{ts}$ is the proportionality constant. The
systemic velocity, $v_{\rm sys}$, and the position of the CM on the
sky are also uniquely determined, because they represent points of
symmetry in the velocity field. However, $V(R')$, $i$ and $v_{tc}$ are
not uniquely determined. Let $\overline{V}(R')$, $\overline{i}$ and
$\overline{v}_{tc}$ be the true values of these quantities. Then any
combination of $V(R')$, $i$ and $v_{tc}$ with
\begin{equation}
\label{vdegenerate}
   V(R') \> = \> 
           \left ( { {\sin \overline{i}} \over {\sin{i}} } \right ) 
               \> \overline{V}(R') +
           { {(v_{tc} - \overline{v}_{tc}) \tan \rho} \over {\sin{i}} }
\end{equation}
for all $R'$ will provide exactly the same predictions both along and
perpendicular to the line of nodes. While the predictions may be very
subtly different at intermediate angles, the quality of realistic
datasets will be insufficient to discriminate between different models
that satisfy equation~(\ref{vdegenerate}). Hence, the component
$v_{tc}$ of the transverse velocity is unconstrained by the observed
velocity field, unless we assume some prior knowledge about the shape
of the rotation curve $V(R')$. Conversely, the rotation curve $V(R')$
cannot be determined unless $v_{tc}$ is known.

\section{The LMC Line-of-Sight Velocity Field Traced by Carbon Stars}
\label{s:data}

The kinematical properties of the LMC have been studied using many
different tracers, including HI (e.g., Kim \etal 1998), star clusters
(Freeman, Illingworth \& Oemler 1983; Schommer \etal 1992), planetary
nebulae (Meatheringham \etal 1988), HII regions and supergiants
(Feitzinger \etal 1977), and carbon stars. The latter have yielded the
largest and most useful datasets in recent years, and we therefore
restrict our analysis to carbon star data. We included and merged two
different radial velocity data sets. The first is the one obtained by
Kunkel \etal (1997), which was analyzed previously by Alves \& Nelson
(2000). This data set covers mostly the periphery of the LMC, and is
distributed fairly homogeneously in position angle. We excluded all
(inter-cloud) stars further than $\rho = 13^{\circ}$ from the LMC
center, because it is not clear to what extent they belong to the LMC
and participate in its kinematics. This leaves a total of 468
stars. The second data set is the one obtained by Hardy, Schommer \&
Suntzeff (2002, in preparation), which was analyzed previously by
Graff \etal (2000). The 573 stars in this data set were selected from
Blanco \etal (1980), and their spatial distribution reflects the
distribution of fields in that survey. The stars sample the inner
parts of the LMC, but with a discontinuous distribution in radius and
position angle. The combined dataset of 1041 stars samples the entire
area of the LMC, but not homogeneously. The same data set, although
with subtly different selection criteria, was studied previously by
Alves (2000); the spatial distribution in the LMC of these carbon
stars with known line-of-sight velocities is shown in his figure~1.

We fitted the velocity field given by equation~(\ref{vlosorth}) to the
data, by minimizing the RMS residual of the fit using a
downhill-simplex routine (Press \etal 1992). The position of the CM
and the parameters $(v_{\rm sys}, v_{ts}, v_{tc}, \Theta)$ were all
treated as free parameters that were optimized in the fit. The
inclination $i$ is essentially degenerate with the amplitude of the
rotation curve $V(R')$, cf.~equation~(\ref{vdegenerate}). It was
therefore, without loss of generality, fixed to the value
\begin{equation}
\label{inclination}
  i = 34.7^{\circ} \pm 6.2^{\circ} 
\end{equation}
determined in Paper~I. The rotation curve $V(R')$ was parameterized as
\begin{equation}
\label{Vcurve}
  V(R') = V_0 \> { {{R'}^{\eta}} \over { {R'}^{\eta} + {R_0}^{\eta} } }
\end{equation}
This corresponds to a velocity that increase as a power-law until some
scale radius $R_0$, after which it flattens to the constant value
$V_0$. The parameters $(V_0, R_0, \eta)$ were also optimized in the
fit. The adoption of a parameterized form for $V(R')$ artificially
removes some of the degeneracy described by
equation~(\ref{vdegenerate}); while infinitely many functions $V(R')$
can fit the data equally well, only some of these can be expected to
be properly described by the adopted parameterization. The
implications of this are addressed below. 

Once the best-fitting model has been identified, we calculate error
bars on the model parameters using Monte-Carlo simulations. Many
different pseudo data sets are created that are analyzed similarly as
the real data set. The dispersions in the inferred model parameters
are a measure of the formal 1-$\sigma$ error bars on the best-fit
model parameters. Each pseudo data set is created by calculating for
each star a velocity $v_{\rm los}$ that is the sum of a random
Gaussian deviate $\Delta v$ and the velocity predicted by the best-fit
model. The Gaussian dispersion of the $\Delta v$ is chosen to
correspond to the RMS residual of the best-fit model.

The data and the best-fitting model are shown in Figure~\ref{f:fit}.
For all radii $1^{\circ} \leq \rho \leq 9^{\circ}$ the fit is about as
good as could be expected. At small radii, $\rho \leq 1^{\circ}$,
there appears to be some systematic discrepancy between the data and
the model. However, the data in the central degree (top left panel of
Figure~\ref{f:fit}) provide too sparse coverage of the full range of
position angles to draw any conclusions from this.  There also appear
to be minor discrepancies in the data-model comparison at radii $\rho
\geq 9^{\circ}$. This could be due to tidal distortions in the LMC
disk and its velocity field at large radii. Either way, these
discrepancies do not affect the primary results of this paper in any
significant way; fits to restricted radial ranges yielded identical
results.

The right ascension $\alpha$ and declination $\delta$ of the CM and
the parameters $v_{\rm sys}$, $w_{ts}$ and $\Theta$ are all determined
unambiguously. The RMS residual of the fit is the (globally averaged)
line-of-sight velocity dispersion $\sigma$ of the LMC. The model fit
yields:
\begin{eqnarray}
\label{fitresults}
\alpha_{\rm CM} & = & 5\chour 27.6\cmin \pm 3.9\cmin    , \nonumber \\
\delta_{\rm CM} & = & -69.87^{\circ} \pm 0.41^{\circ}   , \nonumber \\
v_{\rm sys}     & = & 262.2 \pm 3.4 \kms                , \nonumber \\
w_{ts}          & = & -402.9 \pm 13.0 \kms              , \nonumber \\
\Theta          & = & 129.9^{\circ} \pm 6.0^{\circ}     , \nonumber \\
\sigma          & = & 20.2 \pm 0.5 \kms                 , 
\end{eqnarray}
where the CM position is given in J2000.0 coordinates. 

As discussed in Section~\ref{s:vfieldconcept}, the quantity $v_{tc}$
and the rotation curve $V(R')$ are not uniquely determined by the
line-of-sight velocity field. The parameterized model that best fits
the data has $v_{tc} = 599 \kms$ and the rotation curve $V(R')$ that
is shown as a solid curve in Figure~\ref{f:rotcurve}. However, every
other $v_{tc}$ will provide an equally acceptable fit, provided that
$V(R')$ is modified according to equation~(\ref{vdegenerate}). The
alternative rotation curves thus obtained are shown as dashed curves
in Figure~\ref{f:rotcurve} for several values of $v_{tc}$. The value
of $v_{tc}$ is constrained by the line-of-sight velocity field only
through the fact that some of the rotation curves $V(R')$ needed to
fit the data are unplausible. If $v_{tc}$ is high, then $V(R')$ is
implausibly large for a low-luminosity galaxy such as the LMC. The LMC
is slightly fainter than M33, $M_V = -18.5$ vs. $M_V = -18.9$,
respectively (van den Bergh 2000), and the LMC and M33 have similar
disk scale lengths (Paper~II; Regan \& Vogel 1994). According to the
Tully-Fisher relation, the circular velocity of the LMC should be
lower than for M33. For M33, $V(R') \approx 120 \kms$ at $R' = 11.6
\kpc$ (Corbelli \& Salucci 2000), the largest radius included in our 
study of the LMC. Figure~\ref{f:rotcurve} shows that for $v_{tc} = 600
\kms$, the LMC rotation curve levels off at $V(R') \approx 130 \kms$. 
This makes models with $v_{tc} \gta 600 \kms$ quite implausible. On
the other hand, if $v_{tc}$ is low, less than $\sim 200 \kms$, then
$V(R')$ has a sign change within the radial range out to which its
starlight can be traced. This is never observed in disk
galaxies. Based on these considerations we conclude that 
\begin{equation}
\label{vtcconstraint}
  v_{tc} \> \in \> [200,600] \kms . 
\end{equation}

The results described in this section were found to be very robust. We
experimented with fits to subsets of the data, fits over restricted
radial ranges, fits with different parameterizations for $V(R')$, and
different algorithms for the determination of the parameters that
characterize the velocity field. All of this yielded results that are
statistically equivalent to those described above.

\section{The Proper Motion and Distance of the LMC}
\label{s:litreview}

The analysis of the LMC line-of-sight velocity field has provided
constraints on the velocities $v_{tc}$ and $w_{ts}$
(eqs.~[\ref{fitresults}] and~[\ref{vtcconstraint}]). These velocities
are directly related to the proper motion vector $(\mu_W,\mu_N)$ and
distance $D_0$ of the LMC CM through equations~(\ref{mucsdef}),
(\ref{vtmurel}) and (\ref{wtsdef}). To make further progress it is
therefore useful to review what is already know about these
quantities.

\subsection{Proper Motion}
\label{ss:LMCpropmotion}

Table~\ref{t:propmotion} lists the proper motion measurements that are
available for the LMC. They are from the following sources: Kroupa
\etal (1994), using stars from the PPM Catalogue; Jones \etal (1994),
using photographic plates with a 14 year epoch span; Kroupa \& Bastian
(1997), using Hipparcos data; Drake \etal (2002), using data from the
MACHO project\footnote{The MACHO proper motion measurement is based on
a preliminary analysis of the data. However, the result of the final
analysis is not expected to be considerably different (Drake 2002,
priv.~comm.)}; and Pedreros \etal (2002), using CCD frames with an 11
year epoch span. The measurements of Jones \etal (1994) and Pedreros
\etal (2002) pertain to fields in the outer parts in the LMC
disk. These measurements require corrections for the orientation and
rotation of the LMC disk. The proper motions listed in the table are
based on the corrections that we derive in Appendix~\ref{s:app}, which
are more accurate than the corrections derived in the original
papers. The table does not include the measurement published by
Anguita, Loyola \& Pedreros (2000). Their result for the proper motion
in the declination direction ($\mu_N = 2.9 \pm 0.2 \masyr$) is highly
($10 \sigma$) inconsistent with all other studies. This probably
indicates that the measurement contains some unidentified error; see
Pedreros \etal (2002) and Drake \etal (2002) for some discussion of
this issue.

Figure~\ref{f:pm} shows the data from Table~\ref{t:propmotion} in the
$(\mu_W,\mu_N)$ plane. The ellipses around the data points indicate
the corresponding 68.3\% confidence regions. The proper motion
measurements are in fair agreement with each other, in the sense that
they fall in the same part of the diagram. The weighted
average\footnote{The 1-$\sigma$ error in a weighted average of $N$
measurements $z_i$ is generally given by $[\sum_{i=1}^{N} 1 / (\Delta
z_i)^2 ]^{-1/2}$. However, this is correct only if all the errors in
the data are Gaussian random errors. In the present case there is some
evidence that this is an oversimplification. For example, the
measurements by Jones \etal (1994) and Pedreros \etal (2002) are
mutually inconsistent at the $\sim 3\sigma$ level. In taking the
weighted average we therefore increased all the errors by $25$\%.
This causes the $\chi^2$ that measures the residuals of the data with
respect to their weighted average to be equal to the number of degrees of
freedom.} of the proper motion measurements in
Table~\ref{t:propmotion} is
\begin{equation}
\label{weightB}
   \mu_W = -1.68 \pm 0.16 \masyr , \qquad
   \mu_N =  0.34 \pm 0.16 \masyr .
\end{equation}
This is the LMC proper motion that we will use in the subsequent
discussion. The corresponding 68.3\% confidence region is indicated in
Figure~\ref{f:pm} by a heavy ellipse.

\subsection{Distance}
\label{ss:LMCdistance}

The distance $D_0$ to the LMC is a very important quantity in the
calibration of the cosmological distance ladder. Most of the methods
that are customarily used to study the Hubble constant of the Universe
are ultimately calibrated by the distance to the LMC. The
period-luminosity relation of Cepheids provides the most important,
but by no means the only example. Of course, there now exist methods
that provide information on the Hubble constant without requiring
knowledge of the distance to the LMC. These use, e.g., gravitational
lensing, the Sunyaev-Zeldovich effect, or the Cosmic Microwave
Background radiation. However, the systematic uncertainties in these
methods have not yet been reduced to levels where they can replace the
more traditional techniques. The distance to the LMC therefore remains
a topic of intense interest and debate in modern astronomy.

Many techniques have been used to estimate the LMC distance.
Unfortunately, there continue to be systematic differences between the
results from different techniques that exceed the formal errors. It is
beyond the scope of the present paper to review this topic in
detail. Instead, we refer the reader to the recent reviews by, e.g.,
Westerlund (1997), Gibson \etal (2000) and Freedman \etal (2001). Here
we follow Freedman \etal by adopting
\begin{equation}
\label{distmodulus}
  m - M = 18.50 \pm 0.10 
\end{equation}
for the LMC distance modulus, where $m - M \equiv 5 \log(D_0/\kpc) +
10$. The implied distance is $D_0 = 50.1 \pm 2.5 \kpc$. At this
distance, a proper motion of $1 \masyr$ corresponds to $238 \pm 12
\kms$ (eq.~[\ref{pmtovel}]), and 1 degree on the sky corresponds to 
$0.875 \pm 0.044 \kpc$.

New and improved distance determinations for the LMC continue to
become available at a rapid rate. Recent measurements that were not
yet included in the compilation of Freedman \etal (2001) include,
e.g., measurements using the Tip of the Red Giant Branch (Cioni \etal
2000), eclipsing binaries (Fitzpatrick \etal 2002), RR Lyrae variables
(Benedict \etal 2002) and the Red Clump (Alves \etal 2002, in prep.).
The results from these recent studies are all consistent with the
value adopted in equation~(\ref{distmodulus}).

\section{Precession and Nutation of the LMC Disk}
\label{s:distdidt}

Equation~(\ref{wtsdef}) can be solved for $di/dt$ to obtain
\begin{equation}
\label{didteq}
   di/dt = ( w_{ts} / D_0 ) - \mu_s .
\end{equation}
The quantity $w_{ts}$ is determined by the line-of-sight velocity
field, as is the line-of-nodes position angle $\Theta$
(eq.[\ref{fitresults}]). The proper motion component $\mu_s$ is
determined by the literature average in equation~(\ref{weightB}), with
help of the definition in equation~(\ref{mucsdef}). The distance is
given by the literature average in equation~(\ref{distmodulus}). The
rate of inclination change $di/dt$ is therefore uniquely determined
by the available data. Evaluation of equation~(\ref{didteq}) yields
\begin{equation}
\label{didtresult}
   di/dt = -0.37 \pm 0.22 \masyr = -103 \pm 61 \>\> {\rm degrees}/\Gyr ,
\end{equation}
where a simple Monte-Carlo scheme was used for propagation of errors.

The LMC is the first galaxy for which it has been possible to measure
$di/dt$. It is therefore useful to ask whether the inferred value is
plausible, given our understanding of the LMC and its orbit around the
Milky Way. Weinberg (2000) performed $N$-body simulations to assess
the influence of the Milky Way on the structure and dynamics of the
LMC. He found that the tidal torques from the Milky Way are expected
to induce precession and nutation in the symmetry axis of the LMC disk
plane. His figures~3 and~4 show the variations in the directions of
the LMC symmetry axis for a generic simulation of the LMC (i.e., not
fine-tuned to reproduce all the currently observed features of the
LMC). On average, the rate of directional change in the
three-dimensional angle is $\sim 60^{\circ} / \Gyr$ (i.e., $0.22
\masyr$). However, there are considerable variations with time, and at
certain times the rate is several times larger. Whether this
directional change is seen by an observer as a change in inclination
or as a change in the line-of-nodes position angle depends on the
exact location of the observer. Either way, the observed rate $di/dt$
given by equation~(\ref{didtresult}) has the same order of magnitude
as the generic predictions by Weinberg (2000). We therefore interpret
the observed $di/dt$ as an indication of precession and nutation
induced by tidal torques from the Milky Way.

The line-of-sight velocity field does not determine the second
derivative $d^2 i / dt^2$. It also does not provide any information on
time variations in the position angle $\Theta$ of the line of
nodes. It is therefore not possible to calculate either the past or
the future variation of the orientation of the LMC symmetry axis in an
inertial frame tied to the Milky Way. If one assumes that $d^2 i /
dt^2$ is always equal to zero, then the LMC would make one revolution
every $3.5 \pm 2.1 \Gyr$. However, this number provides no real
insight. The simulations of Weinberg show that LMC symmetry axis is
expected to undergo nutation (i.e., oscillation with $d^2 i / dt^2
\not= 0$) and not tumbling motion (i.e., full revolutions).

\section{Kinematical Properties of the LMC disk}
\label{s:kindisk}

\subsection{Dynamical Center}
\label{ss:center}

Previous authors who have modeled the dynamics of tracers in the LMC
have fixed the dynamical center a priori, generally to coincide with
the kinematical center of the HI gas. Kim \etal (1998) find this HI
kinematical center to be at $\alpha = 5\chour 17.6\cmin$ and $\delta =
-69^{\circ} 1'$ (J2000.0 coordinates), with an accuracy that is
probably better than $\sim 0.2^{\circ}$ in each coordinate. It has
been a well-known result for many years that this position differs by
almost a full degree from the center of the LMC bar (e.g., Westerlund
1997). The data from 2MASS yield for the latter $\alpha = 5\chour
25.1\cmin \pm 0.1\cmin$ and $\delta = -69^{\circ} 47' \pm 1'$
(Paper~II). Our analysis in Section~\ref{s:data} is the first to
provide an accurate measurement of the dynamical center {\it of the
stars} in the LMC, by leaving it as a completely free parameter in the
fit. We obtain $\alpha_{\rm CM} = 5\chour 27.6\cmin
\pm 3.9\cmin$ and $\delta_{\rm CM} = -69^{\circ} 52' \pm 25'$
(eq.~[\ref{fitresults}]). This is consistent with the position of the
center of the bar. It is also consistent with the position of the
center of the outer isophotes of the LMC (corrected for the effect of
viewing perspective), which is at approximately $\alpha = 5\chour
29\cmin$ and $\delta = -69^{\circ} 30'$ (Paper~II). However, the
dynamical center of the stars differs from the dynamical center of the
gas by as much as $1.2^{\circ} \pm 0.6^{\circ}$. These results suggest
that the gas kinematics are quite disturbed, while the stellar
kinematics show little evidence for peculiar behavior. This is not
particularly surprising in view of other knowledge about the gas in
the Magellanic Clouds, which is quite disturbed in general (witness
the Magellanic Bridge and the Magellanic Stream). These results make
it unlikely that any study of the HI in the LMC will ever lead to an
accurate understanding of its structure and dynamics.

\subsection{Rotation Curve}
\label{ss:rotdisp}

For studies of the dynamics and mass distribution of the LMC it is
useful to have an unparameterized description of its rotation and
velocity dispersion profiles. To this end we have performed fits of
the velocity field given by equation~(\ref{vfield}) to individual
rings in the plane of the LMC disk. The rotation velocity $V(R')$ and
the line of sight position angle $\Theta$ were assumed to be constant
over each ring, and were determined so as to best fit the data. The
resulting kinematical profiles are listed in Table~\ref{t:kinematics},
and are shown in Figure~\ref{f:rotcurve}. In the fits we kept the
quantities $(\alpha_{\rm CM}, \delta_{\rm CM}, v_{\rm sys}, w_{ts},
i)$ fixed to their previously determined values
(eqs.~[\ref{inclination}] and~[\ref{fitresults}]). The value of
$v_{tc}$ was fixed to that implied by the LMC proper motion; from
equations~(\ref{mucsdef}, \ref{vtmurel}, \ref{fitresults},
\ref{weightB}, \ref{distmodulus}) one obtains
\begin{equation}
\label{vtcpm} 
   v_{tc} = 253 \pm 52 \kms .
\end{equation}
This is not inconsistent with the weak constraint $v_{tc} \> \in \>
[200,600] \kms$ that was obtained from the analysis of the
line-of-sight velocity field (eq.~[\ref{vtcconstraint}]).

The inferred rotation curve $V(R')$ rises linearly in the central
region and is roughly flat at $V \approx 50 \kms$ for $R' \gta 4 \kpc$
(i.e., $R'/D_0 \gta 0.08$). The error bars in $V(R')$ do not take into
account the errors in $(\alpha_{\rm CM}, \delta_{\rm CM}, v_{\rm sys},
w_{ts}, i, v_{tc})$, which were all kept fixed. In reality, the errors
in $i$ and $v_{tc}$ do add additional uncertainty to the rotation
curve. The uncertainty in $i$ (eq.~[\ref{inclination}]) causes a $\sim
15$\% uncertainty in the normalization of $V(R')$,
cf.~equation~(\ref{vdegenerate}). No other part of our analysis
depends on the exact choice of the inclination. The influence of the
$52 \kms$ error in $v_{tc}$ (eq.~[\ref{vtcpm}]) can be assessed
visually from Figure~\ref{f:rotcurve}.

The innermost data point in Figure~\ref{f:rotcurve} at $R'/D_0 \approx
0.01$ (i.e., $R' \approx 0.5 \kpc$) is clearly discrepant from the
data points at larger radii, both in terms of its kinematic position
angle $\Theta$ and its rotation velocity $V$. Although this could be
attributed to non-circular streaming motions in the region of the bar,
this result has only very low significance. The top left panel of
Figure~\ref{f:fit} shows that the data in this region provide only
very sparse coverage of the full range of position angles. Better data
are needed to address the kinematics in the central kpc with any
confidence.

Alves \& Nelson (2000) previously analyzed the carbon-star kinematics
of the LMC. Like most authors, they didn't use the fully correct
expression for the velocity field of a rotating disk at finite
distance (last term in eq.~[\ref{vfield}]) and they assumed $di/dt =
0$. They corrected for the transverse motion of the LMC using an
assumed proper motion $(\mu_W,\mu_N) = (-1.65 \pm 0.20, -0.17 \pm
0.22) \masyr$. This is an average of the proper motion measurements by
Jones \etal (1994) and Kroupa \& Bastian (1997). This proper motion,
combined with $di/dt = 0$, implies $v_{tc} = 325 \pm 57 \kms$ and
$w_{ts} = -220 \pm 62 \kms$ (cf.~eqs.[\ref{mucsdef},
\ref{vtmurel}, \ref{fitresults}, \ref{weightB}, \ref{distmodulus}). The 
value of $w_{ts}$ is quite different from the value that we have
obtained here from a fit to the velocity field, $w_{ts} = -402.9 \pm
13.0 \kms$ (cf.~eq.~[\ref{fitresults}]). The result of enforcing an
incorrect value of $w_{ts}$ on the solution is that one obtains an
incorrect value for the position angle of the line of nodes. Alves \&
Nelson obtained a kinematic line of nodes that is both twisting and
inconsistent with the geometrically determined line of nodes (see
Section~\ref{s:intro}). Since the rotation curve is determined by the
velocities measured along the line of nodes, one also obtains an
incorrect estimate of the rotation curve. Alves \& Nelson obtained a
rotation curve amplitude of $\sim 70 \kms$, which exceeds ours by $\sim
40$\%.

The analysis by Kim \etal (1998) of the HI velocity field suffered
from similar shortcomings. They corrected for the transverse motion of
the LMC using the proper motion advocated by Jones \etal (1994),
$(\mu_W,\mu_N) = (-1.37 \pm 0.28, -0.18 \pm 0.27) \masyr$. This proper
motion, combined with $di/dt = 0$, implies $v_{tc} = 276 \pm 69
\kms$ and $w_{ts} = -175 \pm 72 \kms$. Again, the value of $w_{ts}$ is 
quite different from the value that we have obtained here from a fit
to the velocity field. So, like Alves \& Nelson, Kim \etal obtained a
kinematic line of nodes that is both twisting and inconsistent with
the geometrically determined line of nodes, as well as a rotation
curve amplitude that exceed ours by $\sim 40$\%.

\subsection{Velocity Dispersion Profile}
\label{ss:disp}

The velocity dispersion profile that we derive for the carbon stars in
the LMC is shown in the middle panel of Figure~\ref{f:rotcurve}. It is
not very different from the velocity dispersion profile derived by
Alves \& Nelson (2000). The dispersion is 20--$22 \kms$ between $1$
and $3.5 \kpc$ from the center, followed by a decline to 16--$17 \kms$
between $3.5$ and $7 \kpc$ from the center, and a subsequent increase
to 21--$22 \kms$ between $7$ and $9 \kpc$.

The different stellar populations in the LMC are not characterized by
a single velocity dispersion. As in the Milky Way, younger populations
have a smaller velocity dispersion than older populations. A summary
of measurements for various populations is given by Gyuk, Dalal \&
Griest (2000). They range from $\sigma \approx 6 \kms$ for the
youngest populations (e.g., supergiants, HII regions, HI gas) to
$\sigma \approx 30 \kms$ for the oldest populations (e.g., old
long-period variables, old clusters). Any discussions in the remainder
of the paper that use our velocity dispersion measurements, e.g.,
those in Section~\ref{ss:height} concerning the LMC scale height,
apply strictly only to carbon stars.  On the other hand, the carbon
stars are part of the intermediate-age population which is believed to
be fairly representative for the bulk of the mass in the LMC. In this
sense, the results inferred for the carbon star population are
believed to be generic for the LMC as a whole.

\subsection{Line of Nodes Position Angle}
\label{ss:nodes}

The fit to the velocity field described in Section~\ref{s:data}
implies a line-of-nodes position angle $\Theta = 129.9^{\circ} \pm
6.0^{\circ}$ (eq.~[\ref{fitresults}]). This agrees within the errors
with the geometrically determined line-of-nodes position angle,
$\Theta = 122.5 \pm 8.3^{\circ}$ (Paper~I). This agreement was not in
any way built into the model, and it therefore provides an important
consistency check on the validity of our approach. This is
particularly important because it resolves the discrepancy that
emerged from previous kinematical analyses of the LMC (see
Section~\ref{s:intro}). It also provides further confirmation of the
finding from Paper II that the position angle of the line nodes is
very different from the position angle of the LMC major axis (${\rm
PA}_{\rm maj} = 189.3^{\circ} \pm 1.4^{\circ}$), and hence that the
LMC is not intrinsically circular.

The bottom panel of Figure~\ref{f:rotcurve} shows that the radial
profile of $\Theta$ is approximately flat (with the exception of the
innermost point, which has very low significance; see
Section~\ref{ss:rotdisp}). The constancy of $\Theta$ as a function of
radius should come as no surprise. Our velocity field model does not
allow for radial twists in $\Theta$. The best-fitting value for
$w_{ts}$ is therefore, by definition, the value that yields the least
possible radial variation in $\Theta$. One might wonder if this is not
the equivalent of forcing a square peg into a round hole. After all,
the projected morphology of the LMC shows a considerable isophotal
twist between the region of the bar and the outer parts of the disk
(Paper~II). While this is certainly a worry, we have found little
evidence that there is a corresponding twist in the kinematic line of
nodes. First, the results of the present paper, which assume an
absence of kinematic twists, provide a highly consistent view of the
LMC. Second, there is no evidence for any isophotal twist at projected
radii $\rho \geq 4.5^{\circ}$, suggesting that at least at these radii
there are probably no twists in the kinematic line of
nodes. Kinematical fits restricted to the range $\rho \geq
4.5^{\circ}$ yielded results that were statistically consistent with
those for the full radial range.

\subsection{Influence of Non-Circular Streamlines}
\label{ss:noncirc}

It was shown in Paper~II that the LMC is not intrinsically circular,
but instead has an ellipticity $\epsilon = 1 - (b/a) = 0.31 \pm 0.01$.
However, the gravitational potential of a mass distribution is always
rounder than the mass distribution, which provides some {\it a priori}
justification for the use of a model in which the streamlines are
circular. There is also some {\it a posteriori} justification from the
fact that the kinematically inferred position angle of the line of
nodes, $\Theta_{\rm kin} = 129.9^{\circ} \pm 6.0^{\circ}$
(eq.~[\ref{fitresults}]), agrees with the geometrically determined
value, $\Theta = 122.5 \pm 8.3^{\circ}$ (Paper~I). Detailed dynamical
models of elliptical disks generally predict a (small) misalignment
between $\Theta_{\rm kin}$ and $\Theta$ (e.g., Schoenmakers, Franx \&
de Zeeuw 1997). The fact that no such misalignment is detectable with
the available statistics ($\Theta_{\rm kin} - \Theta = 7.4^{\circ} \pm
10.2^{\circ}$) suggests that a model with circular streamlines may not
be unreasonable in the present context.  Note, however, that this
argument cannot be reversed. The fact that $\Theta_{\rm kin} - \Theta$
is small does not by itself imply that the streamlines must be nearly
circular (Jalali \& Abolghasemi 2002).

Unfortunately, there are no unambiguous theoretical methods for
calculating the velocity field of an elliptical disk. Some results
have been obtained for gaseous systems (Jalali \& Abolghasemi 2002),
but stellar systems are more complicated because they can have an
anisotropic velocity dispersion (i.e., pressure) tensor. It is
therefore difficult to estimate the systematic errors in our results
due to the simplifying assumption of circular streamlines. It should
be noted though, that even if the streamlines are not circular, this
will not invalidate our approach at a basic level. Even in an
elliptical disk there is generally a position angle $\Upsilon$ along
which the line-of-sight velocity component is zero (Schoenmakers
\etal 1997). Unlike in a circular disk, one will not generally have
that $\Upsilon = \Theta + 90^{\circ}$, where $\Theta$ is the true
geometrical line of nodes.  However, along the position angle
$\Upsilon$ one still has that ${\hat v}_{\rm los} = \pm w_{ts}
\sin \rho$, as in equation~(\ref{vlosperp}). So for an elliptical 
disk one can still, at least in principle, constrain $\Upsilon$ and
$w_{ts}$ from the line-of-sight velocity field. However, the actual
algorithm employed here will not be quite sufficient to yield highly
accurate results. We rely on fitting the velocity field given by
equation~(\ref{vfield}), which for an elliptical disk is not generally
valid at all position angles $\Phi$. It is not clear how important
this is for the present analysis, but it will certainly become more
important as larger samples of carbon star line-of-sight velocities
will become available in the future. This will decrease the formal
error random errors in the modeling, at which point deviations from
circular streamlines may start to play a more important role.

\section{Structure of the LMC Disk}
\label{s:structure}

An important result of the present study is that we find the rotation
curve amplitude of the LMC to be $\sim 40$\% lower than suggested on
the basis of previous analyses (see Section~\ref{ss:rotdisp}). This
changes our understanding of several important structural parameters
and properties of the LMC.

\subsection{Mass}
\label{ss:LMCmass}

Figure~\ref{f:rotcurve} shows that the rotation curve of the LMC is
approximately flat from $R \approx 4 \kpc$ out to the last data point
at $R = 8.9 \kpc$. The weighted mean of the rotation velocity data
points in this radial range is $V = 49.8 \kms$. There are several
uncertainties in this rotation curve amplitude: (i) an error of $\pm
2.1 \kms$ due to the finite number of data points; (ii) an error of
$\pm 7.8 \kms$ due to the error in our knowledge of the LMC
inclination (eq.~[\ref{inclination}]); and most importantly (iii) an
error of approximately $\pm 13.7 \kms$ due the error in our knowledge
of $v_{tc}$ (eq.~[\ref{vtcpm}] and Figure~\ref{f:rotcurve}). Addition
of these errors in quadrature yields $V = 49.8 \pm 15.9 \kms$. The
weighted mean line-of-sight velocity dispersion over the same radial
range is $\sigma = 16.9 \pm 0.5 \kms$.

Due to asymmetric drift, the average azimuthal streaming velocity $V$
is always smaller than the velocity $V_{\rm circ}$ of a tracer on a
circular orbit in the equatorial plane. For the solar neighborhood in
the Milky Way, observations indicate that $V_{\rm circ}^2 \approx V^2
+ 5.5 \sigma_R^2$ (Dehnen \& Binney 1998). For the LMC we expect
similarly that $V_{\rm circ}^2 = V^2 + \kappa \sigma^2$, where
$\sigma$ is the observed line-of-sight velocity dispersion and
$\kappa$ is a model dependent factor. Since the formal error in $V$ is
quite large, there is no particular benefit in making very
sophisticated models to determine $\kappa$. A simple model that is
adequate in the present context considers the equatorial plane of an
axisymmetric system with an isotropic velocity distribution embedded
in an isothermal dark halo. The equations of van der Marel (1991) then
show that $-\kappa$ is equal to the logarithmic slope of the mass
density in the equatorial plane. For an exponential disk this yields
$\kappa = R/R_d$, where $R_d$ is the exponential disk scale
length. From Paper~II we know that the LMC is modestly well described
by an exponential profile with $R_d \approx 1.5 \kpc$. This implies
that at the last measured data point $\kappa \approx 6$ and $V_{\rm
circ} = 64.8 \pm 15.9 \kms$ (the error on $V_{\rm circ}$ does not
include a theoretical error on $\kappa$, given that its influence
would most likely be negligible compared to the formal error in $V$).

The total mass of the LMC inside the last measured data point is
$M_{\rm LMC} (R) = R V_{\rm circ}^2 / G$, where the gravitational
constant $G = 4.3007 \times 10^{-6} \kpc (\kms)^2 \Msun^{-1}$. This
yields $M_{\rm LMC} (8.9 \kpc) = (8.7 \pm 4.3) \times 10^9
\Msun$. Despite its relatively large error bar, our estimate is 
considerably more accurate in a systematic sense than the results of
previous studies. Nonetheless, we find that (after correction for the
fact that different authors tend to quote masses enclosed within
different radii) our result is similar to most other LMC mass
estimates in the literature (e.g., Feitzinger \etal 1977;
Meatheringham \etal 1988; Kim \etal 1998). The reason for this is that
while previous authors generally inferred a higher rotation curve
amplitude $V$ (see Section~\ref{ss:rotdisp}), their general neglect of
the asymmetric drift correction led to similar estimates of $V_{\rm
circ}$. Our mass estimate does differ considerably from that presented
by Schommer \etal (1992). They applied a statistical mass estimator to
the velocities of a sample of 16 old LMC clusters to infer a mass
$M_{\rm LMC} \approx 2 \times 10^{10} \Msun$ within an effective
radius of $\sim 6 \kpc$. This result is inconsistent with the analysis
presented here. Several $N$-body simulation studies of the LMC have
assumed, following Schommer \etal, that $\sim 2 \times 10^{10} \Msun$
is enclosed inside the visible body of the LMC (e.g., Gardiner, Sawa
\& Fujimoto 1994; Gardiner \& Noguchi 1996; Weinberg 2000). It is
likely that these simulations have overestimated the self-gravity of
the LMC compared to the Milky Way tidal force.

\subsection{Tidal Radius}
\label{ss:tidal}

To calculate the tidal radius of the LMC we consider the gravitational
accelerations at a point on the line that connects the Milky Way to
the LMC. Let the point be at a distance $\beta D_0$ from the LMC center
and at a distance $[1-\beta]D_0$ from the Milky Way center. We assume
that $\beta \ll 1$; the calculations can easily be carried out to full
accuracy, but this does not lead to significantly different results.
Let ${\vec a}_{\rm MW}$ be the gravitational acceleration due to the
Milky Way and ${\vec a}_{\rm LMC}$ the gravitational acceleration due
to the LMC. The tidal radius corresponds to the point at which ${\vec
a}_{\rm MW}([1-\beta]D_0) - {\vec a}_{\rm MW}(D_0) = - {\vec a}_{\rm
LMC}(\beta D_0)$.

The traditional assumption in the calculation of the tidal radius is
that both bodies can be approximated to be point masses. In this case
one obtains that
\begin{equation}
\label{taufirst}
  \beta_{\rm tidal} \equiv \beta_1 = 
      [M_{\rm LMC} / 2 M_{\rm MW}]^{1/3} .
\end{equation}
This equation was used, e.g., by Weinberg (2000; his appendix
A). However, this equation is not really appropriate, because there is
strong evidence that the mass of the Milky Way continues to rise
linearly to beyond the distance of the LMC (e.g., Kochanek 1996;
Wilkinson \& Evans 1999). If one assumes instead that the Milky Way
has a flat rotation curve with circular velocity $V_0$, while still
assuming that the LMC is a point mass, then
\begin{equation}
\label{tausecond}
  \beta_{\rm tidal} \equiv \beta_2 =
      [M_{\rm LMC} / M_{\rm MW} (D_0)]^{1/3} .
\end{equation}
where $M_{\rm MW} (D_0)$ is the mass of the Milky Way inside a sphere
of radius $D_0$. This yields a tidal radius that is a factor $2^{1/3}
= 1.26$ larger than the value given by equation~(\ref{taufirst}). If
one assumes that the LMC itself can also not be approximated as a
point mass, but instead has a constant circular velocity $V_{\rm
circ}$ (i.e., mass rising linearly with radius), then
\begin{equation}
\label{tauthird}
  \beta_{\rm tidal} \equiv \beta_3 = V_{\rm circ} / V_0 .
\end{equation}

The kinematical data that are available for the LMC do not extend as
far out as the tidal radius, so we do not know whether the LMC
circular velocity curve at the tidal radius is flat or Keplerian.  The
LMC tidal radius, $r_t \equiv \beta_{\rm tidal} D_0$, must therefore be
in the range $\beta_2 D_0 \lta r_t \lta \beta_3 D_0$. To evaluate
equation~(\ref{tausecond}) we use $M_{\rm LMC} = (8.7 \pm 4.3) \times
10^9 \Msun$ from Section~\ref{ss:LMCmass} and $M_{\rm MW} (D_0) = (4.9
\pm 1.1) \times 10^{11} \Msun$ from Kochanek (1996), which yields
$\beta_2 = 0.26 \pm 0.05$. To evaluate equation~(\ref{tauthird}) we use
$V_{\rm circ} = 64.8 \pm 15.9 \kms$ from Section~\ref{ss:LMCmass} and
$V_0 = 206 \pm 23 \kms$ based on the mass given by Kochanek, which
yields $\beta_3 = 0.31 \pm 0.08$. We combine these results to obtain
for the LMC tidal radius that $r_t = 15.0 \pm 4.5 \kpc$. This
corresponds to an angle on the sky of $17.1^{\circ} \pm 5.1^{\circ}$.

The result thus calculated can be compared to our observational
knowledge of the outer parts of the LMC. Data from the 2MASS survey
show that the isopleths of RGB and AGB stars can be traced out to a
major axis radius of $\sim 9^{\circ}$ and are regular out to this
radius (Paper~II). APM measurements of photographic Schmidt plates
have revealed regular isopleths out to $\sim 11.5^{\circ}$ (Irwin
1991). Observations therefore indicate that $r_t \gta 11.5^{\circ}$,
consistent with value calculated above. Having said this, it must be
kept in mind that the tidal radius is not a particularly well-defined
quantity, especially not for a disk system such as the LMC. The Milky
Way center does not lie in the plane of the LMC disk (Paper~II). As a
consequence, for points in the LMC disk the gravitational
accelerations ${\vec a}_{\rm MW}$ and ${\vec a}_{\rm LMC}$ are not
colinear. So there is no radius in the disk where they add up to
zero. Also, as the LMC orbits the Milky Way, the angle between the
acceleration vectors changes. The effect of the Milky Way tidal force
therefore need not lead to a sharp truncation of the disk, but instead
may lead to thickening of the disk and the formation of an extended
halo of unbound particles (Weinberg 2000).

\subsection{Scale Height}
\label{ss:height}

The observed velocity dispersion profile contains important
information on the scale height of the LMC as function of radius, as
discussed in detail by Alves \& Nelson (2000). They found that the
velocity dispersion profile does not fall steeply enough with radius
to be consistent with a constant scale-height disk. From this they
concluded that the disk must be `flared', with a scale height that
increases radially outward. This can arise naturally as a result of
tidal forces from the Milky Way. These forces become relatively more
important (compared to the LMC self-gravity) as one moves to larger
radii. Indeed, the $N$-body simulations by Weinberg (2000) predict a
considerable thickness for the LMC disk as a result of the Milky Way
tidal forces (although, somewhat surprisingly, the vertical velocity
dispersion in his simulations remains small).

It is straightforward to refit the isothermal flared disk model of
Alves \& Nelson (2000; their eq.~[37]) to our new kinematical results
and preferred LMC disk parameters. For an isothermal disk the vertical
density profile is proportional to ${\rm sech}^2 (z/z_0)$, where $z_0$
can vary with disk radius. The input to the model of Alves \& Nelson
(2000) includes the total disk mass (Section~\ref{ss:LMCmass}), the
radial exponential scale length of the disk (Paper~II), the total mass
of the LMC dark halo (Section~\ref{ss:LMChalo}), the mean density of
the Milky Way's dark halo at the LMC distance (Section~\ref{ss:mass})
and approximate corrections for the finite thickness and finite radial
extent of the disk. The best fit to our velocity dispersion data
(Figure~\ref{f:rotcurve} and Table~\ref{t:kinematics}) yields $z_0 =
0.27 \kpc$ at the LMC center, rising to $z_0 = 1.5 \kpc$ at a radius
of $5.5 \kpc$. If one prefers to model the disk's thickness with an
exponential instead of a ${\rm sech}^2$ function, then the exponential
scale height is approximately half of $z_0$ (Gilmore, King \& van der
Kruit 1990).

It is useful to compare the LMC disk to the disk components of our
Milky Way galaxy\footnote{We use the kinematical properties of the
Milky Way disks as listed in Table 10.4 of Binney \& Merrifield
(1998). The velocity dispersion components $\sigma_R$, $\sigma_\phi$
and $\sigma_z$ were used to calculate the velocity dispersion $\sigma$
that would be observed if the Milky Way were viewed with the same
inclination as the LMC.}. The thin disk of the Milky Way has $V
/\sigma \approx 9.8$ and the thick disk has $V/\sigma \approx
3.9$. For the LMC we have, averaged between $4$ and $9 \kpc$ as in
Section~\ref{ss:LMCmass}, that $V/\sigma \approx 2.9 \pm 0.9$. The
Milky Way thick disk has an exponential vertical scale height $z_d
\approx 1 \kpc$ (Binney \& Merrifield 1998). Its exponential radial
scale length is $R_d \approx 3 \kpc$, if one assumes that the Milky
Way thin and thick disks have the same radial profile. Hence, at the
solar radius ($\sim 3 R_d$), the Milky Way thick disk is characterized
by a relative thickness $z_d / R_d \approx 1/3$. This is
quantitatively similar to the values inferred for the LMC, upon
application of Alves \& Nelson's (2000) flared disk model to our new
data. At $3 R_d \approx 4.5 \kpc$, the model predicts $z_0 = 1.1
\kpc$. The corresponding exponential scale height (see above) is $z_d
\approx 0.5 \kpc$ so that $z_d / R_d \approx 1/3$. So both in terms of
$V/\sigma$ as well as in terms of $z_d / R_d$, the LMC is very similar
to the thick disk of the Milky Way and very different from the thin
disk of the Milky Way.

There are different approaches to estimate the actual shape and
vertical scale height of the LMC disk. A useful alternative to the
approach employed by Alves \& Nelson (2000) is to assume that the LMC
is spheroidal with axial ratio $q$ (for reference, an infinitely thin
disk has $q \ll 1$ whereas the very flattest elliptical galaxies have
$q \approx 0.3$; Binney \& Merrifield 1998). The axial ratio is
directly related to the quantity $V/\sigma$ through hydrostatic
equilibrium. For a self-gravitating system one can use the relation
provided by the tensor virial theorem (Binney \& Tremaine
1987). However, this relation is not accurate for the LMC, which is
embedded in a dark halo (see Section~\ref{ss:LMChalo} below). So we
consider another simple model instead: we assume that the LMC is
embedded in a spherical isothermal halo, and that it has an isotropic
velocity dispersion that is approximately constant as function of
radius. The equations of van der Marel (1991) then yield that
\begin{equation}
\label{axrat}
  q = \left \lbrace 2 + \left [{2 \over \kappa} (V/\sigma)^2) \right ]
      \right \rbrace^{-1/2} ,
\end{equation}
where, as in Section~\ref{ss:LMCmass}, $-\kappa$ is the logarithmic
slope of the mass density in the equatorial plane: $\kappa \approx
R/R_d$ with $R_d \approx 1.5 \kpc$. This simple model suggest that the
axial ratio ranges from $q \approx 0.31$ at $R = 3\kpc$ to $q \approx
0.46$ at $R=9 \kpc$. These results are qualitatively consistent with
those from Alves \& Nelson's (2000) flaring disk model.

The LMC has $V/\sigma$ considerably larger than unity, and it
therefore remains quite justified to regard the LMC as a rotationally
supported disk-like system. However, it is most definitely not a thin
disk. Although the axial ratio values that we have estimated are based
on simple models, and need to be interpreted with care, it is clear
that the LMC is quite thick. While this does not appear to have been
fully appreciated in the literature, it need not come as a surprise.
The inferred thickness is entirely consistent with the predictions
from simulations (Weinberg 2000).
 
\subsection{Dark Halo}
\label{ss:LMChalo}

The total $B$-band magnitude of the LMC is $0.63$ (de Vaucouleurs \&
Freeman 1973), the $B-V$ color\footnote{The discussion in Section 3c
of Bothun \& Thompson (1988) argues that the $B-V$ colors of the LMC
and SMC in their Table 1 are erroneously reversed.} is $0.61 \pm 0.03$
(Bothun \& Thompson 1988), and the V-band extinction is $A_V \approx
0.4$ (Zaritsky 1999). This implies a total intrinsic luminosity $L_V =
3.0 \times 10^9 \Lsun$. Realistic stellar population synthesis and
evolution models for disk galaxies predict a strong correlation
between color and mass-to-light ratio. To fit the extinction-corrected
$B-V$ color, the mass-to-light ratio of the stellar population should
be $M/L_V \approx 0.9 \pm 0.2$ (Bell \& de Jong 2001), with some
residual dependence on the assumed initial mass function and its
low-mass cut-off. This implies a total mass for the visible disk of
the LMC of $M \approx 2.7 \times 10^9 \Msun$. The mass of the neutral
gas in the LMC has been estimated to be $0.5 \times 10^9 \Msun$ (Kim
\etal 1998). The combined mass of the visible material in the LMC is
therefore insufficient to explain the dynamically inferred mass
$M_{\rm LMC} (8.9 \kpc) = (8.7 \pm 4.3) \times 10^9 \Msun$
(Section~\ref{ss:LMCmass}). Consequently, the LMC must be embedded in
a dark halo. This is consistent also with the fact that the observed
rotation curve amplitude is relatively flat as a function of radius.
To determine the properties of the LMC dark halo one must: (i) model
the contributions to the circular velocity from gas, stars and dark
matter; (ii) calculate the asymmetric drift; (iii) perform a detailed
data-model comparison. Such a detailed analysis is beyond the scope of
the present paper. Alves \& Nelson (2000) did perform such modeling,
but their results will need to revised in view of the present paper:
they used an observed rotation curve amplitude that is 40\% higher
than the one that we have inferred here, and they ignored asymmetric
drift.

The observed rotation velocity $V(R)$ listed in
Table~\ref{t:kinematics} rises very slowly; it does not reach its flat
part until $R \approx 4 \kpc$. This is different from the behavior of
the circular velocity curves $V_{\rm circ}(R)$ predicted by realistic
models, which reach velocities near their maximum already at $R
\approx 2 \kpc$ (Alves \& Nelson 2000). This indicates, independent of 
model details, that $V < V_{\rm circ}$ in the central region of the
LMC. Our favored explanation for this is that it is the result of
asymmetric drift. In the central $R \lta 2 \kpc$, $V/\sigma$ is of
order unity. We do not think that it is due to some error in the
rotation curve $V(R)$ inferred from the observations. At small radii,
$V(R)$ is rather insensitive to errors in $v_{tc}$,
cf.~Figure~\ref{f:rotcurve}. There also is no strong dependence on the
assumed position angle of the line of nodes. Alves (2000) also found
that $V < V_{\rm circ}$ at $R \lta 4 \kpc$, despite the use of a very
different line of nodes position angle. Nonetheless, one must not
forget that the central region of the LMC contains a strong bar. So it
is quite likely that the streaming velocity field in this region is
considerably more complicated than the circular streaming model that
we have used in our analysis. A proper understanding of the kinematics
and mass distribution of the LMC in its central few kpc will require
detailed modeling of both bar dynamics and asymmetric drift.

\subsection{Microlensing Optical Depth}
\label{ss:lensing}

The observed microlensing optical depth towards the LMC is $\tau_{\rm
obs} = 12^{+4}_{-3} \times 10^{-8}$ with an additional 20--30\% of
systematic error (Alcock \etal 2000). To interpret this result it is
critically important to know the self-lensing optical depth of the
LMC, $\tau_{\rm self}$. This subject was addressed and reviewed most
recently by Gyuk \etal (2000). For their most favored set of LMC model
parameters\footnote{The relevant estimate to use for comparison to the
Alcock \etal (2000) data is the 30-field average in Table~4 of Gyuk
\etal (2000).} they predict $\tau_{\rm self} = 2.2 \times 10^{-8}$, a
factor of $5.5$ less than the observed value. It is useful to address
whether our new results on LMC structure change this conclusion. 

To lowest order, the LMC self-lensing optical depth depends
exclusively on the observed velocity dispersion and not separately on
either the galaxy mass or scale height (Gould 1995). The values that
we have inferred here for the line-of-sight velocity dispersion of the
LMC are not very different from those used by previous authors. This
argues that, to lowest order, our new findings do not change the
predictions for the LMC self-lensing optical depth. In principle, the
expression derived by Gould (1995) is somewhat idealized. It relies on
the assumptions that: (a) the LMC disk is thin; and (b) the observed
velocity dispersion tracks the scale height as in simple equilibrium
models. The first assumption is suspect based on our discussion in
Section~\ref{ss:height} and the second assumption has been shown to be
violated in the $N$-body simulations of Weinberg (2000). One could
therefore argue that it is prudent to consider more sophisticated
models. However, this was done by Gyuk \etal (2000), and they did not
find results that differ greatly from the simple predictions obtained
directly from the velocity dispersion. Alves \& Nelson (2000)
incorporated the flaring of the LMC disk, and this too was found not
to increase $\tau_{\rm self}$ very significantly. The MACHO survey
fields used to derive $\tau_{\rm obs}$ are mostly situated in the
central 2 kpc of the LMC (Alcock \etal 2000), and at these radii the
flaring is not yet substantial enough to make a big difference (see
Section~\ref{ss:height}). In view of these arguments, it appears
unlikely that the revised understanding of LMC structure that has
emerged from the present study will significantly increase the
predicted LMC self-lensing optical depth.

For LMC self-lensing models to become a viable explanation for the
observed lensing events they will not only have to correctly predict
$\tau_{\rm obs}$, but also the observed spatial and time-scale
distributions of the lensing events (Gyuk \etal 2000; Alcock \etal
2000). Proposed tests of the lensing population based on the spatial
distribution of the events may need to be revisited in light of our
new and coherent understanding of LMC structure. For example, the
tests discussed by Alcock \etal (2000; their section 5.3; see also
Gyuk \etal 2000) are based on an incorrect line of nodes position
angle of $170^{\circ}$. This is quite different from the actual value,
which has now been found using several different methods to be in the
range $120^{\circ}$--$130^{\circ}$ (Papers~I and~II;
Section~\ref{s:data}). This error could become important, particularly
if larger samples of microlensing events become available. The flare
of the LMC disk is also important for self-lensing models. It causes
the spatial distribution to more resemble that expected for LMC halo
lenses.  The event rate will fall off more gradually with radius than
if the disk had a constant thickness, and the near-to-far side event
rate asymmetry will be less. It may be worth investigating new tests
of the spatial event distribution that build on our improved
understanding of the LMC structure (e.g., a test based on the event
distance perpendicular to the line-of-nodes).

\section{The Orbit of the LMC}
\label{s:disc}

\subsection{Systemic Velocity}
\label{ss:vsys}

As mentioned in Section~\ref{s:data}, the kinematical properties of
the LMC have been studied previously using many different tracers.
Each of these studies has yielded an estimate of the systemic
velocity of the LMC. The result that has most often been adopted in
models of the Milky Way dark halo and the Magellanic Stream (e.g.,
Wilkinson \& Evans 1999; Gardiner \etal 1994) is the value $v_{\rm
sys} = 274 \kms$ inferred from HI data by Luks \& Rohlfs (1992). This
value is similar to the value $v_{\rm sys} = 279 \kms$ inferred
subsequently, also from HI data, by Kim \etal (1998). However, the HI
distribution and kinematics in the LMC are quite disturbed. For
example, Luks \& Rohlfs required two separate HI velocity components
to fit their data, and as discussed in Section~\ref{ss:center}, the HI
dynamical center probably does not coincide with the LMC CM. The true
systemic velocity, i.e., the line-of-sight velocity of the CM, is
therefore better estimated from discrete tracers. From our analysis of
1041 carbon stars\footnote{We verified that the two subsets of our
carbon star data set, i.e., those obtained by Kunkel \etal (1997) and
Hardy, Schommer \& Suntzeff (2002, in preparation), respectively,
yield mutually consistent values for $v_{\rm sys}$ (to within the
error bars). So there is no evidence to doubt the velocity calibration
accuracy of either data set.} we have obtained $v_{\rm sys} = 262.2
\pm 3.4 \kms$ (eq.~[\ref{fitresults}]).  The error in the
determination of $v_{\rm sys}$ scales as $1/\sqrt{N}$, where $N$ is
the number of tracers. Previous studies with other tracers generally
had much smaller values of $N$ than available for the present
study. For example, Meatheringham \etal (1988) studied 94 planetary
nebulae; Schommer \etal (1992) studied 83 star clusters; etc. We
conclude that our estimate of $v_{\rm sys}$ is the most accurate one
currently available.

\subsection{Transverse Velocity}
\label{ss:vtransdisc} 

The observed proper motion of the LMC quoted in
equation~(\ref{weightB}) yields an estimate of the transverse velocity
of the LMC. With the help of equations~(\ref{vtmurel})
and~(\ref{distmodulus}) one obtains
\begin{equation}
\label{vtrans}
  v_x = -399 \kms , \qquad 
  v_y =   80 \kms ,
\end{equation}   
where the $x$ and $y$ directions point towards the West and North,
respectively. This result corresponds to a velocity $v_t = 406 \kms$
in the direction of position angle $\Theta_t = 78.7^{\circ}$. The
uncertainty in the distance $D_0$ causes the error $\Delta v$ in the
transverse velocity to be larger in the direction $\Theta_t$ than in
the direction $\Theta_t + 90^{\circ}$:
\begin{equation}
\label{vtranserror}
  \Delta v (\Theta_t)              = 44 \kms , \qquad
  \Delta v (\Theta_t + 90^{\circ}) = 37 \kms .
\end{equation}
The $68.3$\% confidence region on the transverse velocity of the LMC
is shown as a heavy ellipse in Figure~\ref{f:vtrans}.

The analysis of the line-of-sight velocity field also constrains the
transverse velocity of the LMC. This velocity is determined completely
by the orthogonal components $v_{tc}$ and $v_{ts}$, measured along and
perpendicular to the line of nodes, respectively
(Figure~\ref{f:vtrans}). The component $v_{tc}$ is weakly constrained
by the velocity field, $v_{tc} \> \in \> [200,600] \kms$
(eq.~[\ref{vtcconstraint}]). The component $v_{ts} = w_{ts} - D_0
(di/dt)$ (eq.[\ref{wtsdef}]) can be calculated only if $w_{ts}$, $D_0$
and $di/dt$ are all known. While $w_{ts}$ and $D_0$ are rather tightly
constrained, $di/dt$ is unfortunately completely unknown. So in
Section~\ref{s:distdidt} we used this relation to estimate $di/dt$,
using the value of $v_{ts}$ implied by the proper motion data. For the
purpose of visualization it is now useful to reverse this situation
for a moment, and assume that $di/dt$ is known. In particular, let us
assume that $di/dt = 0$. In this case one obtains from the observed
line-of-sight velocity field an estimate of the transverse velocity of
the LMC that is completely independent from the available proper
motion and distance measurements. The 68.3\% confidence region thus
obtained is the trapezoid\footnote{Some brief explanation is in order
as to why the confidence region on the transverse velocity obtained
from the velocity field is a trapezoid. To calculate $v_x$ and $v_y$
it is insufficient to know the components $v_{tc}$ and $v_{ts}$ along
and perpendicular to the line of nodes. One needs to know instead the
components along fixed position angles on the sky. Let these angles be
$\Theta_0$ and $\Theta_0 + 90^{\circ}$, and let the corresponding
velocity components be $v_{tc0}$ and $v_{ts0}$. We choose $\Theta_0 =
129.9^{\circ}$, which is the best estimate of the line-of-nodes
(eq.~[\ref{fitresults}]). Propagation of errors then yields for the
formal error in $v_{ts0}$ that $\Delta v_{ts0} = [ (\Delta v_{ts})^2 +
v_{tc0}^2 (\Delta \Theta)^2 ]^{1/2}$. Here $\Delta \Theta = 0.10$ is
expressed in radians and $\Delta v_{ts} = 13.0 \kms$
(eq.~[\ref{fitresults}]). Because $v_{tc0} > 200
\kms$ (eq.~[\ref{vtcconstraint}]) one finds that the error in $v_{ts0}$ 
is dominated by the error in the position angle of the line of nodes,
and that $\Delta v_{ts0} \approx 0.10 \> v_{tc0}$. Visual
representation of this relation yields a trapezoid.} shown with dashed
lines in Figure~\ref{f:vtrans}. It overlaps (albeit slightly) with the
68.3\% confidence region obtained from the proper motion data. The
agreement can be improved by varying $di/dt$, which shifts the
trapezoid along the $v_{ts}$ direction shown in Figure~\ref{f:vtrans}.
The agreement is optimized for $di/dt = -0.37 \masyr$, the value
derived in Section~\ref{s:distdidt}. The 68.3\% confidence region for
this value is shown as a solid trapezoid in
Figure~\ref{f:vtrans}. Note, that the shift in the $(v_x,v_y)$ plane
induced by this $di/dt$ is fairly small. In addition, it was shown in
Section~\ref{s:distdidt} that the simulations of Weinberg (2000)
actually predict a somewhat smaller absolute value of $di/dt$. So
theory predicts that the true transverse velocity of the LMC should
not be very different from the value predicted from the velocity field
under the assumption that $di/dt = 0$. This is exactly what the proper
motion measurements indicate.

\subsection{Three-Dimensional Space Motion}
\label{ss:orbit} 

Our knowledge of the transverse velocity of the LMC is becoming
steadily more accurate as the number of independent proper motion
measurements increases. We also know now, based on the results shown
in Figure~\ref{f:vtrans}, that the proper motion measurements yield a
picture that is consistent with the results from analysis of the
line-of-sight velocity field. Therefore, there is no reason to believe
that the proper motion measurements are plagued by large systematic
errors. It is therefore prudent to ask what the measurements imply for
the three-dimensional space motion of the LMC. This issue has been
addressed previously by several authors, but the results from the
present paper now allow us to address this with higher confidence.

To determine the three-dimensional motion of the LMC with respect to
the Milky-Way we need to correct for the reflex-motion of the
sun. Following Gardiner \etal (1994) and Kroupa \& Bastian (1997) we
adopt a Cartesian coordinate system $(X,Y,Z)$, with the origin at the
Galactic Center, the $Z$-axis pointing towards the Galactic North
Pole, the $X$-axis pointing in the direction from the sun to the
Galactic Center, and the $Y$-axis pointing in the direction of the
sun's Galactic Rotation. The position and velocity vectors of the sun
are
\begin{equation}
\label{solarvectors}
  {\vec r}_{\odot} = (-R_0, \> 0, \> 0) , \qquad 
  {\vec v}_{\odot} = (U_{\odot}, \> V_0 + V_{\odot}, \> W_{\odot}) .
\end{equation}
The quantities $(R_0,V_0)$ are the distance of the sun from the
Galactic Center, and the circular velocity of the Milky Way at the
position of the sun. We use the standard IAU values $R_0 = 8.5
\kpc$ and $V_0 = 220 \kms$ (Kerr \& Lynden-Bell 1986). The vector
$(U_{\odot}, V_{\odot}, W_{\odot})$ is the sun's velocity with respect
to the Local Standard of Rest for which we use the recent
determination by Dehnen \& Binney (1998): $(U_{\odot}, V_{\odot},
W_{\odot}) = (10.0 \pm 0.4,\> 5.2 \pm 0.6, \> 7.2 \pm 0.4) \kms$.

Let $(l,b)$ be the Galactic coordinates of the LMC. The right
ascension and declination of the dynamical center of the LMC
determined in Section~\ref{s:data} yield, using equation~(2.1) of Binney
\& Merrifield (1998)\footnote{An error in this equation was
corrected; the longitude of the North Celestial Pole is $l_{\rm CP} =
122.932^{\circ}$.}, $l = 280.531^{\circ}$ and $b =
-32.523^{\circ}$. The unit vector from the sun in the direction
towards the LMC is ${\vec u}_1 = (\cos b \cos l, \cos b \sin l, \sin
b)$, so that the position vector of the LMC is ${\vec r}_{\rm LMC} =
{\vec r}_{\odot} + D_0 {\vec u}_1$. For the distance $D_0$ to the LMC
we adopt the value from equation~(\ref{distmodulus}). This implies
\begin{equation}
\label{LMCposition}
  {\vec r}_{\rm LMC} = (-0.78,\> -41.55, \> -26.95) \kpc . 
\end{equation}
By numerical differentiations of the unit vector ${\vec u}_1$ with
respect to right ascension and declination, respectively, one can
calculate the unit vectors ${\vec u}_2$ and ${\vec u}_3$ in the
directions West and North (in the plane of the sky, as seen from the
sun). The unit vectors thus obtained,
\begin{eqnarray}
\label{unitvecs}
  {\vec u}_1 & = & (  0.15410 , \> -0.82897 , \> -0.53764) , \nonumber \\
  {\vec u}_2 & = & (  0.06860 , \>  0.55180 , \> -0.83115) , \\
  {\vec u}_3 & = & ( -0.98567 , \> -0.09120 , \> -0.14190) , \nonumber
\end{eqnarray}
form an orthonormal basis set of the $(X,Y,Z)$ space. The
velocity vector of the LMC corrected for the reflex motion of the sun
is
\begin{equation}
\label{LMCvelocity}
  {\vec v}_{\rm LMC} = {\vec v}_{\odot} + 
                       v_{\rm sys} {\vec u}_1 +
                       v_x {\vec u}_2 + v_y {\vec u}_3 ,
\end{equation}
where $v_{\rm sys}$ is the systemic velocity of the LMC
(eq.~[\ref{fitresults}]) and $v_x$, $v_y$ are the transverse velocity
components of the LMC (eqs.~[\ref{vtrans},\ref{vtranserror}]). The
total velocity of the LMC in the $(X,Y,Z)$ system is $v_{\rm LMC}
\equiv |{\vec v}_{\rm LMC}|$. The radial velocity component is $v_{\rm
LMC,rad} = {\vec v}_{\rm LMC} \cdot {\vec r}_{\rm LMC} / |{\vec
r}_{\rm LMC}|$.  Equation~(\ref{LMCposition}) yields for the distance
of the LMC form the Galactic Center
\begin{equation}
\label{rLMCdef}
   r_{\rm LMC} \equiv |{\vec r}_{\rm LMC}| = 49.53 \kpc .
\end{equation}
The tangential velocity component of the LMC velocity is $v_{\rm
LMC,tan} = (v_{\rm LMC}^2 - v_{\rm LMC,rad}^2)^{1/2}$.  These
equations yield for the three-dimensional space velocity of the LMC:
\begin{eqnarray}
\label{LMCvelres}
  {\vec v}_{\rm LMC} & = & (  -56 \pm 36 , \> 
                             -219 \pm 23 , \>
                              186 \pm 35 ) \kms , \nonumber \\
  v_{\rm LMC} & = & 293 \pm 39 \kms , \quad
  v_{\rm LMC,rad} =  84 \pm  7 \kms , \quad
  v_{\rm LMC,tan} = 281 \pm 41 \kms . 
\end{eqnarray}
The errors in equation~(\ref{LMCvelres}) were obtained by propagation
of the errors in $v_{\rm sys}$, $v_x$ and $v_y$ using a simple
Monte-Carlo scheme. The errors in the model parameters $R_0$, $V_0$,
$U_{\odot}$, $V_{\odot}$, $W_{\odot}$, $D_0$, and the position of the
LMC center on the sky were not explicitly taken into account. However,
it was verified that variation of these parameters within their error
ranges does not change any of the values in equation~(\ref{LMCvelres})
by more than a few km/s. The only exception is $V_0$, which adds
directly into the $Y$-component of ${\vec v}_{\rm LMC}$. This changes
$v_{\rm LMC}$, $v_{\rm LMC,rad}$ and $v_{\rm LMC,tan}$ as well. So a
considerably non-standard value of $V_0$ (e.g., $V_0 = 184 \kms$, as
advocated by Olling \& Merrifield 1998) would change the results in
equation~(\ref{LMCvelres}) by a few tens of $\kms$.

The values in equation~(\ref{LMCvelres}) are not hugely dissimilar
from those calculated by, e.g., Kroupa \& Bastian (1997) and Pedreros
\etal (2002). The main difference is that our error bars
are considerably smaller. This is partly due to the use of a larger
data set, and partly due to a more accurate method of error
propagation.  The results for the LMC in equation~(\ref{LMCvelres})
now provide one of the highest accuracy space velocities of any
Galactic satellite (velocities for other satellites can be found in,
e.g., Wilkinson \& Evans [1999] and Dinescu \etal [2001]).

\subsection{Mass and Extent of the Milky Way}
\label{ss:mass}

Models generally suggest that the Magellanic Stream represents material
torn from the Magellanic Clouds during the previous perigalactic
passage (e.g., Lin \etal 1995). The fact that this is not the first
passage implies that the LMC must be bound to the Milky Way (but see
Shuter 1992 for an alternative view). So the binding energy ${\cal E}
= \Psi - {1\over2} v^2$ must be positive (Binney \& Tremaine 1987).
It is useful to study what this implies for the mass and extent of the
Milky Way.

We adopt a simple model in which the Milky Way dark halo has a
spherically symmetric mass density $\rho(r)$ with an isothermal
profile and a sharp cut-off at $r = r_h$:
\begin{equation}
\label{density}
  \rho = V_0^2 / 4 \pi G r^{2}      , \quad (r \leq r_h) ; \qquad
  \rho = 0                                , \quad (r > r_h) ,
\end{equation}
where, as before, $G$ is the gravitational constant. The velocity
$V_0$ is the (constant) circular velocity of the halo for $r \leq
r_h$, which we assume to be $V_0 = 220 \kms$ as in
Section~\ref{ss:orbit}. There exist other parameterizations for the
mass density that produce a smoother truncation at $r_h$ (e.g., Lin \&
Lynden-Bell 1982; Wilkinson \& Evans 1999), but use of such a profile
would not change the basic argument presented here. The gravitational
potential is
\begin{equation}
\label{potential}
  \Psi = V_0^2 [1 + \ln (r_h/r) ]  , \quad (r \leq r_h) ; \qquad
  \Psi = V_0^2 (r_h/r)             , \quad (r > r_h)    ,
\end{equation}
The enclosed mass is
\begin{equation}
\label{massenclosed}
  M(r) = r   V_0^2 / G                      , \quad (r \leq r_h) ; \qquad
  M(r) = r_h V_0^2 / G \equiv M_{\rm MW}    , \quad (r > r_h)    ,
\end{equation}
where $M_{\rm MW}$ is the total mass of the Milky Way. It is useful
to define some ancillary quantities:
\begin{equation}
\label{etaxi}
  \zeta \equiv r_h / r_{\rm LMC}                , \qquad
  \xi   \equiv {1\over2} (v_{\rm LMC} / V_0)^2  , \qquad
  M'    \equiv r_{\rm LMC} v^2_{\rm LMC} / 2 G  . 
\end{equation}
The condition ${\cal E} > 0$ that the LMC is bound to the Milky Way
implies for the mass enclosed within the present radius of the LMC 
\begin{equation}
\label{Menrlmc}
  M(r_{\rm LMC}) \geq M'
      , \quad (\zeta \leq 1) ; \qquad
  M(r_{\rm LMC}) \geq M' / (1 + \ln \zeta)
      , \quad (\zeta \geq 1) .
\end{equation}
Correspondingly, the total mass of the Milky Way must satisfy
\begin{equation}
\label{Mentot}
  M_{\rm MW} \geq M'
      , \quad (\zeta \leq 1) ; \qquad
  M_{\rm MW} \geq M' \zeta / (1 + \ln \zeta)
      , \quad (\zeta \geq 1) .
\end{equation}
These equations illustrate a basic property of the models that fit the
data: one can decrease the required mass $M(r_{\rm LMC})$ inside
$r_{\rm LMC}$ by increasing $\zeta$, but this increases the total mass
$M_{\rm MW}$. Over the domain $\zeta \geq 1$, the function $h(\zeta)
\equiv \zeta / (1 + \ln \zeta)$ has a minimum $h(\zeta) = 1$ at $\zeta = 1$. 
So if one has no knowledge of the extent of the halo, as quantified by
$\zeta$, then the strict lower limit on the mass of the Milky Way is
\begin{equation}
\label{Mentotstrict}
  M_{\rm MW} \geq M'  .
\end{equation}
In the present situation we do have a limit on $\zeta$. We have assumed
knowledge of $V_0$, and this fixes the normalization of the mass
density (eq.[\ref{density}]). This allows one to calculate the minimum
halo extent $r_h \equiv \zeta r_{\rm LMC}$ for which $M_{\rm MW} \equiv
r_h V_0^2 / G$ satisfies equation~(\ref{Mentot}). The result is:
\begin{equation}
\label{halosize}
  \zeta \geq \xi          , \quad (\xi \leq 1) ; \qquad
  \zeta \geq e^{\xi-1}    , \quad (\xi \geq 1) .
\end{equation}
Substitution of $r_{\rm LMC}$ and $v_{\rm LMC}$ from
equations~(\ref{rLMCdef}) and~(\ref{LMCvelres}) and use of a simple
Monte-Carlo scheme to propagate errors yields
\begin{equation}
\label{halolimits}
  M_{\rm MW} \geq 4.3 \times 10^{11} \Msun   , \qquad
  r_h \geq 39 \kpc                           , \qquad 
  (68.3\% \> {\rm confidence}) . 
\end{equation}

It is interesting to compare these constraints from the assumption that
the LMC is bound to the Milky Way with the results from studies of the
ensemble dynamics of all the Milky Way satellites.  Kochanek (1996)
obtained for the mass of the Milky Way inside $50
\kpc$ that $M (50 \kpc) = (4.9 \pm 1.1) \times 10^{11} \Msun$ and 
Wilkinson \& Evans (1999) obtained $M (50 \kpc) = 5.4^{+0.2}_{-3.6}
\times 10^{11} \Msun$. Both studies indicated that the dark halo extends 
considerably beyond $50 \kpc$ and that $M_{\rm MW}$ exceeds $M (50
\kpc)$ by a factor of a few. These results are fully consistent with
equation~(\ref{halolimits}).

The mass constraint in equation~(\ref{halolimits}) is a factor $\sim
2$ smaller than the limit obtained by Pedreros \etal (2002), also from
the LMC proper motion. Their result assumes that all the mass of the
halo is enclosed within the present radius of the LMC. This is in
contradiction with the results of Kochanek (1996) and Wilkinson \&
Evans (1999), and the results in equation~(\ref{halolimits}) should
therefore be more accurate.

\subsection{Magellanic Stream}
\label{ss:stream}

The properties of the LMC orbit and the Milky Way dark halo are
important parameters in models of the Magellanic Stream. Recent models
generally assume an extended, usually isothermal, dark halo,
consistent with the arguments of Section~\ref{ss:mass}. Models that
fit the generic properties of the Stream suggest that the LMC is just
past pericenter in its orbit. The predicted transverse velocity of the
LMC has generally been found to be in the range $v_t = 400$--$500
\kms$ approximately due East (reviewed by, e.g., Westerlund 1997). 
This is consistent with the observed transverse velocity (see
Figure~\ref{f:vtrans}). In Galactocentric coordinates, all authors
tend to agree that the component $v_{\rm LMC,rad}$ is small. For
$v_{\rm LMC,tan}$, a wide range of physical arguments has yielded
values that include $v_{\rm LMC,tan} = 369 \kms$ (Lin \& Lynden-Bell
1982), $355 \kms$ (Shuter 1992), $352 \kms$ (Heller \& Rohlfs 1994),
$339 \kms$ (Murai \& Fujimoto 1980), $320 \kms$ (Liu 1992) and $285
\kms$ (Gardiner \etal 1994; Gardiner \& Noguchi 1996),
respectively. The observationally derived value is $v_{\rm LMC,tan} =
281 \pm 41\kms$ (equation~(\ref{LMCvelres})), which is consistent with
the lower range of the predicted values. The data are most consistent
with the models of Gardiner \etal (1994) and Gardiner \& Noguchi
(1996), which are some of the most detailed models that have
been constructed for the Magellanic Stream. This agreement is actually
quite remarkable, given the many uncertainties and simplifications
involved in models of the Magellanic Stream. The observational error
on $v_{\rm LMC,tan}$ is almost small enough to start ruling out
specific models.

Lin \etal (1995) modeled the Magellanic Stream using a different
approach than most authors. They fixed the orbit to that implied by
the proper motion measurement of Jones \etal (1994), which was one of
the first measurements to be made, and then explored different halo
potentials. The low Galactocentric velocity $v_{\rm LMC,tan} = 213 \pm
49 \kms$ implied by this proper motion measurement could only be
reproduced with a rather `lean' dark halo. With only $\sim 3 \times
10^{11} \Msun$ inside $50 \kpc$ their mass model is inconsistent with
the results obtained and discussed in Section~\ref{ss:mass}. We
attribute this to the limited accuracy of the Jones \etal (1994)
proper motion measurement. The average proper motion inferred from all
the data in Table~\ref{t:propmotion} is in fact consistent with
canonical (isothermal) models of the dark halo.

The Jones \etal value of $v_{\rm LMC,tan}$ was consistent with the
circular velocity of the Milky Way dark halo. Based on this it has
often been said in the literature that the LMC is on an approximately
circular orbit around the Milky Way. The results presented here do not
support this view. They are consistent with the orbit advocated by
Gardiner \etal (1994) and Gardiner \& Noguchi (1996), which has an
apocenter to pericenter ratio of $\sim 2.5:1$. The perigalactic
distance is $\sim 45 \kpc$ and the present orbital period around the
Milky Way is $\sim 1.5 \Gyr$.

There remains considerable debate about the detailed origin of the
Magellanic Stream. While tidal stripping effects have undoubtedly
played an important role, it remains quite possible that gas dynamical
effects and ram-pressure stripping have also played an important part
in the shaping of the Stream (e.g., Heller \& Rohlfs 1994; Moore \&
Davis 1994). In view of this it is unlikely that any modeling of the
Stream will soon lead to unambiguous predictions for three-dimensional
velocity of the LMC and its orbit. The only thing that can be safely
concluded now is that observational measurements of the
three-dimensional velocity of the LMC are generically consistent with
models for the Magellanic Stream.

\section{The Predicted LMC Proper Motion Velocity Field}
\label{s:propfield}

With future astrometric observations it may become possible to measure
the proper motion velocity field of the LMC. This velocity field can
be calculated with the formulae of Section~\ref{s:vmath} for any
combination of model parameters. We have performed this calculation
for our best model of the LMC structure and kinematics, characterized
by the parameters given in equations~(\ref{inclination},
\ref{fitresults}, \ref{distmodulus}, \ref{didtresult}, \ref{vtrans}). 
The rotation velocity curve $V(R')$ was parameterized as in
equation~(\ref{Vcurve}), with the parameters
\begin{equation}
\label{Vrotparm}
  V_0 = 49.7 \kms   , \qquad 
  R_0/D_0 = 0.055   , \qquad
  \eta = 2.68       ,
\end{equation}
chosen to best fit the points with error bars in
Figure~\ref{f:rotcurve}. In the absence of any observational
information on $d\Theta/dt$, this quantity was assumed to be
zero. Figure~\ref{f:propfield} shows the resulting (residual) proper
motion velocity field (i.e., the proper motion of the LMC CM was
subtracted).

The predicted proper motion velocity field is quite complex, and lacks
any of the simple symmetries that one might have expected. While a
small part of this is due to the fact that a plot of right ascension
versus declination provides a distorted view of the sky, most of it is
due to the fact that there are several different components that
contribute to the velocity field. To provide some insight into the
results, we show the individual components that contribute to the
velocity field in Figure~\ref{f:propcomp}. The top left panel shows
the contribution due to the CM velocity component ${\vec v}_{\rm CM}$
(see eq.~[\ref{vecCM}]). This component dominates the total velocity
field at most positions. It arises from the fact that one observes
different components of the CM velocity vector at different positions
on the sky. In particular, one observes part of the systemic velocity
of the LMC in the $v_2$ component, so that $v_2^2 + v_3^2 \not= v_t^2$
unless one observes at the position of the CM (see eq.~[\ref{vecCM}]);
here $v_2$ and $v_3$ are the velocity components perpendicular to the
line of sight, as defined in Section~\ref{s:vmath}. The top right
panel in Figure~\ref{f:propcomp} shows the contribution due to the
internal rotation component ${\vec v}_{\rm int}$ (see
eq.~[\ref{vecint}]) for the best model. This contribution has the
expected morphology of a clockwise rotation in the plane of the
sky. However, it does have a non-zero radial component ($v_2 \not= 0$)
and a tangential component that varies as a function of position
angle. The contribution of the $di/dt$ component to the velocity field
in Figure~\ref{f:propfield} is small. However, it still is useful to
have some insight into the morphology of this component. The bottom
left panel of Figure~\ref{f:propcomp} shows this component for an
assumed value $(di/dt) = 1 \masyr$, which exceeds the value in our
best fit model by a factor of $\sim 3$. For comparison, the bottom
right panel of Figure~\ref{f:propcomp} shows the contribution of a
$d\Theta/dt$ component of the same size (i.e., $1 \masyr$). As
mentioned above, the latter contribution was set to zero in
Figure~\ref{f:propfield} for lack of observational constraints.

Figures~\ref{f:propfield} and~\ref{f:propcomp} demonstrate a few
important issues. First, as one moves away from the LMC center the
residual proper motion velocity field can be quite significant as
compared to the proper motion vector of the CM. Within the range of
positions shown in Figures~\ref{f:propfield}, the predicted proper
motion can differ by more than $1 \masyr$ from the value for the
CM. Second, the majority of the residual proper motion is due to the
velocity vector of the CM, and not due to the rotation of the LMC
itself. It is therefore crucial that this contribution be properly
included in any interpretation of proper motion measurements that are
not made very close to the CM. And third, there may be contributions
to the proper motion field due to precession and nutation of the LMC
disk plane. These cannot be neglected in studies where high accuracy
is desired.

\section{Prospects for Kinematical Distance Determination of the LMC}
\label{s:distprospects}

As described in Section~\ref{ss:LMCdistance}, there is still
considerable uncertainty and controversy in the determination of the
LMC distance $D_0$. Any new method that can shed additional
information on this quantity is therefore potentially valuable. In
this context it is interesting to note that equation~(\ref{wtsdef})
can be solved for the LMC distance $D_0$ to obtain
\begin{equation}
\label{newdistmethod}
   D_0 = w_{ts} / [\mu_s + (di/dt)] .   
\end{equation}
The quantity $w_{ts}$ is determined by the line-of-sight velocity
field, as is the line-of-nodes position angle $\Theta$
(eq.[\ref{fitresults}]). The proper motion component $\mu_s$ along the
line of nodes can be measured. So if $di/dt$ can be measured, or if
one assumes that it is known, e.g., identical to zero, then this
equation provides a new method for determining the LMC distance. The
method is purely kinematical and bypasses all knowledge of stellar
evolution, with its associated uncertainties. So it is complementary
to most existing methods for determination of the LMC distance.

The underlying idea of equation~(\ref{newdistmethod}) is that the
line-of-sight velocity field yields the transverse velocity vector of
the LMC in km/s, whereas proper motion measurements yield the same
vector in milli-arcsec per year. Combination of the two measurements
yields the distance. It is not a problem that the line-of-sight
velocity field only constrains one component of the transverse
velocity: one only needs one component to determine the distance. This
method is essentially a variant to the `radial velocity gradient'
method which has been used to determine the distance to open clusters
(e.g., Gunn \etal 1988). However, open clusters do not rotate
appreciably, so there is no difficulty in disentangling the dynamics
of the cluster from the apparent solid body rotation induced by the
transverse motion. Gould (2000) was the first to point out that the
method can also be used for the LMC, despite the fact that the LMC is
itself rotating. Our approach is a variation on that proposed by
Gould, but with important differences. First, we have devised a
methodology and software that allow a practical application to real
data. Second, Gould's suggested approach relied on the assumption that
the kinematic line of nodes can be approximated to be the same as the
photometric major axis. This is true for a circular disk, but is now
known to be highly incorrect for the LMC (Paper~II showed that the
major axis position angle is ${\rm PA}_{\rm maj} = 189.3^{\circ} \pm
1.4^{\circ}$, which differs from the kinematic line of nodes by $\sim
60^{\circ}$). We have shown here that that the kinematic line of nodes
can be determined independently from the line-of-sight velocity field,
so that the method can still be used even if the photometric major
axis does not trace the line of nodes.

Two major hurdles must be overcome for the proposed method to ever
become practical: first, accurate independent knowledge must be
obtained on the value of $di/dt$; and second, the formal errors in our
knowledge of $w_{ts}$ and $\mu_s$ must be reduced. These points are
illustrated by Figure~\ref{f:distdidt}. It shows the $68.3$\%
confidence region in the plane spanned by the parameters $di/dt$ and
$m-M$ (the distance modulus corresponding to $D_0$), obtained with the
help of equations~(\ref{mucsdef}, \ref{fitresults}, \ref{weightB}, 
\ref{newdistmethod}). In Section~\ref{s:distdidt} we used 
the independent estimate of the distance given by
equation~(\ref{distmodulus}) to obtain an estimate for $di/dt$
(eq.~[\ref{didtresult}]). If instead the goal is to determine $m-M$,
then one needs an independent estimate of $di/dt$. This estimate must
be accurate to $0.03 \masyr$ to obtain $m-M$ to better than $0.05$
mag. Gould (2000) did not address the influence of $di/dt$, which is
the same as assuming that $di/dt = 0$. This is clearly not acceptable
to obtain an accurate distance estimate. With the present data it
implies that $m-M = 19.08 \pm 0.36$, which is inconsistent with the
generally accepted value (eq.~[\ref{distmodulus}]) at the $1.6 \sigma$
level. Even if $di/dt$ were known with infinite accuracy, the method
wouldn't presently be very competitive. At fixed $di/dt$, the error in
$m-M$ is in the range $0.3$--$0.4$ mag, too large to contribute
meaningfully to the debate over the LMC distance.

The main advance for application of the proposed method will come from
future astrometric space missions such as SIM and GAIA. These can
measure the proper motion of the LMC CM (and thus the component
$\mu_s$ that features in eq.~[\ref{newdistmethod}]) to $1$\% or better
(Gould 2000). This will reduce the width of the confidence band in
Figure~\ref{f:distdidt}. In addition, they will yield information on
the internal proper motion velocity fields of the LMC (for which
predictions were shown in Figures~\ref{f:propfield}
and~\ref{f:propcomp}). The three components $(v_1,v_2,v_3)$ of the
three-dimensional velocity (eq.~[\ref{vonetwothreedef}]) each
constrain a combination of the quantities $D_0$, $di/dt$ and
$d\Theta/dt$, as shown in Section~\ref{s:vmath}. These can be combined
to obtain independent estimates of all three of these
quantities. Information on the rotation of the LMC in the plane of the
sky will also yield an estimate of the rotation curve $V(R')$ in
mas/yr, which can be combined with measurements obtained from
line-of-sight velocities in km/s to obtain another constraint on the
LMC distance. This is called `rotational parallax' (Olling \& Peterson
2002). With sufficient information it may also become possible to
constrain the influence of non-circular streamlines, which will reduce
the effect of systematic errors on the results.

To obtain an accurate distance measurement one will need not only
improved proper motion data (to reduce the uncertaintity in $\mu_s$ in
eq.~[\ref{newdistmethod}]), but also improved information on the
line-of-sight velocities (to reduce the uncertaintity in $w_{ts}$ in
eq.~[\ref{newdistmethod}]). It should be possible to obtain velocities
for $\sim 10$ times more carbon stars than are currently available,
based on the total number of carbon stars in the LMC (Kontizas \etal
2001; Demers, Dallaire \& Battinelli 2002; Alves et al., in prep.).
This will reduce the errors in $w_{ts}$ and $\Theta$ by a factor of
$\sim 3$. The internal proper motion velocity fields that can be
obtained with future astrometric missions can be used to further
reduce the error in $\Theta$.

\section{Summary and Conclusions}
\label{s:conc}

Disk galaxies form an important constituent of the Universe and it is
crucial to have an adequate understanding of their kinematics. The
nearest disk galaxy is our own Milky Way. Our position in the Milky
Way has allowed us to learn many things about it that are inaccessible
for other galaxies. However, there are many things about the
kinematics of our Milky Way that are understood less well than for
disk galaxies in general (e.g., the behavior of the rotation curve
outside the solar radius). This is due to the fact that we reside
inside the Milky Way, so that the Milky Way spreads over the entire
sky. This necessitates the study of Milky Way kinematics in terms of
$(l,b,v)$ diagrams, which are notoriously complicated to interpret
(e.g., Binney \& Merrifield 1998). By contrast, almost all other disk
galaxies, including nearby ones such as M31 and M33, are so far away
that they span $\lta 1^{\circ}$ on the sky. This allows one to model
their kinematics fairly accurately under the assumption that the sky
has no curvature over the area of the galaxy. The LMC is unique in
that it is only disk galaxy that falls between these extremes: it is
not close enough to cover the whole sky, yet close enough to have its
observables influenced by the curvature of the sky.

To fully understand the implications of the large extent of the LMC on
the sky, we have derived general expressions for the velocity fields
of the LMC, both along the line-of-sight and in the plane of the
sky. We have analyzed these expressions to understand which quantities
are uniquely constrained by the data, and which other quantities are
degenerate. While we are not aware of any previous presentation in the
literature of our general expressions for the LMC velocity field, this
does not mean that previous authors have failed to see the importance
of the large angular extent of the LMC. The most importance
consequence is that the transverse motion of the LMC introduces a
spurious solid-body rotation component in the observed line-of-sight
velocity field, and this has certainly been realized by nearly all
authors in the past decades. However, most authors have treated this
component as a `contamination', which needs to be subtracted before
the internal kinematics of the LMC can be studied (e.g., Kim \etal
1998; Alves \& Nelson 2000). While this is a valid viewpoint, it
requires an accurate measurement of the LMC proper motion, which has
not generally been available. In addition, this approach fails to
realize that there is important information in the velocity field that
can be extracted from the data without requiring prior knowledge of
the proper motion. While this has been recognized previously by some
authors, interpretations were always accompanied by the assumption
that the line of nodes coincides with the photometric major axis of
the LMC (e.g, Feitzinger \etal 1977; Meatheringham \etal 1988; Gould
2000). While this may have seemed reasonable (it is true for a
circular disk), this is now known to be invalid (Papers~I and~II).
The present paper represents the first exposition of what assumptions
are actually necessary and/or reasonable, and what can actually be
learned from the observed line-of-sight velocity field with and
without these assumptions.

We have fitted the general expression for the LMC line-of-sight
velocity field to data for 1041 carbon stars. Several things can be
inferred even without any knowledge of the LMC transverse velocity and
its associated solid-body rotation component. Most importantly, the
position angle of the line of nodes is uniquely constrained at $\Theta
= 129.9^{\circ} \pm 6.0^{\circ}$. This is consistent with the value
$\Theta = 122.5 \pm 8.3^{\circ}$ determined geometrically in Paper~I.
The observed drift in the center of the isophotes at large radii is
also consistent with these estimates, when interpreted as a result of
viewing perspective (as discussed in Paper~II). With three independent
arguments there can now be little remaining doubt that the position
angle of the line of nodes is in fact quite different from the major
axis position angle, ${\rm PA}_{\rm maj} = 189.3^{\circ} \pm
1.4^{\circ}$, with the corollary that the LMC must be intrinsically
elongated (Paper~II).

The LMC dynamical center is determined by the data to $\sim
0.4^{\circ}$ accuracy in each coordinate. The inferred position is
consistent with the center of the bar and the center of the outer
isophotes. However, it is offset by $1.2^{\circ} \pm 0.6^{\circ}$ from
the kinematical center of the HI. The most plausible interpretation of
this is that the structure and kinematics of the HI in the LMC are
highly disturbed. This is quite consistent with our general knowledge
of the gas morphology in the Magellanic Clouds system. Almost all
previous studies of the kinematics of the LMC, even the ones using
stellar tracers (carbon stars, planetary nebulae, etc.), have
generally fixed the kinematical center a priori to coincide with the
HI kinematical center. The results of the present study indicate that
this not appropriate.

The LMC rotation curve can only be inferred with knowledge of
$v_{tc}$, the transverse velocity component along the line of
nodes. This component is itself not (strongly) constrained by the
velocity field. By contrast, the orthogonal component $v_{ts}$ is
constrained by the velocity field, but only through the quantity
$w_{ts} \equiv v_{ts} + D_0 (di/dt)$. We find that $w_{ts} = -402.9
\pm 13.0 \kms$. This provides a non-trivial constraint on a combination 
of three apparently unrelated quantities: the transverse velocity
component of the LMC in the direction perpendicular to the line of
nodes, the LMC distance, and the rate of inclination change $di/dt$.

To further interpret the results, we have compiled various LMC proper
motion measurements from the literature. For two of the measurements,
the ones by Jones \etal (1994) and Pedreros \etal (2002), we have
calculated an improved correction for the offset of the field under
study from the LMC CM. The weighted average proper motion, combined
with an estimate for the LMC distance, yields both $v_{tc}$ and
$v_{ts}$.

The $v_{ts}$ inferred from the proper motion data can be combined with
$w_{ts}$ inferred from the velocity field analysis and an estimate of
the LMC distance to determine the rate of inclination change of the
LMC disk: $(di/dt) = -0.37 \pm 0.22 \masyr = -103 \pm 61$
degrees/Gyr. This is not inconsistent with what can reasonably be
explained as a result of precession and nutation of the LMC disk due
to tidal torques from the Milky Way (Weinberg 2000).

The $v_{tc}$ inferred from the proper motion data can be combined with
the results from the velocity field analysis to yield the LMC rotation
curve. We find that it flattens out at $V = 49.8 \pm 15.9 \kms$. The
error is dominated by our knowledge of the LMC proper motion. The
rotation curve amplitude inferred here is $\sim 40$\% lower than what
has been indicated by {\it all} previous analyses of the LMC
kinematics, including, e.g., the carbon star study of Alves \& Nelson
(2000) and the HI study of Kim \etal (1998). This is due to the fact
that these studies used or imposed values of $v_{tc}$, $v_{ts}$,
$w_{ts}$, $di/dt$ and/or $\Theta$ that are inconsistent with the data
and the analysis presented here.

Upon correction for asymmetric drift, the circular velocity of the LMC
is estimated to be $V_{\rm circ} = 64.8 \pm 15.9 \kms$. This implies
an enclosed mass inside the last measured data point $M_{\rm LMC} (8.9
\kpc) = (8.7 \pm 4.3) \times 10^9 \Msun$. This corrects some previous 
estimates, which have ranged to values as high as $2 \times 10^{10}
\Msun$. The inferred mass exceeds the sum of the estimated masses
of the stellar disk ($\sim 2.7 \times 10^9 \Msun$) and the gaseous
disk ($\sim 0.5 \times 10^9 \Msun$). This implies that the LMC is
embedded in a dark halo, consistent with the relative flatness of the
observed rotation curve out to the last measured data point.

To estimate the tidal radius of the LMC we note that the traditional
formulae for calculating this quantity are incorrect for galaxies that
are embedded in a dark halo. We derive alternative formulae that are
appropriate for the Milky Way -- LMC system, which yield an LMC tidal
radius $r_t = 15.0 \pm 4.5 \kpc$ (i.e., $17.1^{\circ} \pm
5.1^{\circ}$). As expected, this exceeds the radii out to which the
LMC isopleths are observed to be regular.

The carbon star analysis yields an average line-of-sight velocity
dispersion $\sigma = 20.2 \pm 0.5 \kms$. The radial profile of the
velocity dispersion does not fall steeply enough with radius to be
consistent with a constant scale-height disk. Application of the
isothermal flared disk model formalism of Alves \& Nelson (2000), in
which the vertical density profile is proportional to ${\rm sech}^2
(z/z_0)$, implies that $z_0$ increases from $0.27 \kpc$ at the LMC
center to $1.5 \kpc$ at a radius of $5.5 \kpc$. More generally, the
importance of rotational support and the disk scale height are both
directly related to the quantity $V/\sigma$. For the LMC carbon stars
we find $V/\sigma \approx 2.9 \pm 0.9$. Compared to the Milky Way disk
components, this is smaller than the values for both the thin disk ($V
/\sigma \approx 9.8$) and the thick disk ($V/\sigma \approx
3.9$). Simple arguments for models stratified on spheroids indicate
that the LMC disk could have an (out-of-plane) axial ratio of $\sim
0.3$ or larger. These results indicate that the LMC is a much thicker
disk system than has been assumed previously. This is consistent with
the predictions of $N$-body simulations by Weinberg (2000) that
address the Milky Way tidal influence on the LMC.

To lowest order, the LMC self-lensing optical depth depends
exclusively on the observed velocity dispersion and not separately on
either the galaxy mass or scale height. Although we find a different
rotation curve amplitude for the LMC than previous authors (which
affects the mass and scale height of the LMC), we do find the same
velocity dispersion. It is therefore not expected that the new
insights from the present paper will substantially alter existing
estimates for the LMC self-lensing optical depth.

The proper motion data yield an estimate of the transverse velocity of
the LMC. This velocity is similar to what can be estimated completely
independently from the line-of-sight velocity field (for reasonable
values of $di/dt$). This provides confidence in the proper motion
data, and indicates that there is no reason to believe that they are
plagued by large systematic errors. The transverse velocity is $v_t =
406 \kms$ in the direction of position angle $\Theta_t =
78.7^{\circ}$, with errors of $\sim 40 \kms$ in each coordinate. This
can be combined with the line-of-sight velocity inferred from the
carbon star data, $v_{\rm sys} = 262.2 \pm 3.4 \kms$, to determine the
LMC velocity ${\vec v}_{\rm LMC}$ in the Galactocentric rest frame.
After correction for the reflex motion of the sun we find that $v_{\rm
LMC} = 293 \pm 39 \kms$, with radial and tangential components $v_{\rm
LMC,rad} = 84 \pm 7\kms$ and $v_{\rm LMC,tan} = 281 \pm 41
\kms$, respectively. This provides one of the most accurate space
velocities of any Milky Way satellite.  The result is at the low end
of the range of velocities that has been predicted by models for the
Magellanic Stream. The data are fully consistent with the models of
Gardiner \etal (1994) and Gardiner \& Noguchi (1996), which are some
of the most detailed models that have been constructed. The implied
orbit of the LMC has an apocenter to pericenter distance ratio $\sim
2.5:1$. The perigalactic distance is $\sim 45 \kpc$ and the present
orbital period around the Milky Way is $\sim 1.5 \Gyr$. The constraint
that the LMC is bound to the Milky Way provides a robust limit on the
minimum mass and extent of the Milky Way dark halo. For an isothermal
halo model truncated at $r_h$ one finds that $M_{\rm MW} \geq 4.3
\times 10^{11} \Msun$ and $r_h \geq 39
\kpc$ (68.3\% confidence).

In Figures~\ref{f:propfield} and~\ref{f:propcomp} we have presented
predictions for the LMC proper motion velocity field, based on our
best fit model. This velocity field is dominated by the influence of
the CM velocity (the proper-motion components of which vary with
position in the LMC), and not by the internal rotation velocity
component. The proper motions due to $di/dt$ and $d\Theta/dt$ play a
minor, but non-negligible role.

The results of the line-of-sight velocity field analysis (in
particular $w_{ts}$ and the position angle of the line of nodes) can
be combined with proper motion data to obtain an estimate of the LMC
distance. We argue that this method has promise, but show that it will
not be competitive until: (a) $di/dt$ can be accurately determined
from independent data (e.g., measurement of the LMC proper motion
velocity field); (b) the proper motion of the LMC CM is determined
more accurately; and (c) a larger set of stellar line-of-sight
velocities is obtained for analysis.

The next advance in our understanding of LMC structure and kinematics
will have to come from more accurate proper motion data. Such
measurements will be possible with future space missions (e.g., SIM,
GAIA), and can possibly be achieved with other means as well (e.g., a
study by Alcock \etal is underway with the Advanced Camera for Surveys
on Hubble Space Telescope). The uncertainties in our knowledge of many
of the LMC's structural parameters are dominated by the uncertainties
in the presently available proper motion measurements
(Section~\ref{ss:LMCpropmotion}). This is true, e.g., for the LMC
rotation curve $V(R')$, for its mass and tidal radius, and for
$di/dt$. Our knowledge of these quantities will improve significantly
with a more precise measurement of the proper motion of the LMC CM.
This will also determine the space motion of the LMC to high accuracy.
Combined with models for the Magellanic Stream this should allow the
extraction of improved constraints on the density distribution of the
Milky Way dark halo. Measurements of the proper motion velocity field
of the LMC will yield better constraints on the LMC distance, by
providing independent constraints on $di/dt$, $d\Theta/dt$, and
possible deviations from circular streamlines. In addition, it will
allow application of the method of rotational parallax, which compares
the rotation amplitude in $\masyr$ inferred from the proper motion
velocity field with the same quantity in km/s inferred from the
line-of-sight velocity field.


\acknowledgments
We are grateful to Andrew Drake, Mario Pedreros and Mir Abbas Jalali
for discussions and/or sharing results prior to publication. We
dedicate this paper to the memory of Robert Schommer. He made many
important contributions to the understanding of the structure and
kinematics of the LMC, and he was a valued collaborator on the
acquisition of carbon star velocity data analyzed in the present
paper.

\clearpage


\appendix

\section{The Proper Motion of the LMC Center of Mass from 
Off-Center Measurements}
\label{s:app}

Some of the proper motion measurements that have been reported in the
literature for the LMC pertain to fields that are at a considerable
distance from the LMC CM. This includes in particular the measurements
of Jones \etal (1994) and Pedreros \etal (2002). These measurements
must be corrected for the orientation and rotation of the LMC disk to
obtain an estimate of the proper motion of the LMC CM. These
corrections have usually been made with the aid of kinematical models
for the LMC that are less sophisticated than those presented
here. Among other things, the position angle of the line of nodes that
has generally been assumed is quite different from the value derived
here and in Paper~I. In view of this, we rederived the corrections
that must be made to the measurements of Jones \etal (1994) and
Pedreros \etal (2002) to obtain estimates of the proper motion of the
LMC CM. The results are more accurate than the estimates quoted in
these respective papers.

The field studied by Jones \etal (1994) has J2000 coordinates $\alpha
= 97.491^{\circ}$ and $\delta = -64.166^{\circ}$. The measured proper
motion was $\mu_{W} = -1.26 \pm 0.28$ mas/yr and $\mu_{N} = 0.26 \pm
0.27$ mas/yr. Given the CM position derived in Section~\ref{s:data}
(eq.~[\ref{fitresults}]), the field center has $\rho = 8.298^{\circ}$
and $\Phi = 54.195^{\circ}$. With the inclination and line of nodes
position angle given by equations~(\ref{inclination})
and~(\ref{fitresults}), one obtains from equation~(\ref{Racc}) that
$R'/D_0 = 0.16$. At this position in the LMC disk $V(R')
\approx 50 \kms$, according to Figure~\ref{f:rotcurve}. We take 
$v_{\rm sys}$ from equation~(\ref{fitresults}) and $D_0$ from
equation~(\ref{distmodulus}). The formulae of Section~\ref{s:vmath}
then allow calculation of the proper motions $(\mu_W,\mu_N)$ at the
Jones \etal field center, for any assumed proper motion
$(\mu_{W,0},\mu_{N,0})$ of the LMC CM. At any value of
$(\mu_{W,0},\mu_{N,0})$ we choose $di/dt$ so as to reproduce the value
$w_{ts}$ that was inferred from the line-of-sight velocity field
(eq.~[\ref{fitresults}]). We determined those values of
$(\mu_{W,0},\mu_{N,0})$ that reproduce the Jones \etal measurement,
which yielded $\mu_{W,0} = -1.36 \pm 0.28$ mas/yr and $\mu_{N,0} =
-0.16 \pm 0.27$ mas/yr.

The field studied by Pedreros \etal (2002) has J2000 coordinates
$\alpha = 74.989^{\circ}$ and $\delta = -64.378^{\circ}$. The measured
proper motion was $\mu_{W} = -1.8 \pm 0.2$ mas/yr and $\mu_{N} = 0.3
\pm 0.2$ mas/yr. We proceeded similarly as for the Jones \etal data. 
The field center has $\rho = 6.111^{\circ}$ and $\Phi =
330.698^{\circ}$, and $R'/D_0 = 0.11$. At this position in the LMC
disk $V(R') \approx 50 \kms$, according to Figure~\ref{f:rotcurve}.
The proper motion of the LMC CM implied by the Pedreros \etal
measurement is $\mu_{W,0} = -1.83 \pm 0.20$ mas/yr and $\mu_{N,0} =
0.66 \pm 0.20$ mas/yr.

The sizes of the corrections that must be applied to the Jones \etal
(1994) and Pedreros \etal (2002) data to obtain an estimate of the
proper motion of the LMC CM are $0.43$ mas/yr and $0.36$ mas/yr,
respectively. At the distance of the LMC this corresponds to $103
\kms$ and $86 \kms$, respectively, which considerably exceeds the
rotation amplitude of the LMC. The reason for this is that the
residual proper motion field (i.e., the proper motion field with the
proper motion of the CM subtracted) is dominated by the velocity
component of the LMC CM, and not by the velocity component due to
internal rotation. This was discussed in
Section~\ref{s:propfield}. Because only a small part of the
corrections is due to the rotation of the LMC, the corrections are
quite insensitive to possible errors in the parameters that
characterize the model. As a result, the formal errors in
$(\mu_{W,0},\mu_{N,0})$ are dominated by the formal errors in the
observed $(\mu_{W},\mu_{N})$.

The corrections that were derived assume that the position angle of
the line of nodes is not time dependent. If instead $d\Theta/dt \not=
0$, then the $v_3$ term in equation~(\ref{vecpn}) has an extra
component. However, the estimated $(\mu_{W,0},\mu_{N,0})$ do not
depend sensitively on $d\Theta/dt$. If $d\Theta/dt = 1 \masyr$, then
$(\mu_{W,0},\mu_{N,0})$ changes by $(-0.076,0.109) \masyr$ for the
Jones \etal measurement and by $(-0.093,-0.053)$ for the Pedreros
\etal measurement. These corrections are small compared to the formal
errors in the data.  The $N$-body simulations of Weinberg (2000) show
that it is unlikely that the size of $d\Theta/dt$ is much larger
than $1 \masyr = 277.78^{\circ} / \Gyr$.


\ifsubmode\else
\baselineskip=10pt
\fi


\clearpage


\ifsubmode\else
\baselineskip=14pt
\fi


\newcommand{\figcapdrawskyview}{Illustration of the 
projected view of the sky. All vectors and angles lie in the plane of
the paper (perspective was not used in the drawing). All quantities
are rigorously defined in the text. The galaxy center of mass (CM)
with coordinates $(\alpha_0,\delta_0)$ is chosen to be the origin
${\cal O}$ of an $(x,y,z)$ coordinate system. The $z$-axis (not drawn)
points vertically out of the paper, towards the observer. The angles
$\rho$ and $\Phi \equiv \phi-90^{\circ}$ define the projected position
on the sky of a tracer with coordinates $(\alpha,\delta)$. The angle
$\Gamma$ is the angle between the (proper motion) vectors $\mu_W$ and
$\mu_N$ towards the local directions of West and North, and the
velocity components $v_2$ and $v_3$ defined in the
text.\label{f:drawskyview}}

\newcommand{\figcapdrawsideview}{Schematic `side view' of the 
observer-galaxy system. All quantities are rigorously defined in the
text. The galaxy center of mass (CM) is the origin of an $(x,y,z)$
coordinate system. The $z$-axis points towards the observer. In
reality, the $x$-axis points vertically out of the paper but is shown
here using drawing perspective. The distance from the observer to the
CM is $D_0$; the distance from the observer to a tracer is $D$. The
angle from the CM to the tracer, as seen by the observer, is
$\rho$. The velocity of the tracer can be decomposed into three
orthogonal components, $v_1$, $v_2$ and $v_3$, as described in the
text. These components are shown schematically using drawing
perspective. The component $v_1 \equiv v_{\rm los}$ lies along the
line of sight, and points away from the
observer.\label{f:drawsideview}}

\newcommand{\figcapdrawviewang}{Illustration of the 
observer's view of the galaxy disk. All quantities are rigorously
defined in the text. The cartesian $(x,y,z)$ coordinate system is the
same as in Figure~\ref{f:drawskyview}, with the $z$-axis (not drawn)
pointing vertically out of the paper, towards the observer. The
$(x',y',z')$ system is a second cartesian coordinate system, the axes
of which are shown with drawing perspective. The $(x',y')$ plane is
the plane of the galaxy disk, and the $z'$-axis is its symmetry
axis. The plane of the galaxy disk is titled diagonally out of the
paper. The inclination $i$ is the angle between the $(x,y)$ plane of
the sky, and the $(x',y')$ plane of the galaxy disk. The $x'$-axis is
defined to lie along the lines of nodes (the intersection of the
$(x,y)$ plane of the sky, and the $(x',y')$ plane of the galaxy
disk). The angle $\Theta \equiv \theta-90^{\circ}$ is the position
angle of the line of nodes in the plane of the
sky.\label{f:drawviewang}}

\newcommand{\figcapdrawvel}{Illustration of the quantities used to describe 
the transverse velocity of the LMC CM. The transverse velocity vector
has size $v_t$ and is directed along position angle $\Theta_t =
\theta_t - 90^{\circ}$. It has components $v_x$ and $v_y$ 
towards the West and North, respectively. For analysis of the velocity
field of the LMC it useful to decompose the transverse velocity vector
into a sum of two orthoghonal vectors, as illustrated. One vector lies
along the position angle $\Theta$ of the line of nodes, and has length
$v_{tc}$.  The other lies along the position angle
$\Theta-90^{\circ}$, and has length $-v_{ts}$ (i.e., it has length
$v_{ts}$ in the direction $\Theta+90^{\circ}$). The velocity
components shown in the figure correspond to the proper motions listed
in parenthesis (upon division by the LMC distance
$D_0$).\label{f:drawvel}}

\newcommand{\figcapfit}
{Carbon star line-of-sight velocity data from Kunkel
\etal (1997) and Hardy \etal (2002, in prep.), as function of position
angle $\Phi$ on the sky. The displayed range of the angle $\Phi$ is
$0^{\circ}$--$720^{\circ}$, so each star is plotted twice. Each panel
corresponds to a different range of angular distances $\rho$ from the
LMC center, as indicated. The curves show the predictions of the
best-fit model (calculated at the center of the radial range for the
given panel).\label{f:fit}}

\newcommand{\figcaprotcurve}
{{\bf (Top Panel)} Rotation velocity $V$ in the plane of the LMC disk
as function of the cylindrical radius $R'$ (expressed in units of
$D_0$, the LMC distance; $D_0 \approx 50 \kpc$). The solid curve
corresponds to the best fit to the data (see Figure~\ref{f:fit}) for a
model parameterized by equation~(\ref{Vcurve}). However, each of the
dashed curves provides the same acceptable fit, but for a different
assumed value of the component $v_{tc}$ of the transverse velocity
along the line of nodes. The values of $v_{tc}$ (in km/s) for the
different curves are labeled. The radial range that is displayed
corresponds to the radial range $\rho \leq 13^{\circ}$ for which
kinematical data were included in the fit. The exponential disk scale
length of the LMC is $R'_d \approx 1.4 \kpc$ (Paper~II), which
corresponds to $R'_d/D_0 \approx 0.028$. The data points with error
bars show the results of model fits to individual rings in the LMC
disk plane, as described in Section~\ref{ss:rotdisp}. These fits had
$V$ and $\Theta$ as only free parameters; the quantity $v_{tc}$ was
fixed to the value $253 \kms$ indicated by proper motion data
(eq.~[\ref{vtcpm}]), and the other model parameters were fixed to the
values (eq.~[\ref{fitresults}]) determined for the best-fit model
shown in Figure~\ref{f:fit}. {\bf (Middle Panel)} Radial profile of
the line-of-sight velocity dispersion. {\bf (Bottom Panel)} Radial
profile of the line-of-nodes position angle
$\Theta$.\label{f:rotcurve}}

\newcommand{\figcappm}
{The $(\mu_W,\mu_N)$ plane spanned by the proper motion components of
the LMC center of mass. The data points with error bars show available
proper motion measurements, as listed in Tabel~\ref{t:propmotion}.
Dotted ellipses are the corresponding 68.3\% confidence regions (note
that the 1-$\sigma$ error bars for a two-dimensional Gaussian only
contain $39.3$\% of the cumulative probability). The heavy solid
ellipse is the 68.3\% confidence region that corresponds to the
weighted average of the data.\label{f:pm}}

\newcommand{\figcapvtrans}
{The $(v_x,v_y)$ plane spanned by the transverse velocity components
of the LMC center of mass. The heavy solid ellipse is the 68.3\%
confidence region obtained from the proper motion data listed in
Table~\ref{t:propmotion} and shown in Figure~\ref{f:pm}. The dotted
trapezoid is the 68.3\% confidence region obtained from the LMC
line-of-sight velocity field under the assumption that $di/dt=0$.
Along the long axis of the trapezoid all points are equally probable;
along the short axis the probability distribution is Gaussian. The
solid trapezoid shows the corresponding result if $di/dt$ is chosen to
best agree with the proper motion data. The arrows indicate the
directions along which the components $v_{tc}$ and $v_{ts}$ of the
transverse velocity are measured. The former is measured along the
line of nodes, and the latter is measured perpendicular to the line of
nodes. A change in the assumed $di/dt$ corresponds to a shift of the
trapezoid along the $v_{ts}$ direction.\label{f:vtrans}}

\newcommand{\figcappropfield}
{The proper motion velocity field predicted by our best model for the
LMC structure and kinematics. At each position on in a grid in right
ascension and declination we have calculated the predicted proper
motion components $(\mu_W,\mu_N)$ using the formulae of
Section~\ref{s:vmath}. The proper motion vector is shown as a line
segment, which starts at the grid point for which it was calculated
(thin solid dots). The proper motion of the LMC CM, which is shown as
a heavy line segment starting at the LMC CM (heavy dot), was
subtracted before plotting (in other words, the actual predicted
proper motion at a given point is the vector sum of the line segment
shown at that point and the heavy line segment). Each segment is shown
with the $\mu_W$ component horizontally to the right and the $\mu_N$
component vertically upwards, with equal scales in mas/yr in both
directions. The LMC CM proper motion vector in the model has length
$1.71 \masyr$, which determines the scale of all line
segments.\label{f:propfield}}

\newcommand{\figcappropcomp}
{Vector components of the predicted LMC proper motion velocity field,
displayed as in Figure~\ref{f:propfield}. {\bf (Top Left)} The proper
motions due to the CM velocity component ${\vec v}_{\rm CM}$ (see
eq.~[\ref{vecCM}]) for the model shown in Figure~\ref{f:propfield}.
{\bf (Top Right)} The proper motions due to the internal rotation
component ${\vec v}_{\rm int}$ (see eq.~[\ref{vecint}]) for the model
shown in Figure~\ref{f:propfield}. {\bf (Bottom Left)} The proper
motions due to the disk precession/nutation velocity component ${\vec
v}_{\rm pn}$ (see eq.~[\ref{vecpn}]) for $(di/dt) = 1 \masyr$ and
$(d\Theta/dt) = 0$. {\bf (Bottom Right)} The proper motions due to the
disk precession/nutation velocity component ${\vec v}_{\rm pn}$ for
$(di/dt) = 0$ and $(d\Theta/dt) = 1 \masyr$. For comparison, the model
shown in Figure~\ref{f:propfield} has $(di/dt) = -0.37
\masyr$ and $(d\Theta/dt) = 0$.\label{f:propcomp}}

\newcommand{\figcapdistdidt}
{The distance modulus of the LMC $m-M$ (in magnitudes) versus the rate
of inclination change $di/dt$. The analysis of the line-of-sight
velocity field and the available proper motion data constrain a
combination of these parameters. The heavy solid curve shows the best
fit, while the thinner surrounding curves indicate the $1\sigma$
confidence region. The dashed horizontal line indicates the canonical
distance modulus of the LMC inferred from other studies: $m-M = 18.50$
(see Section~\ref{ss:LMCdistance}); the uncertainty on that distance
modulus is $\sim 0.1$ mag.\label{f:distdidt}}


\ifsubmode
\figcaption{\figcapdrawskyview}
\figcaption{\figcapdrawsideview}
\figcaption{\figcapdrawviewang}
\figcaption{\figcapdrawvel}
\figcaption{\figcapfit}
\figcaption{\figcaprotcurve}
\figcaption{\figcappm}
\figcaption{\figcapvtrans}
\figcaption{\figcappropfield}
\figcaption{\figcappropcomp}
\figcaption{\figcapdistdidt}

\clearpage
\else\printfigtrue\fi

\ifprintfig


\clearpage
\begin{figure}
\epsfxsize=0.8\hsize
\centerline{\epsfbox{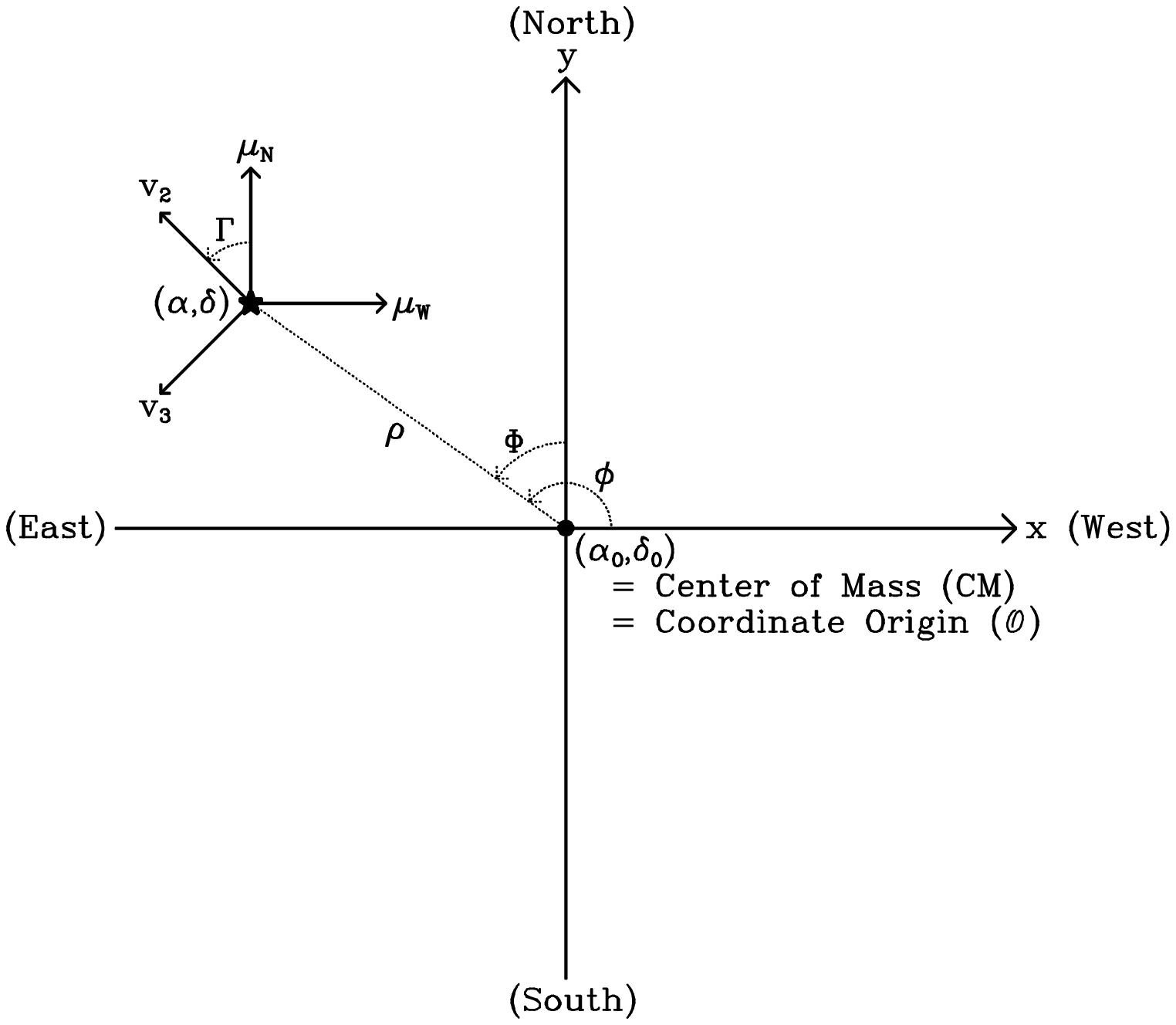}}
\ifsubmode
\vskip3.0truecm
\setcounter{figure}{0}
\addtocounter{figure}{1}
\centerline{Figure~\thefigure}
\else
\figcaption{\figcapdrawskyview}
\fi
\end{figure}


\clearpage
\begin{figure}
\epsfxsize=0.8\hsize
\centerline{\epsfbox{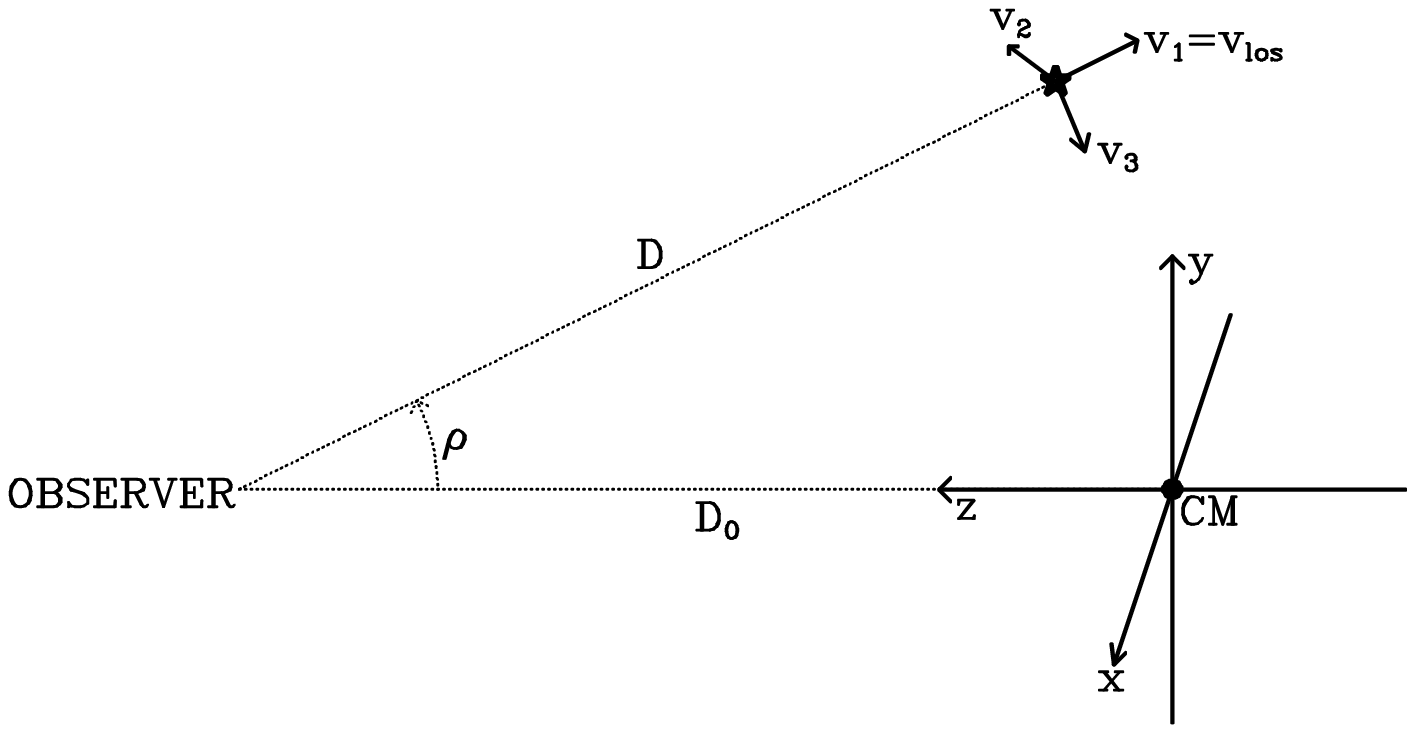}}
\ifsubmode
\vskip3.0truecm
\addtocounter{figure}{1}
\centerline{Figure~\thefigure}
\else
\figcaption{\figcapdrawsideview}
\fi
\end{figure}


\clearpage
\begin{figure}
\epsfxsize=0.8\hsize
\centerline{\epsfbox{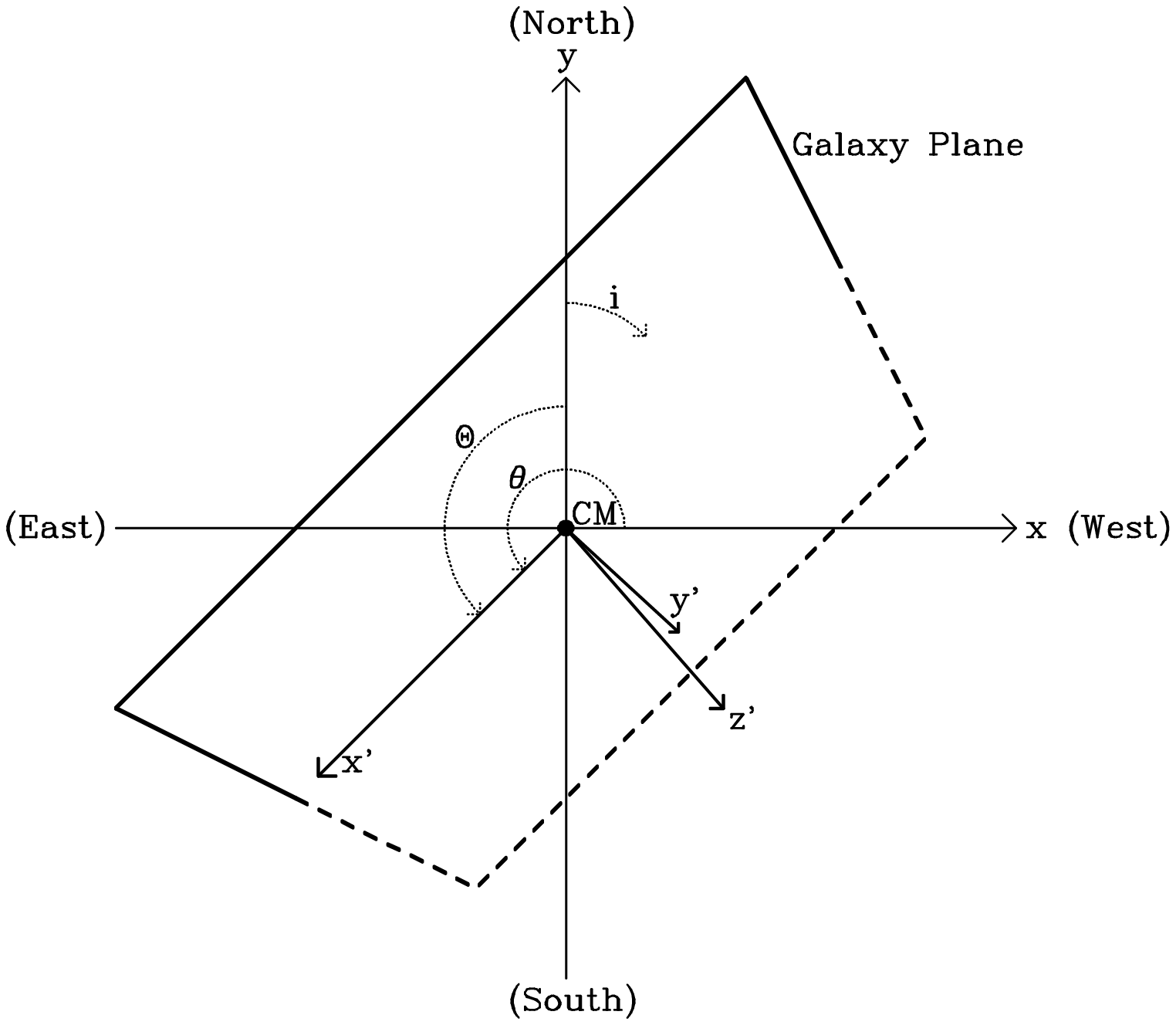}}
\ifsubmode
\vskip3.0truecm
\addtocounter{figure}{1}
\centerline{Figure~\thefigure}
\else
\figcaption{\figcapdrawviewang}
\fi
\end{figure}


\clearpage
\begin{figure}
\epsfxsize=0.8\hsize
\centerline{\epsfbox{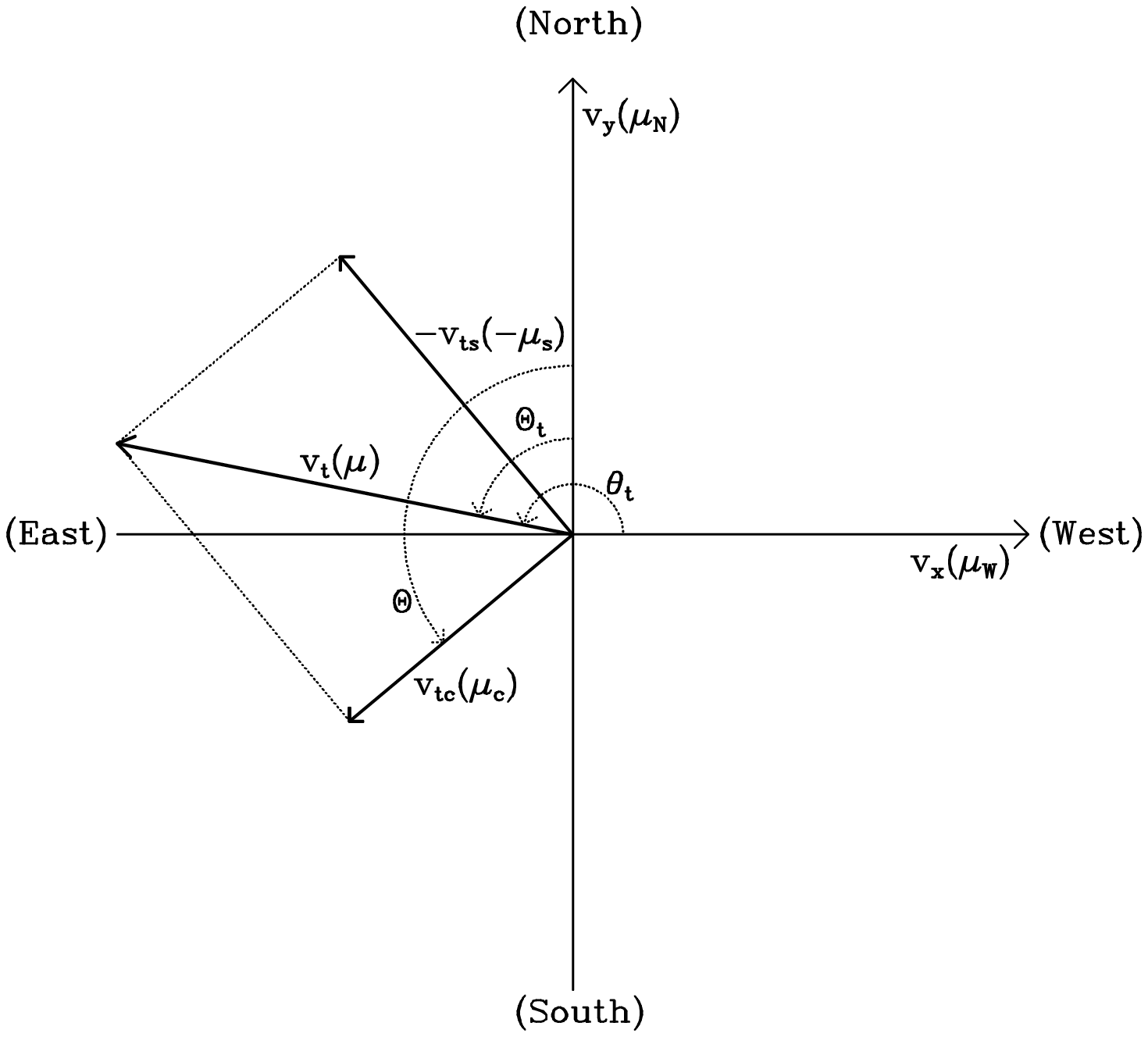}}
\ifsubmode
\vskip3.0truecm
\addtocounter{figure}{1}
\centerline{Figure~\thefigure}
\else
\figcaption{\figcapdrawvel}
\fi
\end{figure}


\clearpage
\begin{figure}
\epsfxsize=0.8\hsize
\centerline{\epsfbox{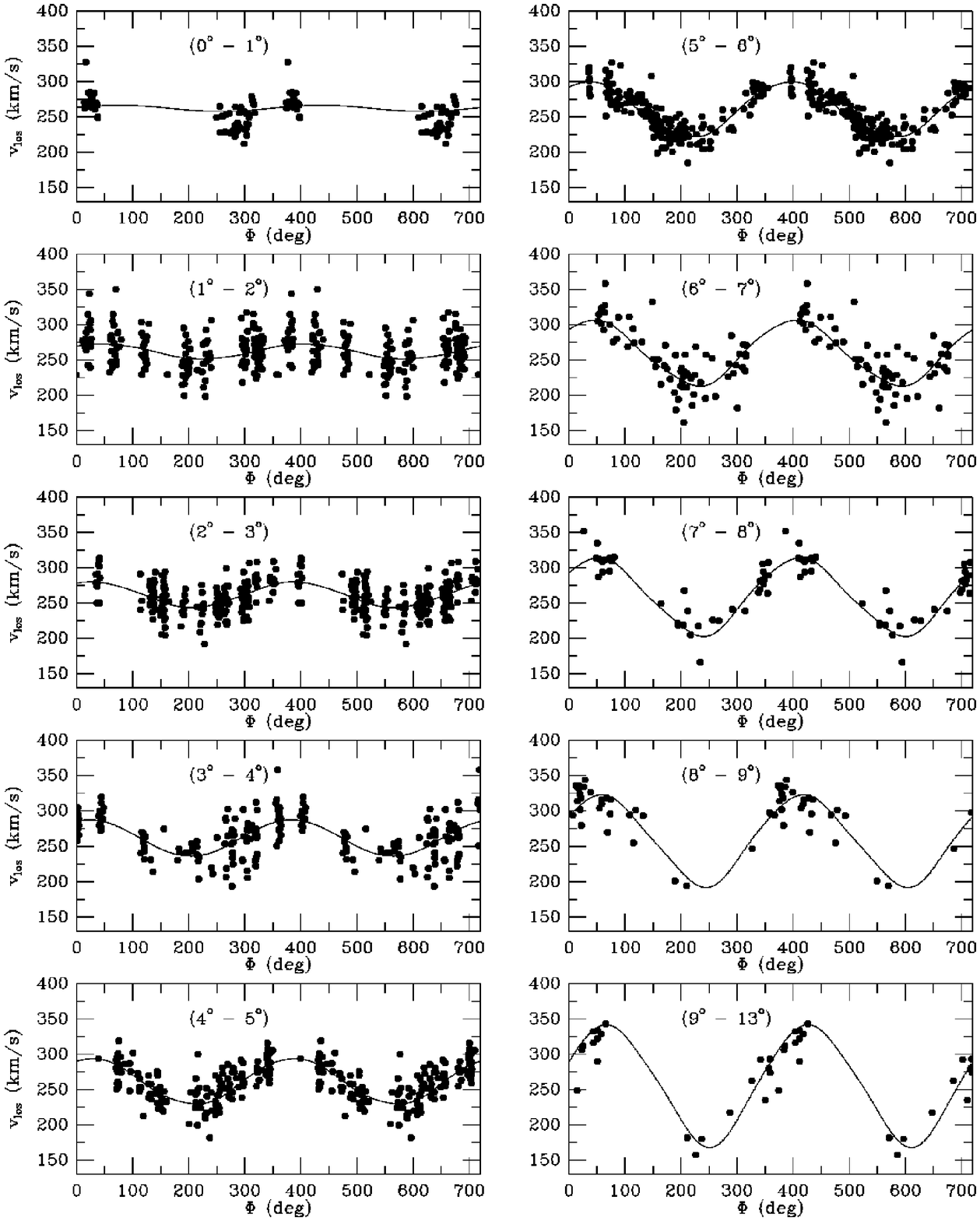}}
\ifsubmode
\vskip3.0truecm
\addtocounter{figure}{1}
\centerline{Figure~\thefigure}
\else
\figcaption{\figcapfit}
\fi
\end{figure}


\clearpage
\begin{figure}
\epsfxsize=0.8\hsize
\centerline{\epsfbox{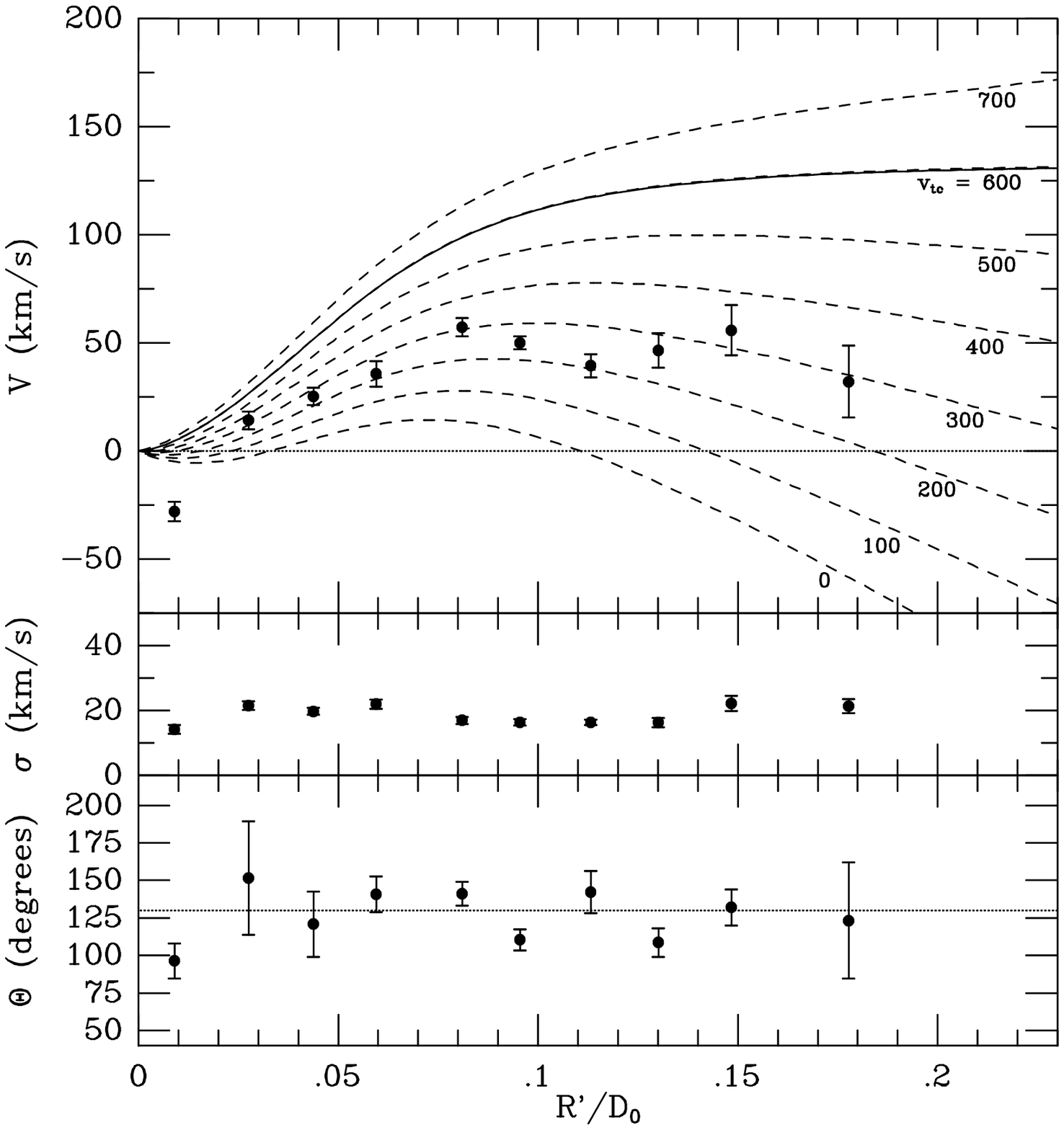}}
\ifsubmode
\vskip3.0truecm
\addtocounter{figure}{1}
\centerline{Figure~\thefigure}
\else
\figcaption{\figcaprotcurve}
\fi
\end{figure}


\clearpage
\begin{figure}
\epsfxsize=0.8\hsize
\centerline{\epsfbox{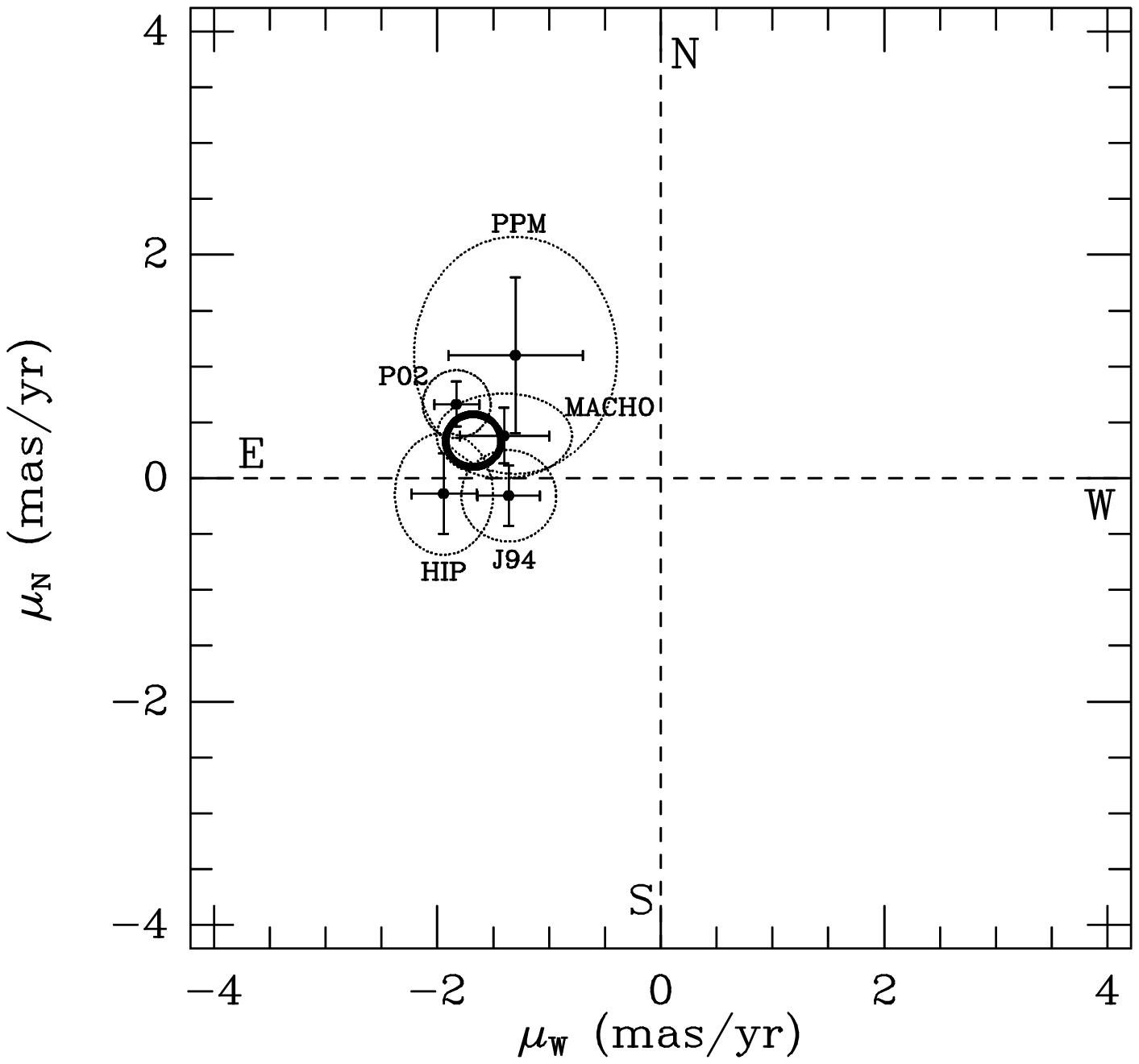}}
\ifsubmode
\vskip3.0truecm
\addtocounter{figure}{1}
\centerline{Figure~\thefigure}
\else
\figcaption{\figcappm}
\fi
\end{figure}


\clearpage
\begin{figure}
\epsfxsize=0.8\hsize
\centerline{\epsfbox{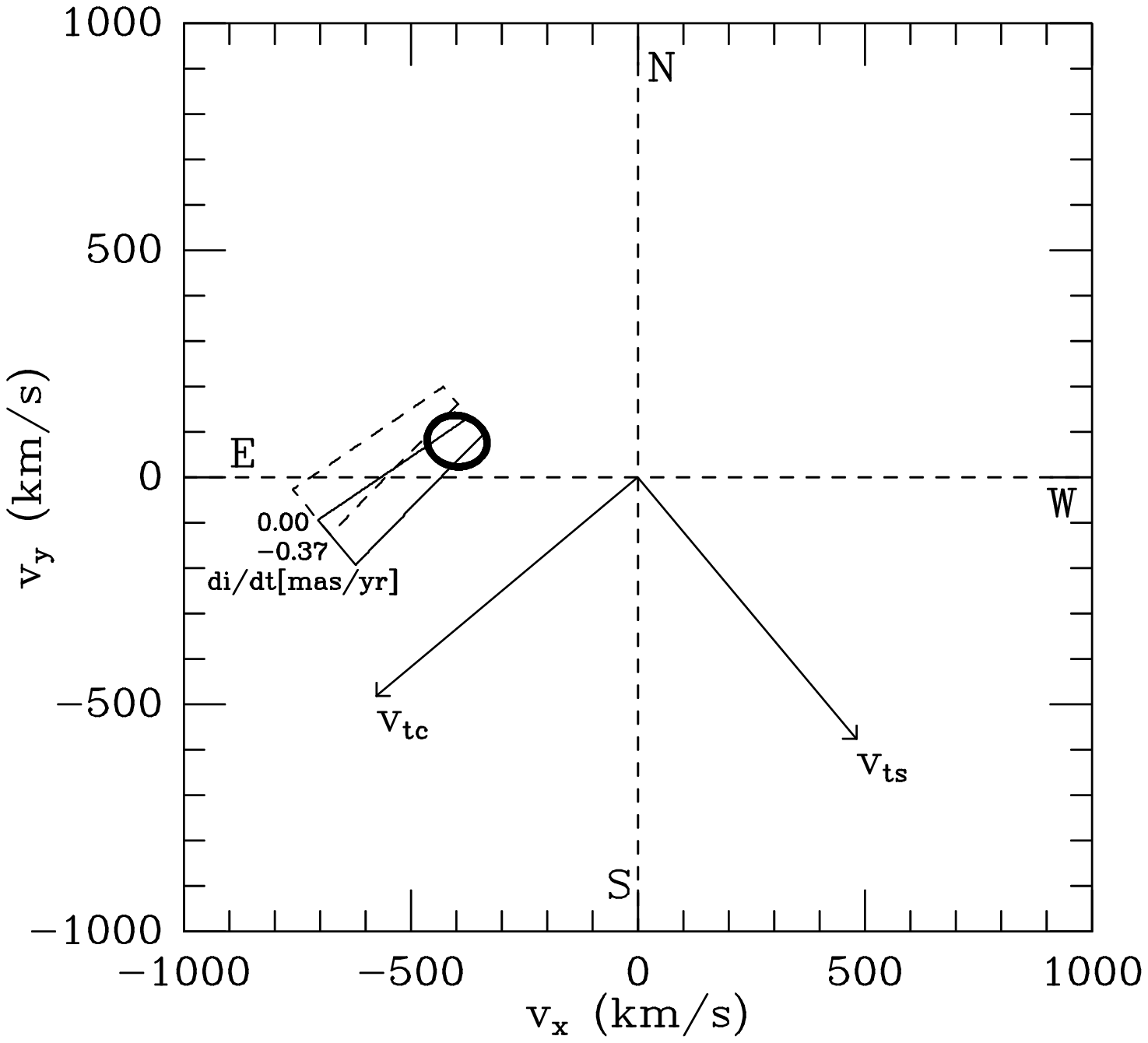}}
\ifsubmode
\vskip3.0truecm
\addtocounter{figure}{1}
\centerline{Figure~\thefigure}
\else
\figcaption{\figcapvtrans}
\fi
\end{figure}


\clearpage
\begin{figure}
\epsfxsize=0.8\hsize
\centerline{\epsfbox{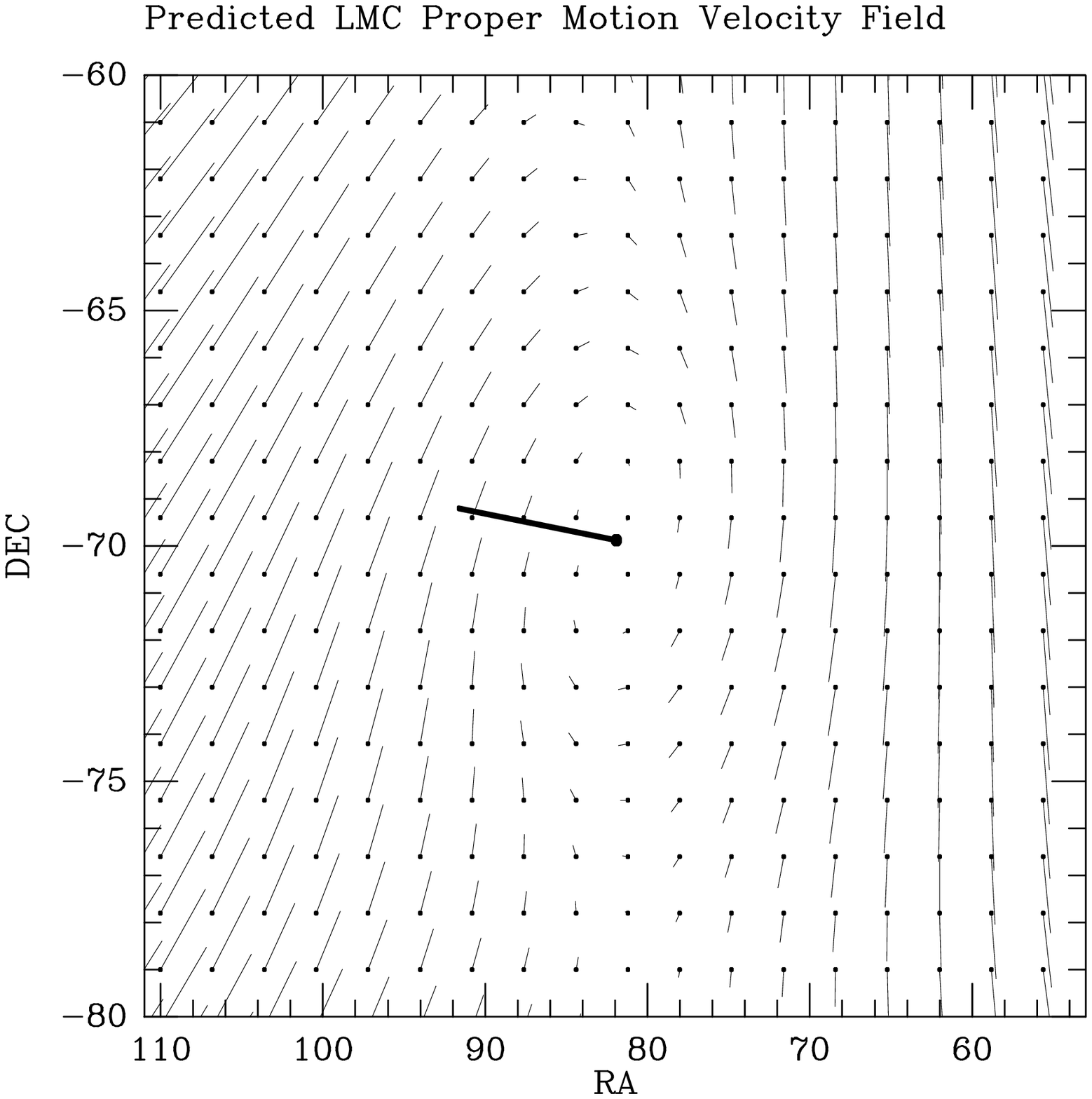}}
\ifsubmode
\vskip3.0truecm
\addtocounter{figure}{1}
\centerline{Figure~\thefigure}
\else
\figcaption{\figcappropfield}
\fi
\end{figure}


\clearpage
\begin{figure}
\centerline{%
\epsfxsize=0.48\hsize
\epsfbox{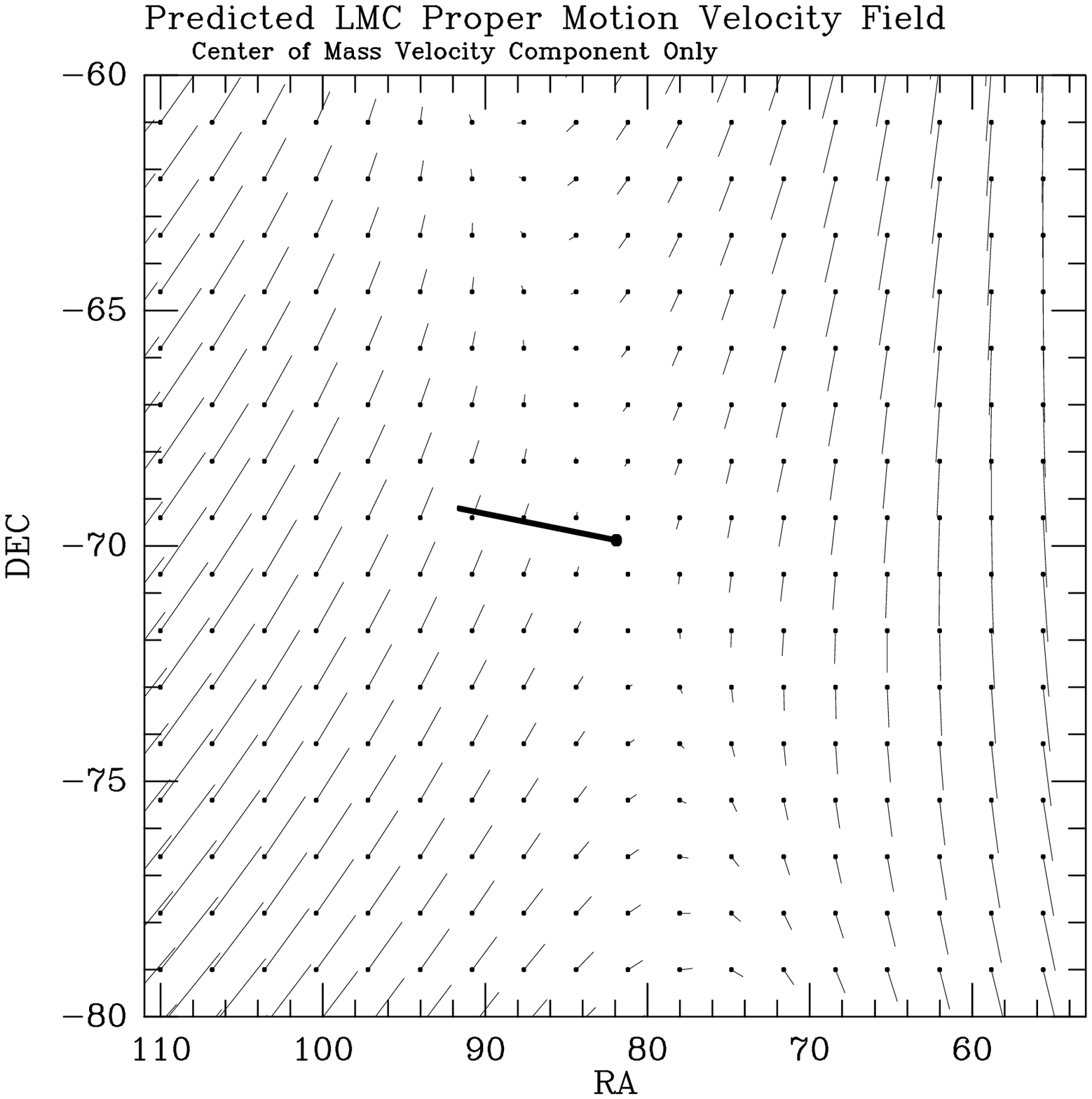}
\qquad
\epsfxsize=0.48\hsize
\epsfbox{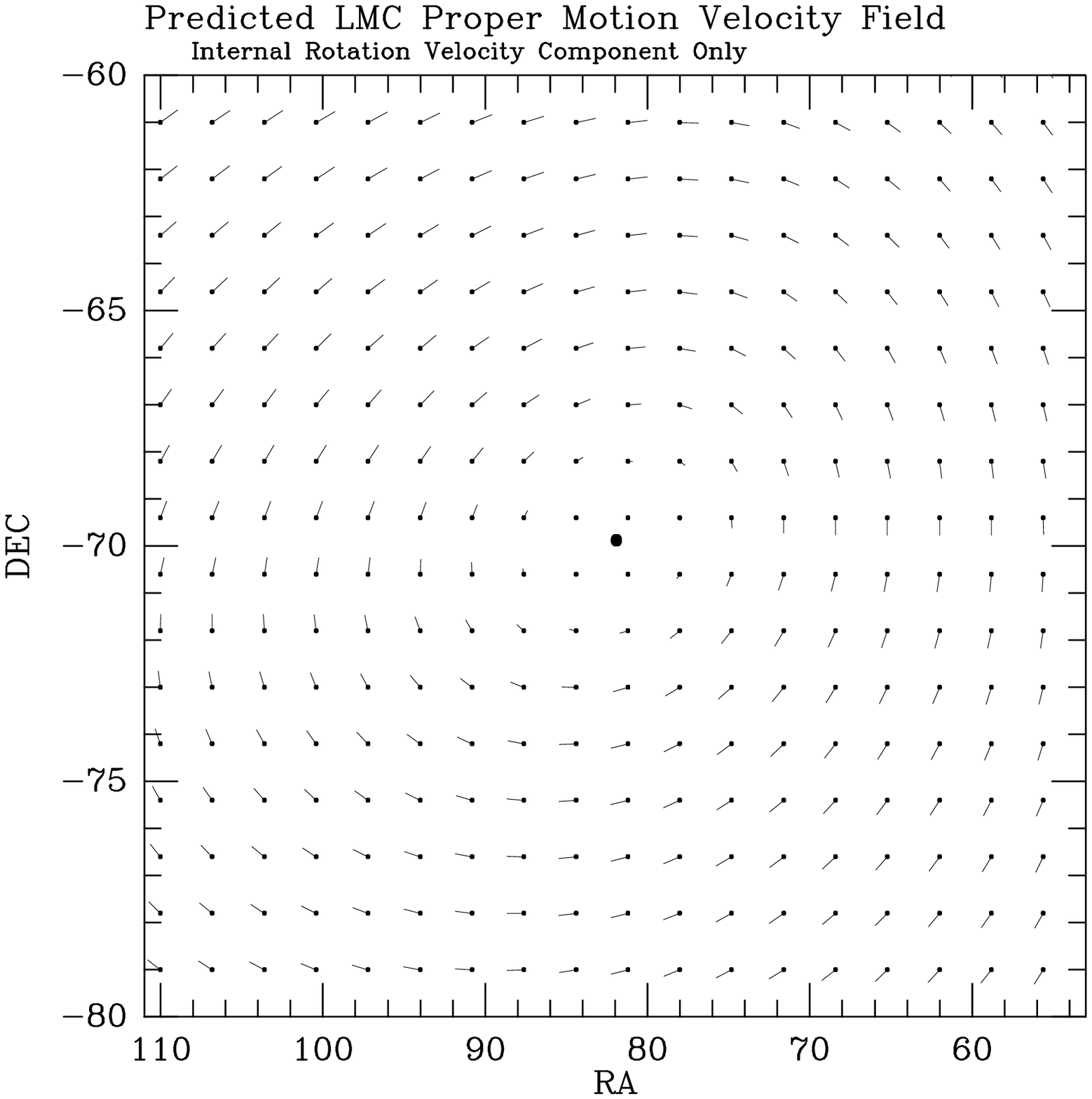}
}
\medskip
\centerline{%
\epsfxsize=0.48\hsize
\epsfbox{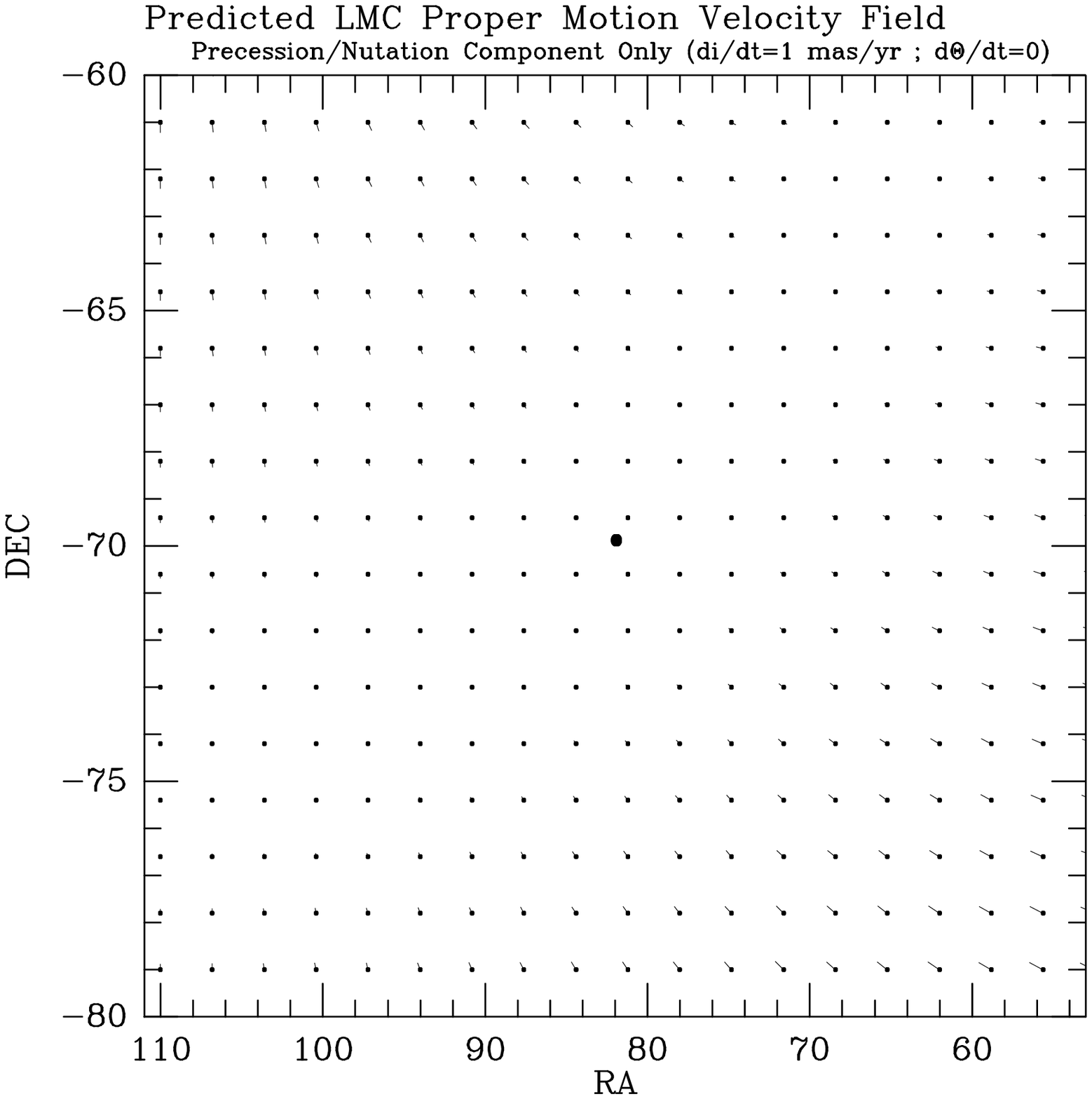}
\qquad
\epsfxsize=0.48\hsize
\epsfbox{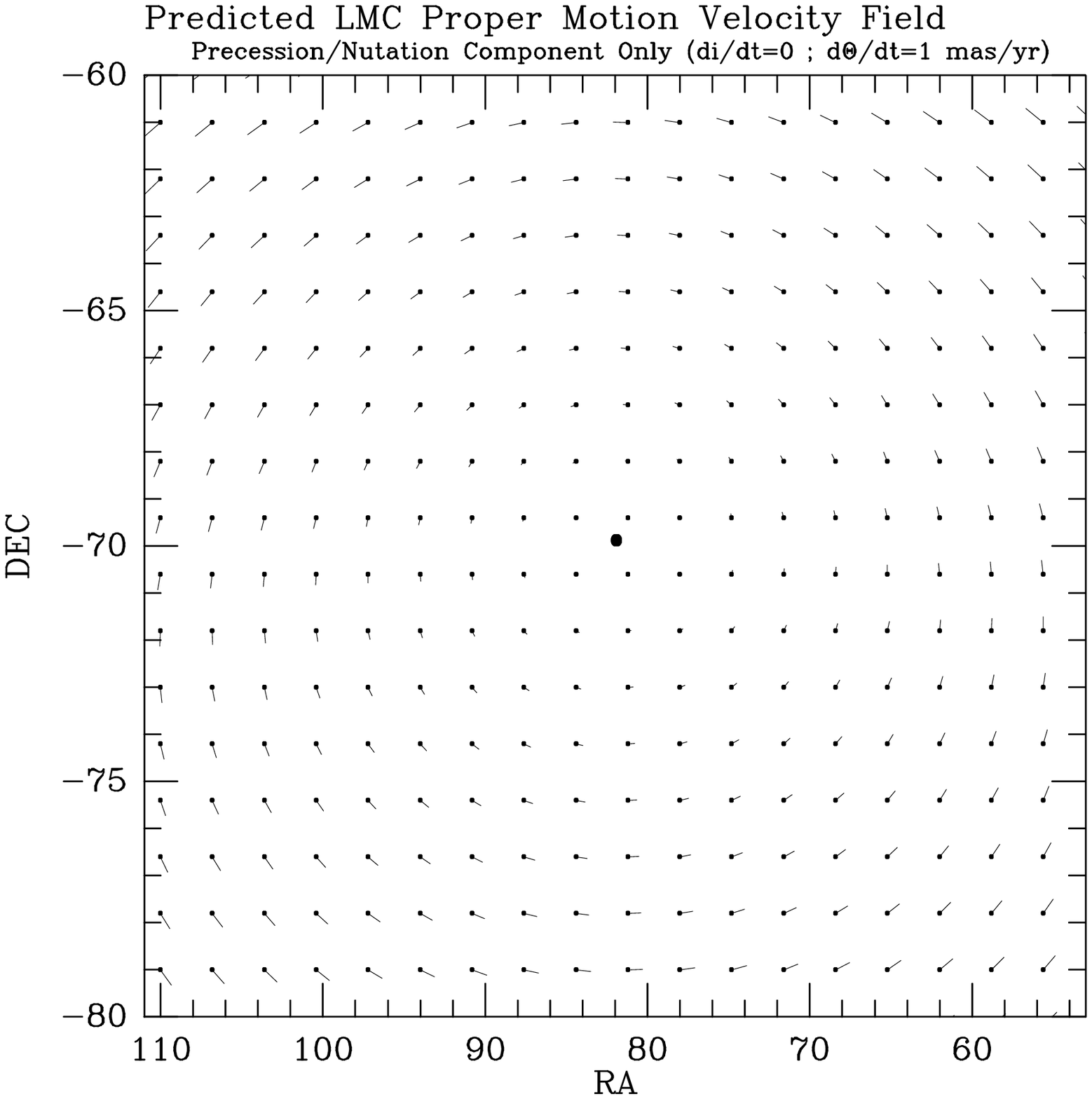}
}
\ifsubmode
\vskip3.0truecm
\addtocounter{figure}{1}
\centerline{Figure~\thefigure}
\else
\figcaption{\figcappropcomp}
\fi
\end{figure}


\clearpage
\begin{figure}
\epsfxsize=0.8\hsize
\centerline{\epsfbox{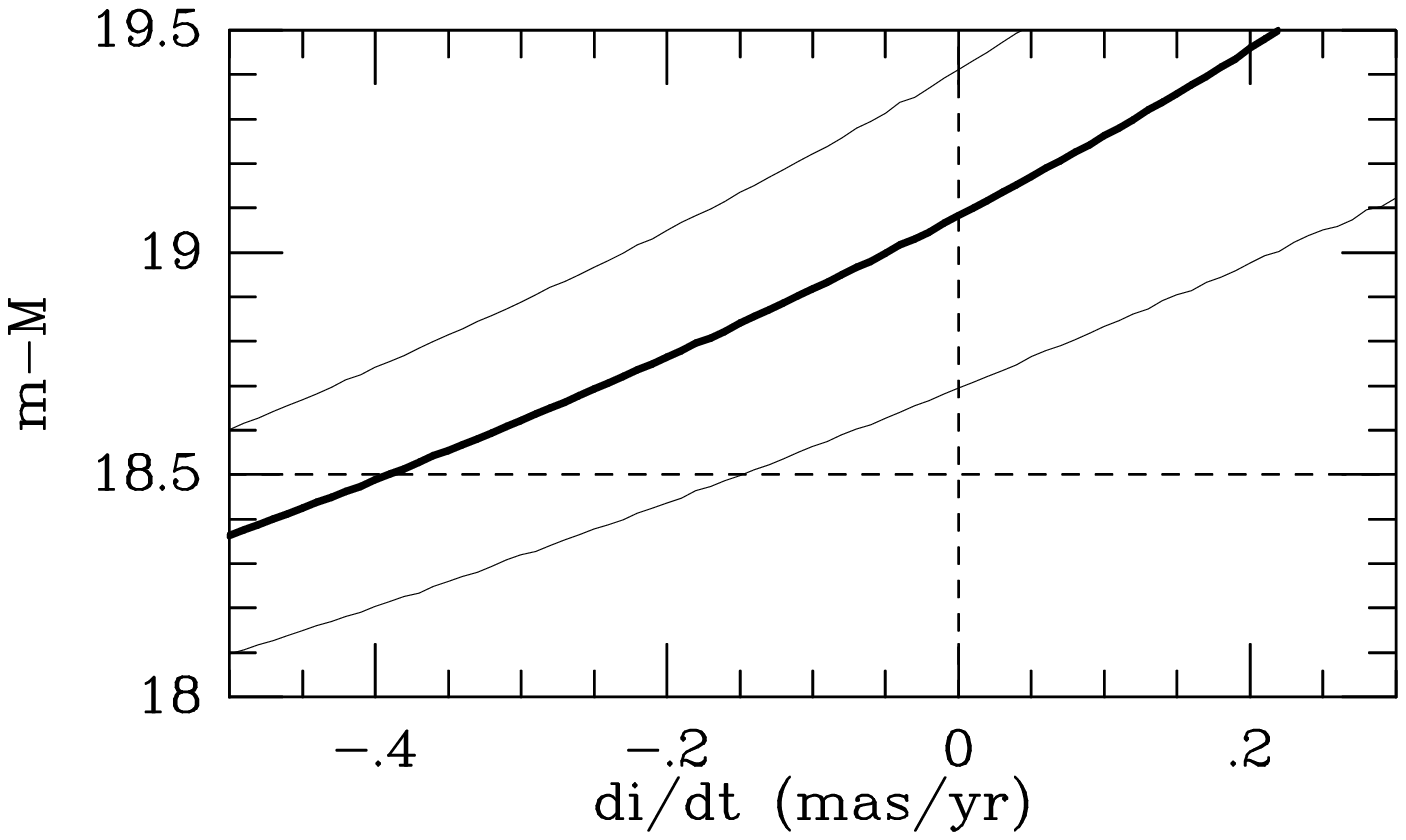}}
\ifsubmode
\vskip3.0truecm
\addtocounter{figure}{1}
\centerline{Figure~\thefigure}
\else
\figcaption{\figcapdistdidt}
\fi
\end{figure}


\fi


\clearpage
\ifsubmode\pagestyle{empty}\fi


\begin{deluxetable}{llrrl}
\tablecaption{Measurements of the Proper Motion of the LMC Center of Mass\label{t:propmotion}} 
\tablehead{
\colhead{paper} & \colhead{label} &
\colhead{$\mu_W$} & \colhead{$\mu_N$} & \colhead{comments} \\
 & & \colhead{($\masyr$)} & \colhead{($\masyr$)} & \\ }
\startdata
Kroupa \etal (1994)      & PPM   & $-1.30 \pm 0.60$ & $ 1.10 \pm 0.70$ & \\
Jones \etal (1994)       & JKL   & $-1.36 \pm 0.28$ & $-0.16 \pm 0.27$ & 
   see Appendix~\ref{s:app} \\
Kroupa \& Bastian (1997) & HIP   & $-1.94 \pm 0.29$ & $-0.14 \pm 0.36$ & \\
Pedreros \etal (2002)    & P02   & $-1.83 \pm 0.20$ & $ 0.66 \pm 0.20$ & 
   see Appendix~\ref{s:app} \\
Drake \etal (2002)       & MACHO & $-1.40 \pm 0.40$ & $ 0.38 \pm 0.25$ & \\
\enddata
\tablecomments{Measurements of the proper motion of the LMC CM
from various sources, as listed in column~(1). The label used to
indicate the measurements in Figure~\ref{f:pm} is listed in
column~(2). The proper motions listed in columns~(3) and~(4) are
characterized by the components in the directions towards the West
($\mu_W \equiv - \mu_{\alpha} \cos \delta$) and North ($\mu_N \equiv
\mu_{\delta}$), respectively.  Listed uncertainties are formal
$1\sigma$ measurement errors. Jones \etal (1994) and Pedreros \etal
(2002) reported proper motion data for fields at many degrees from the
LMC CM. These measurements were used to estimate the proper motion of
the LMC CM as discussed in the Appendix. These estimates are more
accurate than those quoted in the original papers.}
\end{deluxetable}


\begin{deluxetable}{ccccccc}
\tablecaption{Internal Kinematics of the LMC\label{t:kinematics}} 
\tablehead{
\colhead{$R'/D_0$} & \colhead{$V$} & \colhead{$\Delta V$} &
\colhead{$\sigma$} & \colhead{$\Delta \sigma$} &
\colhead{$\Theta$} & \colhead{$\Delta \Theta$} \\
 & \colhead{($\kms$)} & \colhead{($\kms$)} & \colhead{($\kms$)} &
   \colhead{($\kms$)} & \colhead{(deg)} & \colhead{(deg)} \\ }
\startdata
0.009 & -27.9 &  4.5 & 14.2 &  1.3 &  96.4 & 11.8 \\
0.028 &  14.2 &  4.1 & 21.5 &  1.4 & 151.6 & 37.7 \\
0.044 &  25.2 &  4.0 & 19.7 &  1.1 & 120.9 & 21.8 \\
0.060 &  35.7 &  5.9 & 22.0 &  1.4 & 140.8 & 12.0 \\
0.081 &  57.3 &  4.2 & 17.0 &  1.1 & 141.2 &  7.9 \\
0.096 &  50.0 &  3.1 & 16.3 &  0.9 & 110.5 &  7.0 \\
0.113 &  39.4 &  5.3 & 16.3 &  0.9 & 142.4 & 14.0 \\
0.130 &  46.6 &  8.0 & 16.3 &  1.4 & 108.7 &  9.5 \\
0.148 &  55.8 & 11.6 & 22.1 &  2.3 & 132.0 & 12.1 \\
0.178 &  32.1 & 16.5 & 21.3 &  2.2 & 123.2 & 38.7 \\
\enddata
\tablecomments{Radial profiles for the LMC of the rotation velocity 
$V(R')$, line-of-sight velocity dispersion $\sigma(R')$, and
kinematical line-of-nodes position angle $\Theta(R')$, as calculated
in Section~\ref{s:kindisk} and shown in Figure~\ref{f:rotcurve}.  All
quantities are listed with formal $1\sigma$ random errors. The radius
$R'$ is the cylindrical radius in the disk, which in column~(1) is
expressed in units of the LMC distance $D_0$ (which equals $50.1
\kpc$ for $m-M = 18.50$). The quantity $V$ is the in-plane velocity as
defined in Section~\ref{s:vmath}.}
\end{deluxetable}



\end{document}